\documentclass[aps,prc,amsfonts,preprintnumbers,superscriptaddress,showpacs,nofootinbib]{revtex4-1}
\usepackage{epsfig}
\usepackage{graphics}
\usepackage{amsmath}
\usepackage{amssymb}
\usepackage{dsfont}
\usepackage{hyperref}
\usepackage{bm}
\usepackage{color}
\newcommand{\beq}{\begin{equation}}
\newcommand{\eeq}{\end{equation}}
\newcommand{\beqa}{\begin{eqnarray}}
\newcommand{\eeqa}{\end{eqnarray}}
\newcommand{\nn}{\nonumber \\ }

\begin{document}
\title{Nucleon-nucleon interaction in chiral EFT with a finite cutoff: explicit perturbative renormalization at next-to-leading order}

\author{A.~M.~Gasparyan}
\email[]{Email: ashot.gasparyan@rub.de}
\affiliation{Ruhr-Universit\"at Bochum, Fakult\"at f\"ur Physik und
        Astronomie, Institut f\"ur Theoretische Physik II,  D-44780
        Bochum, Germany}
\affiliation{NRC “Kurchatov Institute” - ITEP, B. Cheremushkinskaya 25, 117218 Moscow, Russia}
      
\author{E.~Epelbaum}
\email[]{Email: evgeny.epelbaum@rub.de}
\affiliation{Ruhr-Universit\"at Bochum, Fakult\"at f\"ur Physik und
        Astronomie, Institut f\"ur Theoretische Physik II,  D-44780
        Bochum, Germany}
        
\begin{abstract}
We present a study of two-nucleon scattering
in chiral effective field theory with a finite
cutoff to next-to-leading order in the chiral expansion.
In the proposed scheme, the contributions of the lowest-order
interaction to the scattering amplitude are summed up to
an arbitrary order, while the corrections beyond leading order are iterated only once.
We consider a general form of the regulator for the leading-order potential
including local and non-local structures.
The main objective of the paper is to address formal aspects of
renormalizability within the considered scheme. In particular, we provide
a rigorous proof, valid to all orders in the iterations of the
leading-order potential, that power-counting breaking terms originating from the integration regions with momenta of the order
of the cutoff can be absorbed into the renormalization of the
low energy constants of the leading contact interactions.
We also demonstrate that the cutoff dependence of the scattering
amplitude can be reduced by perturbatively subtracting the regulator 
artifacts  at next-to-leading order. 
The obtained numerical results for phase shifts in $P$- and higher
partial waves confirm the applicability of our scheme for  
nucleon-nucleon scattering.
\end{abstract}

\maketitle

\section{Introduction}
Few-nucleon systems were first considered in the framework of chiral effective 
field theory by Weinberg in Refs.~\cite{Weinberg:1990rz,Weinberg:1991um}. 
For processes involving two and more nucleons, 
Weinberg suggested to apply the systematic power counting to the effective potential 
(instead of the amplitude as done for 
purely pionic or single-nucleon systems), which is
defined as a sum of all $n$-nucleon-irreducible diagrams.
To obtain the scattering amplitudes or other observables, 
one needs to solve the Lippmann-Schwinger or Schr\"odinger equation.
Since then, there has been enormous progress in the realization of these ideas using, 
however, quite different approaches and philosophies.
So far, there is no consensus in the community 
as to  which approach is most efficient and suits best the principles of effective
field theory. The descriptions of various schemes and methods 
can be found, e.g., in Refs.~\cite{Ordonez:1992xp,Kaiser:1997mw,
Park:1998cu,Gegelia:1998xr,
Birse:1998dk,Birse:2005um,Lutz:1999yr,Frederico:1999ps,
Higa:2003jk,Higa:2003sz,Epelbaum:1998ka,Epelbaum:1999dj,Kaplan:1998tg,
Kaplan:1998we,Fleming:1999ee,Beane:2000wh,Beane:2001bc,Kaiser:2001pc,PavonValderrama:2003np,PavonValderrama:2005uj,Nogga:2005hy,Epelbaum:2006pt,
Djukanovic:2007zz,Long:2007vp,Yang:2009pn,Valderrama:2009ei,Epelbaum:2009sd,
Timoteo:2010mm,Valderrama:2011mv,Albaladejo:2011bu,Long:2011xw,Epelbaum:2012ua,Gasparyan:2012km,
Albaladejo:2012sa,Ren:2016jna,Entem:2016ipb,Behrendt:2016nql,PavonValderrama:2016lqn,Reinert:2017usi,Entem:2017gor,Epelbaum:2017byx,Epelbaum:2018zli,Kaplan:2019znu,Epelbaum:2020maf}.
For review articles and references see
Refs.~\cite{Bedaque:2002mn,Epelbaum:2008ga,Machleidt:2011zz,
Epelbaum:2012vx,Epelbaum:2019kcf,Hammer:2019poc}.

Chiral effective theory is an effective field theory (EFT)
of QCD. It is based on the expansion of physical quantities around 
the chiral and zero-energy limits.
The expansion parameter $Q$ is given by the ratio of the soft scale 
(pion mass $M_\pi$ and $3$-momenta of interacting particles $p$) denoted by $q$ 
and hard scale $\Lambda_{b}$: $Q=\frac{q}{\Lambda_{b}}$. 
The hard chiral symmetry breaking scale can be roughly identified 
with the mass of the lightest meson (excluding Goldstone bosons -- pions), namely $\rho$-meson: $M_\rho\sim 770$ MeV.
The main ingredient of chiral EFT is 
the most general effective Lagrangian containing pion and  nucleon fields 
(we restrict ourselves to the case of two-flavor QCD)
and consistent with symmetries of the underlying
theory. It can be organized as a series of local terms with an increasing 
number of derivatives and/or powers of the quark mass insertions.
Low-energy constants (LECs) accompanying the interaction terms in the
effective Lagrangian account for effects of higher-energy physics.

In the purely pionic or single-nucleon sectors, the $S$-matrix is obtained 
from the effective Lagrangian by means of perturbation theory.
In the two-nucleon case, a strictly perturbative expansion of the scattering 
amplitude is not applicable as otherwise the deuteron bound state
could not be described. As argued by Weinberg \cite{Weinberg:1990rz,Weinberg:1991um}, the reason for the
failure of perturbation theory can be attributed to the enhancement 
of two-nucleon (2N) reducible diagrams by factors of $m_N/q$, where $m_N$ is the nucleon mass, 
as compared to the irreducible graphs with the same vertices and the
number of loops. From the naive dimensional analysis,  
the leading contributions at order $Q^0$ are the one-pion exchange and 
non-derivative four-nucleon contact terms. The 2N-irreducible one-loop graphs 
with the leading-order (LO) vertices appear at the chiral order $Q^2$, 
whereas the 2N-reducible one-loop diagrams are of order $m_N
Q^2/q=m_N q/\Lambda_{b}^2$.
In few-nucleon applications, 
the latter factor is often counted as $\mathcal{O}(Q^0)$ (i.e. $q/m_N\sim Q^2$), which is justified, 
in particular, by large numerical factors typically accompanying it.
This means that all multiple iterations of the leading-order diagrams 
with intermediate two-nucleon states are of the same order and
must be resummed in a non-perturbative manner. The contributions of
higher-order terms in the chiral expansion can be formally treated 
perturbatively or non-perturbatively depending on the chosen scheme.

One should mention that within the so-called  
KSW approach~\cite{Kaplan:1996xu,Kaplan:1998tg,Kaplan:1998we,Savage:1998vh},
an attempt was made to resum
only the iterations of the leading contact terms, 
whereas the iterations of the one-pion exchange were treated perturbatively.
However, the convergence  of such an approach 
is not satisfactory~\cite{Gegelia:1998ee,Gegelia:1999ja,Cohen:1999iaa,Fleming:1999ee},
which is a strong indication that also the leading-order one-pion 
exchange should be taken into account non-perturbatively if one aims at
an efficient and convergent scheme. For a related recent work see Ref.~\cite{Kaplan:2019znu}.

The series of the resummed leading-order contributions 
(in particular, of the one-pion exchange) contains an infinite number 
of divergent terms with increasing power of divergence. Therefore, one needs to introduce an 
infinite number of counter terms in the effective Lagrangian in order to absorb them.
In other words, the leading-order nucleon-nucleon amplitude 
defined this way  is not renormalizable in the standard sense.
A natural solution is 
to regularize this infinite series by introducing a regulator in the form of a cutoff.

Introducing a regulator requires special care.
First of all, a regulator may break the symmetries  
(Lorentz, gauge, chiral) of the effective Lagrangian and, 
therefore, of the underlying theory.
To avoid it, one can, in principle, introduce a regulator 
in the form that formally preserves all the symmetries using, 
e.g., the method of higher covariant derivatives
\cite{Slavnov:1971aw}, see
Refs.~\cite{Djukanovic:2006mc,Long:2016vnq,Behrendt:2016nql} for
applications in the two-nucleon sector of chiral EFT.
One should also keep in mind that the introduction of a cutoff may
lead to regulator artifacts. 

One proposed strategy of eliminating cutoff dependence and artifacts
amounts to taking the cutoff much larger than the hard scale $\Lambda_b$ or even
infinitely large,  see e.g.~\cite{Nogga:2005hy}.
In such an approach, one has to introduce a contact term at leading order 
in each attractive spin-triplet partial wave where the
one-pion-exchange contribution is regarded as non-perturbative 
in order to remove a strong cutoff dependence 
(and even more contact terms at higher orders \cite{Long:2011xw,Long:2011qx}).
This scheme has been criticized in \cite{Epelbaum:2009sd,Epelbaum:2018zli,Epelbaum:2020maf}, 
in particular, because it does not provide an explicit transition 
to the regime with a perturbative one-pion exchange
where one would need an infinite number of counter terms 
to absorb all divergent terms in the amplitude if the cutoff is infinite, or
to remove the terms proportional to positive powers of $\Lambda$ if the cutoff is finite.
Moreover, in such an approach, in attractive spin-triplet channels, 
a large (or infinite) number of spurious bound states is generated,
which is at odds with the spectrum of the underlying theory.
Therefore, we will not consider this scheme in the present work.

An alternative way is to set the cutoff value of the order of or smaller
than the hard scale.
In what follows, we will use a term ``finite cutoff'' assuming
this prescription.
The advantage of the choice of a cutoff to be of order 
$\Lambda_{b}$~\cite{Lepage:1997cs,Gegelia:1998iu,Gegelia:2004pz,Epelbaum:2006pt} 
is that all positive powers of $\Lambda$ appearing
in the iterations of the leading order potential 
are compensated by $\Lambda_{b}$ in the form
$(\Lambda / \Lambda_{b})^n$ and do not blow up.
The drawback of the ``small'' cutoff  is that it distorts the analytic structure of the amplitude
and there is a rather narrow window for its acceptable values if one wants to avoid artifacts.
In this work, we propose a scheme where the effect of the regulator is
systematically 
removed order by order in perturbation theory, so that smaller values of the cutoff
can be used.

In calculations beyond leading order, a finite cutoff generates
contributions that violate power counting.
This is caused by the fact that typical integrals converge at momenta $p\sim\Lambda$ 
so that the terms such as $p^2/\Lambda_{b}^2$ become of order
$\sim\Lambda^2/\Lambda_{b}^2\sim Q^0$ instead of $Q^2$.
Usually, when working with a finite cutoff, one iterates
subleading contributions  by solving the Lippmann-Schwinger or Schr\"odinger equation 
(see Refs.~\cite{Reinert:2017usi,Entem:2017gor,Reinert:2020mcu} for the most recent 
high-accuracy-level implementations) and \emph{assumes} that power-counting breaking terms can be 
absorbed by renormalizing the relevant contact interactions.
The renormalization is performed implicitly by adjusting the strength of the contact terms numerically.
However, it is important to show whether such a renormalization can be realized 
explicitly and the power counting is indeed restored, so that the
observed good agreement with experimental data is not accidental given
the large number of adjustable parameters.
It is also important to understand under which conditions the renormalization is possible.
Such an understanding is necessary, in particular, if one aims at extending
the calculations beyond the two-nucleon system, e.g.~to few-nucleon
dynamics and/or few-nucleon processes involving electroweak probes.

The usual requirement for a theory to be renormalizable is a possibility 
to absorb all ultraviolet divergences appearing in the $S$-matrix into a redefinition (renormalization) of the parameters
of the effective Lagrangian without modifying its structure, i.e.~respecting all
symmetries of the theory. Obviously, introducing a finite cutoff
automatically removes all ultraviolet divergences from the scattering amplitude.
Nevertheless, as stated above,
the problem is shifted to the appearance of positive powers of the cutoff
in places where the power counting implies positive powers of the soft scale.
Therefore, it is natural to extend the notion of renormalizability for
the case at hand by requiring that all above mentioned power-counting
breaking terms are absorbable by shifts (renormalization)
of the low-energy constants of lower orders.

In this paper, we address the issue of the renormalizability 
in the above sense at next-to-leading (NLO) chiral order, i.e.~$\mathcal{O}(Q^2)$.
We prove the renormalizability to all orders in the leading-order interaction $V_0$
by means of the Bogoliubov-Parasiuk-Hepp-Zimmermann (BPHZ)
\cite{Bogoliubov:1957gp,Hepp:1966eg,Zimmermann:1969jj}  subtraction procedure in the $S$-waves. 
In higher partial waves, the renormalization at the considered chiral order
works automatically.
This result is useful in those channels where the series in $V_0$
converges (albeit possibly much slower than the chiral expansion).
We allow for a rather general form of the regulator for $V_0$,
including both local and non-local ones. Moreover, we also allow for the inclusion 
into the leading order interaction of contact terms
quadratic in momenta in addition to the one-pion-exchange potential
and derivativeless contact interactions. This feature may be
particularly useful for approaches relying on a perturbative
inclusion of higher-order corrections, where the unitarity of the
scattering amplitude is maintained only approximately.  

It should be emphasized that providing 
the renormalizability of the amplitude to all orders in $V_0$ 
does not necessarily imply the renormalizability in the
non-perturbative regime, where the series in $V_0$ does not converge.
This means that our results are not directly applicable in the 
channels that feature strong
non-perturbative effects, or where the perturbative expansion
is not expected to be efficient.
In particular, this happens in 
$^1S_0$, $^3S_1-{^3D}_1$ and $^3P_0$ partial waves.
Nevertheless, such a perturbative treatment is very instructive also in understanding the 
non-perturbative case and will serve as a crucial ingredient in our analysis of the non-perturbative 
renormalization in a subsequent publication.

Our paper is organized as follows.
In Sec.~\ref{Sec:formalism}, we describe our formalism based on the effective 
Lagrangian, introduce the effective potential and discuss the
calculation of the scattering amplitude.
In Sec.~\ref{Sec:Power_counting} we demonstrate the validity of the power counting
for the LO and NLO interactions in higher partial waves.
The case of the $S$-waves and the problem of renormalization
and the power-counting restoration
is addressed separately in Sec.~\ref{Sec:perturbarive_NLO_Swave}.
The illustration of our formal findings for the case of the realistic nucleon-nucleon interaction is presented in Sec.~\ref{Sec:results}. 
The paper ends with a summary of the main findings.
Useful bounds on the effective potential and their derivation are collected in Appendix.

\section{Formalism}
\label{Sec:formalism}

\subsection{Effective Lagrangian and power counting}
\label{Sec:Lagrangian}
An effective field theory is based on the most general effective
Lagrangian consistent with the symmetries of the underlying theory \cite{Weinberg:1978kz}.
In chiral EFT, the relevant Lagrangian can be written as the sum
of the pure pionic terms, the single nucleon terms, the two-nucleon interactions,
etc.:
\begin{align}
\mathcal{L}_{\rm eff} = \mathcal{L}_{\pi}^{(2)} + 
\mathcal{L}_{\pi}^{(4)} + \mathcal{L}_{\pi N}^{(1)} +
\mathcal{L}_{\pi N}^{(2)} +
\mathcal{L}_{NN}^{(0)} +\mathcal{L}_{NN}^{(2)}+\dots \,, 
\label{Eq:effective_Lagrangian}
\end{align}
where the superscripts refer to the chiral order, i.e. 
to the number of derivatives and/or insertions of the pion mass. 
The explicit form of the terms in the effective Lagrangian
can be found in Ref.~\cite{Gasser:1983yg} (for the pionic part),
in Ref.~\cite{Fettes:2000gb} (for the pion-nucleon part)
and in Ref.~\cite{Girlanda:2010ya} (for the nucleon-nucleon part).
For the present study, the detailed structure of the Lagrangian
and its particular realization is irrelevant.

The Lagrangian can be split into the (renormalized) kinetic and
interaction terms:
\beqa
\mathcal{L}_{\rm eff} (x) &=:&\mathcal{L}_\text{kin}^N(x)+
\mathcal{L}_\text{kin}^\pi(x)+\mathcal{L}_\text{int}(x)\,,\nonumber\\
 \mathcal{L}_\text{kin}^N(x) &= & N^\dagger(x) \Big(i \partial_0
 +\frac{\vec\nabla^2}{2m_N} \Big) N(x)\,,\quad 
 \mathcal{L}_\text{kin}^\pi(x)=\frac{1}{2}\partial_\mu{\bm \pi}(x)
   \cdot\partial^\mu{\bm \pi}(x) -{\frac{1}{2}} M_\pi^2\, {\bm \pi}(x)^2,
\eeqa
where $N(x)$ is the large component of the nucleon field (we use the
non-relativistic form of the Lagrangian) and ${\bm \pi}(x)$ is the pion field.

For estimating various contributions to the nucleon-nucleon scattering amplitude we adopt the standard Weinberg's power counting \cite{Weinberg:1991um} as a starting point. However, to be more general, we do not
exclude possible promotions of formally higher-order contributions to 
lower orders.
In Weinberg's power counting, the chiral order
for a potential contribution, i.e. the one corresponding
to two-nucleon-irreducible diagrams (or more precisely to diagrams that do not possess the two-nucleon unitarity cut) is given by
\begin{align}
 D=2L+\sum_{i}\left(d_i+\frac{n_i}2-2\right)\,,
\end{align}
where $L$ is the number of loops, the sum runs over 
all vertices of the diagram, $n_i$ is the number of nucleon lines 
in vertex $i$ and $d_i$ is the number of derivatives and the pion-mass
insertions at vertex $i$.

For the 2N-reducible amplitude with the potential components
$\mathcal{V}^{(D_1)}$, $\mathcal{V}^{(D_2)}$,\dots,$\mathcal{V}^{(D_k)}$
of orders $D_1$, $D_2$,\dots,$D_k$,
\begin{align}
 \mathcal{V}^{(D_1)}\mathcal{G}\mathcal{V}^{(D_2)}\mathcal{G}\dots
 \mathcal{G}\mathcal{V}^{(D_k)} \equiv
 \int\left[\prod_{i=1}^{k-1} \frac{d^4p_i}{(2\pi)^4}\mathcal{G}(p_i^{\alpha_i},P^\rho)\right]
 \mathcal{V}^{(D_1)}(p'^\nu,p_1^{\alpha_1})\mathcal{V}^{(D_2)}(p_1^{\alpha_1},p_2^{\alpha_2})
 \dots \mathcal{V}^{(D_k)}(p_{k-1}^{\alpha_{k-1}},p^\mu)\,,
\end{align}
where, $P^\rho$ is the total $4$-momentum of the two-nucleon system
and $p^\mu$, $p'^\nu$, $p_i^{\alpha_i}$ are the $4$-momenta of the first nucleon in the 
initial, final and intermediate state, respectively,
the chiral order is given by the sum
\begin{align}
 D=D_1+D_2+\dots+D_k\,.
\end{align}
This enhancement is due to the infrared (pinch) singularity of 
the two-nucleon propagators $\mathcal{G}(p_i,P)$  \cite{Weinberg:1991um}
appearing when both nucleons are on shell.
Therefore, the leading-order potential $\mathcal{V}^{(0)}$ has to be 
iterated to an infinite order,
\begin{align}
 \mathcal{V}^{(0)}+\mathcal{V}^{(0)}\mathcal{G}\mathcal{V}^{(0)}+\mathcal{V}^{(0)}\mathcal{G}\mathcal{V}^{(0)}\mathcal{G}\mathcal{V}^{(0)}+\dots\,,
\end{align}
whereas the subleading-order potentials can be treated perturbatively.
This is the scheme that we adopt in the present work.

Typically, the non-perturbative resummation of the potential contributions
is performed using static (i.e.~independent of the zeroth components of the momenta)
potentials. Moreover, because of an infinite number of divergent
diagrams appearing in the course of such a resummation, 
a regulator has to be introduced.

In order to include the regularized static potential contribution
on the level of the Lagrangian, we add and subtract  the 
non-local potential part to $\mathcal{L}_\text{int}$
\begin{align}
 \mathcal{L}_\text{int}=\mathcal{L}_V+\left(\mathcal{L_\text{int}}- \mathcal{L}_V\right)\,,
\end{align}
with 
\begin{align}
\mathcal{L}_V(x)&=-\int  d\vec y \,d\vec y\,'
\frac{1}{2} N^{\dagger}_{j_1}(x_0,\vec x-\vec y\,'/2) 
N^{\dagger}_{j_2}(x_0,\vec x+\vec y\,'/2)
V(\vec y\,',\vec y)_{j_1,j_2;i_1,i_2}
N_{i_2}(x_0,\vec x+\vec y/2)N_{i_1}(x_0,\vec x-\vec y/2)\,,
\label{Eq:L_V}
\end{align}
where $i_1,i_2,j_1,j_2$ are the combined spin and isospin indices of the 
corresponding nucleons.
The potential function $V(\vec x\,',\vec x)$ is the Fourier
transform of the center-of-mass (c.m.) plane-wave potential $V(\vec p\,',\vec p)$:
\begin{align}
V(\vec x\,',\vec x)=\int\frac{d^3p}{(2\pi)^3}\frac{d^3p'}{(2\pi)^3}
V(\vec p\,',\vec p)e^{i\vec p\cdot \vec x}e^{-i\vec p\,'\cdot \vec x\,'}\,,
\end{align}
where $\vec p$ ($\vec p\,'$) is the c.m. momentum of the initial (final) nucleon.
The case of a local potential corresponds to
\begin{align}
 V(\vec p\,',\vec p)=V(\vec q=\vec p\,'-\vec p)\,,\quad 
 V(\vec x\,',\vec x)=V(\vec x)\delta(\vec x\,'-\vec x)\,.
\end{align}
The potential $V$ can be regularized using various types of regulators:
local, non-local, semilocal or even a lattice regularization.

Now, we can organize the potential in accordance with the chiral expansion:
\begin{align}
 V=V^{(0)}+V^{(2)}+V^{(3)}+V^{(4)}+\dots
\end{align}
Bare potentials $V^{(i)}$ can be split into the renormalized parts $V_i$ and 
the counter terms $\delta V_i$:
\begin{align}
V^{(i)}=V_i+\delta V_i\,, 
\quad \delta V_i=\delta V_i^{(2)}+\delta V_i^{(3)}+\delta V_i^{(4)}+\dots
 \label{Eq:counter_terms}
\end{align}
The counter terms $\delta V_i^{(j)}$ ($j>i$) absorb the power counting violating terms
appearing at order $\mathcal{O}(Q^j)$. 
They have the form of polynomials in momenta (of the same power as present in $V_i$) and are either unregulated or regulated with
an arbitrarily large cutoff in order to make all integrals well defined.

One can easily iterate the renormalized leading order potential $V_0$
performing the integrations over the zeroth components of the intermediate
momenta (we suppress the spin and isospin indices):
\begin{align}
 V_0\sum_{n=0}^\infty\left(\mathcal{G}V_0\right)^n
=V_0\sum_{n=0}^\infty\left(G V_0\right)^n
=V_0(\vec p\,',\vec p)+
\sum_{n=1}^\infty\int\bigg[ \prod_{i=1}^n\frac{d^3p_i}{(2\pi)^3}
G(p_i;p_\text{on}) \bigg]
V_0(\vec p\,',\vec p_1)\dots V_0(\vec p_n,\vec p)\,,
\end{align}
since the potential $V_0$ is static.
The two-nucleon propagator $G$ is given by 
\begin{align}
G(p_i; p_\text{on})=\frac{m_N}{p_\text{on}^2-p_i^2+i \epsilon}\,,
\end{align}
where $p_i$ and $p_\text{on}$ are the absolute values of the
intermediate and on-shell $3$-momenta, respectively.

The potential terms $V^{(i)}$ correspond to the
two-nucleon-irreducible contributions $\mathcal{V}^{(i)}$ to the
scattering amplitude 
constructed from the original Lagrangian $\mathcal{L}_\text{int}$
so that we can write
\begin{align}
 \mathcal{V}^{(i)}=V^{(i)}+\delta\mathcal{V}^{(i)}\,.
\end{align}
The difference $\delta\mathcal{V}^{(i)}$ can originate from 
the regulator corrections (the difference of the regulated and unregulated potential, see below),
from explicit $1/m_N$ corrections and
from the non-static contributions:
\begin{align}
 \delta\mathcal{V}^{(i)}=\delta_\Lambda V^{(i)}
+\delta_{1/m_N} V^{(i)}
+\delta_\text{n.s.} V^{(i)}\,.
\end{align}
The non-static corrections are defined to vanish on-shell 
($p_0=\vec p\,^2/(2m_N)$); thus, they cancel the pinch
singularities in the nucleon propagators and contribute to higher orders
due to the absence of the infrared enhancement.
Therefore, all corrections contained  in
$\left(\mathcal{L_\text{int}}- \mathcal{L}_V\right)$ can be treated
perturbatively.
Note that the counter terms $\delta V_i^{(j)}$ do not contribute to $\delta\mathcal{V}^{(i)}$
because they are static (or can be chosen to be static using equations of motion) and unregulated.

In fact, non-potential contributions stemming from 
$\left(\mathcal{L_\text{int}}- \mathcal{L}_V\right)$,
i.e. the ones that cannot be cast into the form of the static $NN$ potential as in Eq.~\eqref{Eq:L_V},
can be eliminated from the nucleon-nucleon scattering amplitude at low energies.
This can be shown, e.g., by the method of unitary transformations~\cite{Epelbaum:1998ka,Epelbaum:1999dj}
or applying time-ordered perturbation theory~\cite{Pastore:2009is}.
Notice also a close connection of the considered matching to 
the methods of the potential non-relativistic QED and QCD
\cite{Caswell:1985ui,Pineda:1997bj,Pineda:1998kn,Brambilla:1999xf,Brambilla:2004jw}.
As long as we stay well below the pion production threshold,
all effects of radiation pions can be reduced to 
the contact interactions with a non-analytic $M_\pi$ dependence \cite{Mehen:1999hz,Mondejar:2006yu}.
In this work, we do not focus on the $M_\pi$ dependence of the nucleon-nucleon 
amplitude and can, therefore, ignore the $M_\pi$ dependence of the parameters in
the effective Lagrangian.

For illustration, we show how the non-static corrections to the one-pion-exchange
potential 
\begin{align}
\delta_\text{n.s.}\mathcal{V}^{(0)}_{1\pi}=
\mathcal{V}^{(0)}_{1\pi}-V^{(0)}_{1\pi}
=\mathcal{V}^{(0)}_{1\pi}-\mathcal{V}^{(0)}_{1\pi}
\Big|_{p_0=p'_0=\frac{p_\text{on}^2}{2m_N}} \,,
\end{align}
where we have ignored, for simplicity, all other corrections ($1/\Lambda$, $1/m_N$),
 are combined into various potential terms of higher
 orders. 
The non-static corrections start to appear at next-to-leading order, i.e. at order $\mathcal{O}(Q^2)$:
\begin{align}
T_\text{n.s.}^{(2)}=\sum_{m,n=0}^\infty(V^{(0)}_{1\pi} \mathcal{G})^m
\left(\delta_\text{n.s.}\mathcal{V}^{(0)}_{1\pi}
+\delta_\text{n.s.}\mathcal{V}^{(0)}_{1\pi}\mathcal{G}\,\delta_\text{n.s.}\mathcal{V}^{(0)}_{1\pi}\right)
 (\mathcal{G}V^{(0)}_{1\pi} )^n\,.
 \label{Eq:nonstatic_1pi}
\end{align}
The term quadratic in $\delta_\text{n.s.}\mathcal{V}^{(0)}_{1\pi}$
in Eq.~\eqref{Eq:nonstatic_1pi} is also of order $\mathcal{O}(Q^2)$
because there are terms where both factors $\delta_\text{n.s.}\mathcal{V}^{(0)}_{1\pi}$ cancel the pinch singularity in the same propagator $\mathcal{G}$.
Therefore, it is convenient to rearrange the corrections due to 
$\delta_\text{n.s.}\mathcal{V}^{(0)}_{1\pi}$ in the following way.
Summing all orders in $\delta_\text{n.s.}\mathcal{V}^{(0)}_{1\pi}$, we
first obtain the
exact formal expression for the non-static amplitude:
\begin{align}
 T_\text{n.s.}=\mathcal{V}^{(0)}_{1\pi}
\frac{1}{1-\mathcal{G}\mathcal{V}^{(0)}_{1\pi}}=
\left(V^{(0)}_{1\pi}+\delta_\text{n.s.} \mathcal{V}^{(0)}_{1\pi}\right)
\frac{1}{1-\mathcal{G}\left(V^{(0)}_{1\pi}+\delta_\text{n.s.} \mathcal{V}^{(0)}_{1\pi}\right)}\,.
\label{Eq:Tnonstatic}
\end{align}
For the on-shell amplitude, we get the fully static approximation
also if instead of replacing $\mathcal{V}^{(0)}_{1\pi}$ with 
$V^{(0)}_{1\pi}$, we replace $\mathcal{G}$ with 
\begin{align}
 \hat G(p^\mu,P^\nu)=G(p;p_\text{on})2\pi
\delta\left(p_0-\frac{\vec p\,^2}{2m_N}\right)\,.
\end{align}
The on-shell static amplitude is then given by
\begin{align}
T_\text{st}=V^{(0)}_{1\pi}
\frac{1}{1-\mathcal{G} V^{(0)}_{1\pi}}
=\mathcal{V}^{(0)}_{1\pi}
\frac{1}{1-\hat{G}\mathcal{V}^{(0)}_{1\pi}}\,.
\label{Eq:static_G}
\end{align}
This equivalence is an obvious consequence of the fact that integrals over the zeroth component can be easily performed when the integrand involves only static potentials.
Expanding Eq.~\eqref{Eq:Tnonstatic} in $\delta G = \mathcal{G}-\hat G$,
we obtain:
\beqa
  T_\text{n.s.}&=&
\mathcal{V}^{(0)}_{1\pi}
\left\{
\left[1-\delta\mathcal{G}\mathcal{V}^{(0)}_{1\pi}
\left(1-\hat{G}\mathcal{V}^{(0)}_{1\pi}\right)^{-1}\right]
\left(1-\hat{G}\mathcal{V}^{(0)}_{1\pi}\right)
\right\}^{-1} \nn 
&=&\mathcal{V}^{(0)}_{1\pi}\left(1-\hat{G}\mathcal{V}^{(0)}_{1\pi}\right)^{-1}
\sum_{n=0}^\infty
\left[
\delta\mathcal{G}\mathcal{V}^{(0)}_{1\pi}
\left(1-\hat{G}\mathcal{V}^{(0)}_{1\pi}\right)^{-1}\right]^n\nonumber\\
&=&T_\text{st}+\left(1-V^{(0)}_{1\pi}\hat G\right)^{-1}
\mathcal{V}^{(0)}_{1\pi}
\delta\mathcal{G}\mathcal{V}^{(0)}_{1\pi}
\left(1-\hat{G}V^{(0)}_{1\pi}\right)^{-1}
+\left(1-V^{(0)}_{1\pi}\hat G\right)^{-1}
\mathcal{V}^{(0)}_{1\pi}
\delta\mathcal{G}\mathcal{V}^{(0)}_{1\pi}
\delta\mathcal{G}\mathcal{V}^{(0)}_{1\pi}
\left(1-\hat{G}V^{(0)}_{1\pi}\right)^{-1}\nonumber\\
&&{}+\left(1-V^{(0)}_{1\pi}\hat G\right)^{-1}
\mathcal{V}^{(0)}_{1\pi}
\delta\mathcal{G}
\mathcal{V}^{(0)}_{1\pi}
\hat G 
\left(1-V^{(0)}_{1\pi}\hat{G}\right)^{-1}
\mathcal{V}^{(0)}_{1\pi}
\delta\mathcal{G}\mathcal{V}^{(0)}_{1\pi}
\left(1-\hat{G}V^{(0)}_{1\pi}\right)^{-1}+\dots\nonumber\\
&=&T_\text{st}+\sum_{m,n=0}^\infty\left(V^{(0)}_{1\pi}G\right)^{m}
\left(V_{2\pi}^{(2)}+V_{3\pi}^{(4)}+\dots\right)
\left(G V^{(0)}_{1\pi}\right)^{n}+
\sum_{m,n,k=0}^\infty\left(V^{(0)}_{1\pi}G\right)^{m}
V_{2\pi}^{(2)}
G\left(V^{(0)}_{1\pi}G\right)^k
V_{2\pi}^{(2)}
\left(G V^{(0)}_{1\pi}\right)^{n} \nn
&&{}+\dots\,,
\eeqa
where the quantities
\begin{align}
V_{2\pi}^{(2)}=\mathcal{V}^{(0)}_{1\pi}
\delta\mathcal{G}
\mathcal{V}^{(0)}_{1\pi}\Big|_{p_0=p'_0=\frac{p_\text{on}^2}{2m_N}}\,,\quad
V_{3\pi}^{(4)}=\mathcal{V}^{(0)}_{1\pi}
\delta\mathcal{G}
\mathcal{V}^{(0)}_{1\pi}
\delta\mathcal{G}
\mathcal{V}^{(0)}_{1\pi}\Big|_{p_0=p'_0=\frac{p_\text{on}^2}{2m_N}}
\end{align}
can be identified (up to $1/m_N$-terms) with the static potentials corresponding to the $2\pi$- and $3\pi$-exchange box diagrams.

\begin{figure}[tb]
\includegraphics[width=7cm]{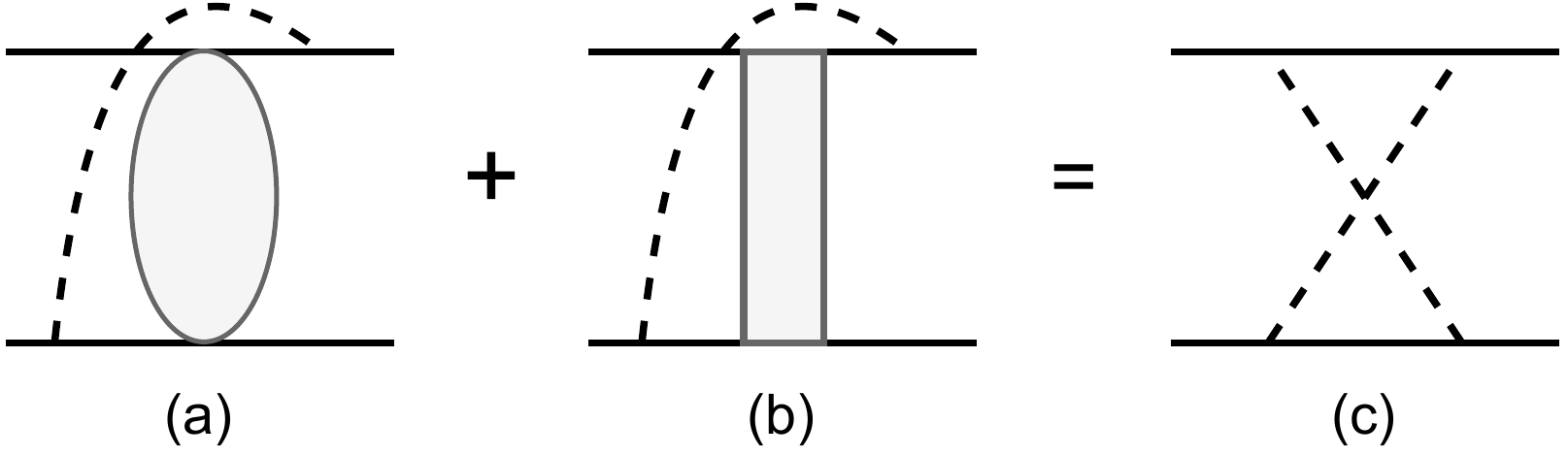}
\caption{An example of a diagram with the insertion of ${V}^{(0)}_{1\pi}$ (denoted by the ellipse) 
without an infrared enhancement and the same diagram with  the insertion of $\delta\mathcal{V}^{(0)}_{1\pi}$ 
(denoted by the rectangle). Dashed and solid lines refer to pions and
nucleons, respectively. The sum of these diagrams is the crossed box two-pion-exchange diagram. }
\label{Fig:Diagram}
\end{figure}

One should emphasize that the leading-order potential
$V^{(0)}$ (or, more precisely, the corresponding term in the Lagrangian)
is iterated only in  two-nucleon reducible diagrams,
whose contributions are enhanced.
When there is no infrared enhancement, 
$V^{(0)}$ needs not be iterated.
Moreover, in such a case, $V^{(0)}$ is not enhanced with respect to $\delta\mathcal{V}^{(0)}$.
Therefore, one has to take into account the whole contribution $V^{(0)}+\delta\mathcal{V}^{(0)}=\mathcal{V}^{(0)}$.
Consider, e.g., diagram~$(a)$ in Fig.~\ref{Fig:Diagram} with the insertion of the static
one-pion exchange potential $V^{(0)}_{1\pi}$, which is not
enhanced. Notice that the somewhat unusual type of diagrams in the left-hand side
of the depicted equation reflects the appearance of the nonlocal
``vertex'' $\mathcal{L}_V$ in the Lagrangian $\mathcal{L}_{\rm int}$.
The diagram~$(b)$ with the insertion of $\delta\mathcal{V}^{(0)}_{1\pi}$ is of the same order
and has to be included on the same footing.
The sum of diagrams~$(a)$ and~$(b)$ results in diagram~$(c)$, which is a regular $2\pi$-exchange crossed box diagram.

Eventually, the part of the effective Lagrangian that is relevant
for nucleon-nucleon amplitude is given by 
\begin{align}
 \mathcal{L}(x)=\mathcal{L}_\text{kin}^N(x)+\mathcal{L}_V(x)
+\mathcal{L}_{\delta_\Lambda V}(x)\,,
\end{align}
where $\mathcal{L}_{\delta_\Lambda V}(x)$ is the part corresponding
to the potential $\delta_\Lambda V$, which represents the correction
due to a finite cutoff:
\begin{align}
 \delta_\Lambda V=\sum_i\delta_\Lambda V^{(i)}\,,\quad \delta_\Lambda V^{(i)}:=V^{(i)}_{\Lambda=\infty}-V^{(i)}_\Lambda\,.
\end{align}
In $V^{(i)}_{\Lambda=\infty}$, one can set $\Lambda$ to be 
arbitrarily large but finite to make integrals mathematically well-defined.
If one keeps the $\delta_\Lambda V$ term in the Lagrangian,
the whole Lagrangian remains formally independent of both the form of
the regulator and the cutoff value. Of course, the nucleon-nucleon
amplitude calculated at any finite order in chiral EFT does not
feature this property.
This is caused by the fact that only the leading order potential is iterated
in a (possibly) non-perturbative manner, whereas the subleading
contributions are treated perturbatively.
Different choices of the regulator can lead to different
non-perturbative regimes as reflected by a different number of (quasi)bound states, etc..
Perturbative corrections at higher orders can obviously not change such a regime.
Therefore, one should choose the regulator such that the
non-perturbative regime matches empirical observations.
In particular, large values of the cutoff 
for the leading-order potential as considered e.g.~in Ref.~\cite{Nogga:2005hy} 
are excluded in our scheme
since they would lead to the appearance of spurious bound states
in attractive spin-triplet channels.
For low partial waves, this typically happens 
for cutoffs 
$\Lambda\gtrsim\Lambda_{b}$.

Nevertheless, taking into account the $\delta_\Lambda V$ terms will supposedly
make the remaining cutoff dependence within the chosen non-perturbative regime weaker.
In particular, if the leading-order potential is local and of Yukawa type,
the left-hand singularities originating from the regulator are always perturbative \cite{Martin:1961}. Thus, they will be shifted further away from the physical region
after each correction due to $\delta_\Lambda V$.
We will study numerically the effect of including the $\delta_\Lambda V$ terms
in connection with the cutoff dependence in Sec.~\ref{Sec:results}.
A formal analysis of such contributions and its interplay with the renormalization
of the Lagrangian parameters requires the treatment of terms in the
amplitude higher than NLO,
which is beyond the scope of this paper.

If one chooses the cutoff of order $\Lambda_{b}$, one can formally
expand $\delta_\Lambda V$ in $1/\Lambda$ and absorb the expanded terms
into the nucleon-nucleon contact interactions, although the symmetries
of the original Lagrangian will, in general, be lost.
Note that this cannot be done for a non-local regulator of 
the long-range part of the potential (such as the one-pion-exchange potential)
because of the mixture of long-range and 
short-range contributions.
Moreover, even in the cases when such an expansion is possible,
it might turn out to be more efficient to keep $\delta_\Lambda V$ unexpanded if 
the expansion converges slowly.
In particular, one may benefit from keeping the $\delta_\Lambda V$ term
if the cutoff is chosen to be smaller than $\Lambda_{b}$.
One can even perform the expansion of the amplitude in $\delta_\Lambda V$
independently from the chiral expansion in $1/\Lambda_{b}$.
The possibility to go to smaller values of $\Lambda$ is a welcome
feature, e.g., for many-body applications.

In the present study, we analyze the leading
and next-to-leading order nucleon-nucleon amplitude
 $T_0$ and  $T_2$ using the following notation:
\begin{align}
 &T_0=\sum_{n=0}^\infty T_{0}^{[n]}\,, \quad T_{0}^{[n]}=V_0 K^n\,,\nonumber\\
&T_2=\sum_{m,n=0}^\infty T_2^{[m,n]}\,,\quad T_2^{[m,n]}=\bar K^m V_2 K^n\,,\nonumber\\
&K=GV_0\,,\quad \bar K=V_0 G\,, \quad V_0=V^{(0)}\,,\quad V_2=V^{(2)}+\delta_\Lambda V_0\,, 
\label{Eq:T0_T2}
\end{align}
where we consider $\delta_\Lambda V_0$ to be of the same order as $V^{(2)}$.
The explicit expressions for $V_0$ and $V_2$ will be given in Secs.~\ref{Sec:LOpotential},~\ref{Sec:NLOpotential}.

We assume that the series for $T_0$ and $T_2$ in Eq.~\eqref{Eq:T0_T2}
converge even though one might need to take into account several
terms in the expansion to obtain an accurate result.
The truly non-perturbative treatment of the leading-order 
interaction based on the findings of the present work
will be considered in a subsequent publication.
The above convergence assumption implies that we should exclude from 
the numerical analysis the partial waves with strong non-perturbative
effects, namely 
$^1S_0$, $^3S_1$ and $^3P_0$ (although the formal treatment is
applicable to all partial waves).

Note that the analysis of the power counting in this section was 
so far based on formal arguments. Many contributions to the amplitude were
claimed to be of a certain (high) order even though they are mathematically
not well defined (i.e.~divergent).
Such a procedure can only be justified if one can prove that the renormalized
amplitude, i.e.~the one expressed in terms of the renormalized quantities, indeed
obeys the formal power counting.
One of the main objectives of our study is to prove that the amplitudes
$T_0$ and $T_2$ defined in Eq.~\eqref{Eq:T0_T2} obey the assumed
power counting meaning that all power-counting breaking contributions
in $T_2$ stemming from the integrations over large momenta can be absorbed into contact interactions.
For $T_2$, the only counter terms of a lower order are the leading-order $S$-wave momentum-independent
contact interactions. Therefore, in addition to $T_2$ defined in Eq.~\eqref{Eq:T0_T2}, 
there is a contribution from the counter terms $\delta V_0^{(2)}=\sum_{s,t=0}^\infty \delta V_0^{(2),[s,t]}$:
\begin{align}
 \delta T_2=\sum_{m,n=0}^\infty \delta T_2^{[m,n]}\,,\quad \delta T_2^{[m,n]}=\sum_{s=0}^m\sum_{t=0}^n \bar K^{m-s} 
 \delta V_0^{(2),[s,t]} K^{n-t}\,.
\end{align}
The explicit form of the counter terms will be specified in
Sec.~\ref{Sec:perturbarive_NLO_Swave}, see Eqs.~\eqref{Eq:R_operation1},~\eqref{eq:BPHZ_with_dots}.

\subsection{Leading-order potential}
\label{Sec:LOpotential}
The leading-order potential $V_0(\vec p\,',\vec p)$  
 that we consider is the sum of the regulated static
one-pion-exchange potential and the short-range part:
\begin{align}
 &V_0(\vec p\,',\vec p \, )=V^{(0)}_{1\pi,\Lambda}(\vec p\,',\vec p)+V^{(0)}_{\text{short},\Lambda}(\vec p\,',\vec p)\,.
\end{align}
\subsubsection{Short-range leading-order potential}
As already pointed out in the introduction, the short-range part of
the leading-order potential can be chosen to include 
the
momentum-independent contact interactions $V_{C_S}$, $V_{C_T}$ and
contact terms quadratic in momenta $V_{C_{1, \ldots , 7}}$ multiplied
by the power-like non-local form factor of an appropriate power $n$:
\begin{align}
&V^{(0)}_{\text{short},\Lambda}(\vec p\,',\vec p)=\sum_{i=S,T,1,..,7}
C_i \, V_{C_i}\, F_{\Lambda_i, n_i}( p\,', p)\,,
\label{Eq:short_range_nonlocal}
\end{align}
with
\begin{align}
&F_{\Lambda, n}( p\,', p)=F_{\Lambda, n}( p\,')F_{\Lambda, n}(p)
\,, \quad F_{\Lambda, n}(p)=\left[ F_{\Lambda}(p)\right]^n\,,\quad
F_{\Lambda}(p)=\frac{\Lambda^2}{p^2+\Lambda^2}\,.
\label{Eq:nonlocal_formfactor}
\end{align}
One can also introduce a regulator of a Gaussian form by replacing
$F_{\Lambda, n}(p)$ with 
\begin{align}
F_{\Lambda, \text{exp}}(p)=\exp{(-p^2/\Lambda^2)}\,. 
\label{Eq:F_exp}
\end{align}

The explicit form of the contact interactions is given
by~\cite{Epelbaum:2004fk}
\beqa
\{V_{C_S}, V_{C_T} \} &=& \{1\,,\; \vec{\sigma}_1 \cdot
\vec{\sigma}_2\} \nn
\{V_{C_1}, \ldots , V_{C_7} \} &=& \{q^2 \,, \; 
k^2 \,, \; 
q^2 ( \vec{\sigma}_1 \cdot \vec{\sigma}_2)\,, \; 
k^2  ( \vec{\sigma}_1 \cdot \vec{\sigma}_2) \,, \; 
\frac{i}{2}( \vec{\sigma}_1 + \vec{\sigma}_2)
\cdot ( \vec{k} \times \vec{q} \,)\,, \; 
(\vec{q}\cdot \vec{\sigma}_1 )(\vec{q}\cdot \vec{\sigma}_2 )\,, \; 
(\vec{k}\cdot \vec{\sigma}_1 )(\vec{k}\cdot \vec{\sigma}_2) \}\,,
\label{Eq:C_i}
\eeqa
where $\vec q = \vec p\,'-\vec p$ and $\vec k = ( \vec p\,'+\vec
p)/2$. 
The contact terms can be equivalently expressed in terms of the partial-wave 
projectors $P_i$, $i= {^1S_0}$, $^3S_1$,  $^3S_1\to{^3D_1}$, $^1P_1$, $^3P_1$, $^3P_0$, $^3P_2$~\cite{Epelbaum:2004fk}.

Alternatively, one could introduce local short-range interactions (for the terms that depend only on $\vec q$, except for the spin-orbit term)
using a different basis \cite{Gezerlis:2014zia}
\begin{align}
&V^{(0)}_{\text{short},\Lambda}(\vec p\,',\vec p \, )=\sum_{i=S,T,1,..,7}
\tilde C_i \, V_{\tilde C_i}\, F_{q,\Lambda_i, n_i}( \vec p\,', \vec p)\,,
\label{Eq:short_range_local}
\end{align}
with
\beqa
V_{\tilde C_i}&=&V_{C_i}\,,\quad \text{ for }\,  i=S,T,1,3,5,6\,, \nn
\{V_{\tilde C_2}, \; V_{\tilde C_4}, \; V_{\tilde C_7} \} &=& \{
q^2 \, {\bm \tau}_1 \cdot {\bm \tau}_2\,, \;
q^2 ( \vec{\sigma}_1 \cdot \vec{\sigma}_2) \, {\bm \tau}_1 \cdot {\bm
  \tau}_2 \,,\; 
(\vec{q}\cdot \vec{\sigma}_1 )(\vec{q}\cdot \vec{\sigma}_2 ) {\bm
  \tau}_1 \cdot {\bm \tau}_2 \}
\label{Eq:C_i_tilde}
\eeqa
and the  local regulator
\begin{align}
&F_{q,\Lambda,n}( \vec p\,', \vec p\, )=[F_\Lambda(q)]^n=
\left(\frac{\Lambda^2}{q^2+\Lambda^2}\right)^{n}\,,
\label{Eq:local_formfactor}
\end{align}
or with the regulator in the Gaussian form $F_{\Lambda,
  \text{exp}}(q)$.

\subsubsection{One-pion-exchange potential}
We split the one-pion-exchange potential into the 
triplet, singlet, and contact parts
\beq
V^{(0)}_{1\pi}= -\bigg(\frac{g_A}{2F_\pi}\bigg)^2 \bm\tau_1 \cdot \bm \tau_2\, 
\frac{\vec{\sigma}_1 \cdot\vec{q}\,\vec{\sigma}_2\cdot\vec{q}}{q^2 + M_\pi^2} 
=:V^{(0)}_{1\pi,\text{t}}+V^{(0)}_{1\pi,\text{s}}+V^{(0)}_{1\pi,\text{ct}}\,,
\label{Eq:V_1pi_0}
\eeq
with 
\beqa
V^{(0)}_{1\pi,\text{s}}&=&\bigg(\frac{g_A}{2F_\pi}\bigg)^2 \bm\tau_1 \cdot \bm \tau_2\,
\frac{(\vec{\sigma}_{1} \cdot\vec{\sigma}_{2}-1 )}{4}\frac{M_{\pi}^{2}}{q^{2}+M_{\pi}^{2}} \,,\nn
V^{(0)}_{1\pi,\text{ct}}&=&-\bigg(\frac{g_A}{2F_\pi}\bigg)^2 
\bm\tau_1 \cdot \bm \tau_2\,  \frac{(\vec{\sigma}_{1} \cdot\vec{\sigma}_{2}-1 )}{4}\,.
\label{Eq:V_1pi}
\eeqa
All three parts can be regularized separately.
The contact part $V_{1\pi,\text{ct}}$ can be absorbed by the
leading-order $^1S_0$ contact term and thus needs not be considered separately.
The triplet and singlet potential can be regularized by means
of the non-local form factor (see Eq.~\eqref{Eq:nonlocal_formfactor}):
\begin{align}
V^{(0)}_{1\pi,\Lambda}(\vec p\,',\vec p \,
  )=V^{(0)}_{1\pi,\text{s}}(\vec p\,',\vec p\, )F_{\Lambda_{\text{s}}, n_\text{s}}( p\,', p)
+V^{(0)}_{1\pi,\text{t}}(\vec p\,',\vec p)F_{\Lambda_{\text{t}}, n_\text{t}}( p\,', p)\,,
\label{Eq:V_1pi_nonlocal}
\end{align}
or by means of the local regulator: 
\begin{align}
V^{(0)}_{1\pi,\Lambda}(\vec p\,',\vec p \,
  )=V^{(0)}_{1\pi,\text{s}}(\vec p\,',\vec p \,
  )F_{q,1\pi,\{\Lambda_{\text{s}}\}}(\vec p\,',\vec p \, )
+V^{(0)}_{1\pi,\text{t}}(\vec p\,',\vec p \,
  )F_{q,1\pi,\{\Lambda_{\text{t}}\}}(\vec p\,',\vec p \, )\,,
  \label{Eq:V_1pi_local}
\end{align}
with
\begin{align}
 &F_{q,1\pi,\{\Lambda_k\}}(\vec p\,',\vec p\, )=\prod_{k}
\frac{\Lambda_k^2-M_\pi^2}{q^2+\Lambda_k^2}\,.
\label{Eq:local_formfactor_1pi}
\end{align}
The local regulator for the one-pion-exchange potential is introduced in a
rather general form as a product of dipole form factors.
Since a single dipole form factor is sufficient to make the leading-order 
Lippmann-Schwinger equation regular, 
one can choose $\Lambda_1\sim\Lambda_b$ and
the cutoffs $\Lambda_k$ for $k>1$ 
can be chosen arbitrarily large.
(In the case of the non-local regulators, all cutoffs 
must be chosen to be of order $\Lambda_b$ to maintain the power counting in the NLO amplitude.
This is caused by a more singular ultraviolet behavior of the subtracted non-local LO potential,
see Sec.~\ref{Sec:nonlocal_formfactor_bounds}.)
The local Gaussian regulator for the one-pion-exchange potential is  
possible as well:
\begin{align}
F_{q,1\pi,\text{exp},\Lambda}(\vec p\,',\vec p \, )=\exp{\left[-(q^2+M_\pi^2)/\Lambda^2\right]}\,. 
\label{Eq:1pi_exchange_local_gaussian_regulator}
\end{align}

The reasons for an individual regularization of the parts of the one-pion-exchange potential are the following.
First of all, the singlet one-pion-exchange potential is regular in the sense that
its  iterations  in the spin-singlet partial waves do not require  a
regularization provided the LO potential consists solely of the
  one-pion exchange. Regularization is then only necessary to render the 
NLO contributions finite. However, the corresponding cutoff
can be chosen very large or even infinite in contrast to the spin-triplet partial waves.
Nevertheless,  for practical reasons, it may still be advantageous to regularize the one-pion-exchange potential
in such spin-singlet channels.
 
Removing the contact part from the singlet one-pion-exchange potential
as done e.g.~in Ref.~\cite{Reinert:2017usi}
allows one to avoid regularization artifacts in the case of a local form factor.
If we regularize $V_{1\pi,\text{ct}}$ together with $V_{1\pi,\text{s}}$
or simply the whole $V_{1\pi}$ by multiplying it with $F_\Lambda(q)$,
this will act as an ``antiregularization'' in the spin-singlet partial
waves beyond the $^1S_0$ one,
because the range of the potential in momentum space becomes equal to $\Lambda$ instead of $M_\pi$ with 
its strength (coupling constant) being fixed.  
Such an effect is most pronounced in the $^1P_1$ partial wave, where for 
reasonable values of the cutoffs one observes resonance-like structures close to threshold
in the leading order amplitude.

Note that such a splitting of the one-pion-exchange potential can be realized on the
level of the effective Lagrangian and also on the level of the local Lagrangian density
if one regards the potential as a sum of the exchanges of a single
pion and of heavier particles
with different spins.

\subsection{Next-to-leading-order potential}
\label{Sec:NLOpotential}
The next-to-leading-order potential $V_2(\vec p\,',\vec p)$ consists of the short-range part, the two-pion-exchange potential and
the corrections due to the presence of the regulators in the leading-order potential:
\begin{align}
 &V_2(\vec p\,',\vec p)=V^{(2)}_{2\pi}(\vec p\,',\vec
   p)+V^{(2)}_{\text{short}}(\vec p\,',\vec p)
   +\delta_\Lambda V^{(0)}(\vec p\,',\vec p)\,.
\end{align}

The short-range part of the next-to-leading-order potential is given by the sum of contact terms (see Eq.~\eqref{Eq:C_i}):
\begin{align}
&V^{(2)}_{\text{short}}(\vec p\,',\vec p)=\sum_{i=S,T,1,..,7}
C_{2,i} \, V_{C_i}\,.
\label{Eq:V2_short_range}
\end{align}
Some (or all) of these structures appear also in the LO potential, see Sec.~\ref{Sec:LOpotential}.
How these terms are split into the LO and NLO contributions depends on the chosen scheme.
  
The non-polynomial part of the two-pion-exchange potential
is given by~\cite{Epelbaum:2004fk} (note that, for convenience, we have subtracted some polynomial terms from $V^{(2)}_{2\pi}$ as compared to Ref.~\cite{Epelbaum:2004fk})
\begin{align}
V^{(2)}_{2\pi}(\vec p\,',\vec p)&= - \frac{\bm \tau_1 \cdot \bm \tau_2}{384 \pi^2 F_\pi^4}\,
\tilde L(q) \, \biggl[4M_\pi^2 (5g_A^4 - 4g_A^2 -1)
+ q^2(23g_A^4 - 10g_A^2 -1)
+ \frac{48 g_A^4 M_\pi^4}{4 M_\pi^2 + q^2} \biggr]\nonumber\\
&+ \frac{\bm \tau_1 \cdot \bm \tau_2}{8 \pi^2 F_\pi^4}\frac{g_A^4 M_\pi^2 q^2}{4 M_\pi^2 + q^2}
-\frac{3 g_{A}^{4}}{64 \pi^{2} F_{\pi}^{4}}\tilde  L(q)
\left[\vec{\sigma}_{1} \cdot \vec{q} \vec{\sigma}_{2} \cdot \vec{q}-q^{2}
 \vec{\sigma}_{1} \cdot \vec{\sigma}_{2}\right]\,,
\end{align}
where
\begin{align}
\tilde L(q) :=L(q)-L(0)=L(q)-1\,, \quad 
L(q)=\frac{1}{q} \sqrt{4 M_{\pi}^{2}+q^{2}} \log \frac{\sqrt{4 M_{\pi}^{2}+q^{2}}+q}{2 M_{\pi}}\,.
\end{align}

For the sake of brevity, we do not show explicitly possible regulators for the next-to-leading order potential. Nevertheless, all our estimates work for both unregulated and regulated versions of the potential, see Appendix~\ref{Sec:Bounds_NLO}. The regulator can be a combination of any local or non-local forms.
One can also use a spectral function regularization for the two-pion-exchange
potential by introducing a finite upper limit in the dispersion representation
of $\tilde L(q)$:
\begin{align}
 \tilde L(q)=q^2\int_{2M_\pi}^{\Lambda_\rho} \frac{d\mu}{\mu^2}\frac{\sqrt{\mu^2-4M_\pi^2}}{q^2+\mu^2}\,.
\end{align}

\subsection{Partial-wave decomposition and contour rotation}
It is convenient to express the nucleon-nucleon amplitude
in the $lsj$ basis~\cite{Erkelenz:1971caz}, in which all
momentum integrations become one-dimensional.
The potentials in the partial wave basis become $n_\text{PW}\times n_\text{PW}$  
matrices, where $n_\text{PW}=1$ ($n_\text{PW}=2$) for the uncoupled (coupled)
partial waves.

In order to avoid difficulties related with the principal value 
part of the two-nucleon propagator when
estimating the momentum-space integrals,
we perform all
integrations over the loop momenta in the partial-wave basis
along the complex contour $\mathcal{C}$, e.g.
\begin{align}
 V_0K=V_0G V_0
 =\int_0^\infty \frac{p''^2 dp''}{(2\pi)^3}
  V_0(p',p'')\frac{m_N^2}{p_\text{on}^2-p''^2+i\epsilon} V_0(p'',p)
  =\int_\mathcal{C}\frac{p''^2 dp''}{(2\pi)^3}
 V_0(p',p'')\frac{m_N^2}{p_\text{on}^2-p''^2}V_0(p'',p)\,,
\end{align}
where the contour $\mathcal{C}$ is given by $p''=|p''|e^{-i\alpha_{\mathcal{C}}}$,
$0\le|p''|<\infty$.
This method has been used in the literature, e.g, for handling three-body singularities \cite{Hetherington:1965zza,Aaron:1966zz,Cahill:1971ddy}.
This also applies to the integrals of the form
$V_0 K^n$, $\bar K^m V_2 K^n$, etc.

\begin{figure}[tb]
\includegraphics[width=5cm]{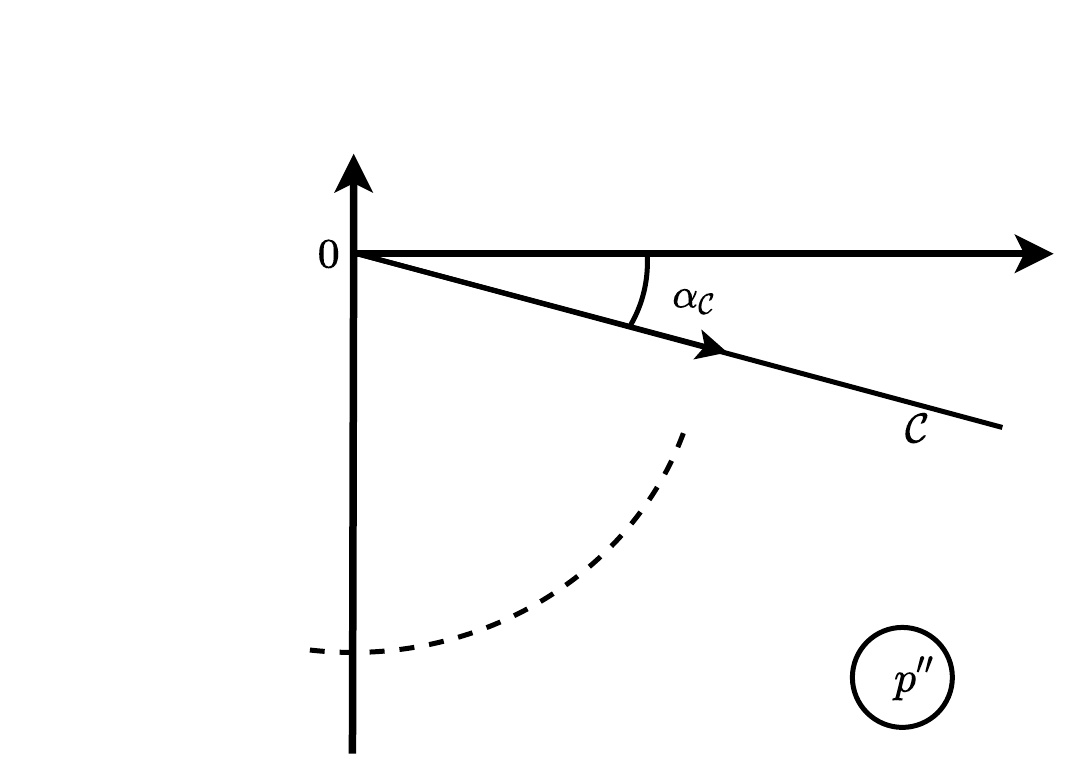}
\caption{Contour rotation in the $p''$ complex plane. 
The dashed arc indicates the singularities of the partial-wave one-pion-exchange potential.}
\label{Fig:Contour}
\end{figure}

The momentum $p$ ($p'$) can either be on shell $p=p_\text{on}$ or lie
on the complex contour. 
The contour-rotation angle $\alpha_{\mathcal{C}}$ is determined by the nearest
singularity of the integrand.
It originates from the one-pion-exchange
potential in the configuration when one momentum is on-shell
and the other momentum lies on $\mathcal{C}$.
If the maximal considered on-shell momentum is $\left(p_\text{on}\right)_\text{max}$, 
then $\tan(\alpha_{\mathcal{C}})<\frac{M_\pi}{\left(p_\text{on}\right)_\text{max}}$.
We can, therefore, choose
\begin{align}
 \alpha_{\mathcal{C}}=\frac{1}{2}\arctan\frac{M_\pi}{\left(p_\text{on}\right)_\text{max}}\,,
\end{align}
which is shown schematically on Fig.~\ref{Fig:Contour}.

\section{Power counting: Leading and next-to-leading orders}
\label{Sec:Power_counting}
In this section, we prove that the LO amplitude in all partial
  waves and the NLO contribution in $P$- and higher partial waves,
  which both do not
require renormalization, satisfy the chiral power counting.

\subsection{Dimensionless constants}
In all our estimates, various dimensionful quantities will be
expressed in terms of dimensionless constants of order one
with respect to momenta and hard and soft scales.
We denote them by $\mathcal{M}_i$, where the index $i$ 
indicates the origin of the constant or a quantity that it depends upon:
angular momentum, order of expansion, etc.
When such dimensionless constants appear in an equation, we
will implicitly 
assume them
to be of order one.
In some cases, we provide explicit values or upper bounds for those constants.
We do, however, not attempt to give the best estimate for such bounds
because they often depend on the details of a particular scheme,
whereas we want to keep the presentation general.
In practice, the estimates of the relevant constants can  in each
particular practical case  be most easily performed numerically.
We further emphasize that in certain cases, 
additional factors of $\pi$, $4\pi$, etc., are introduced to trace back angular and momentum integrations.
Finally, in some cases, we use the same names for constants in similar equations, which leads to no contradiction
because one can always choose the largest constant for all considered equations.

\subsection{Bounds on the potentials and the propagator}
The two-nucleon propagator $G(p{;p_\text{on}})=m_N/(p_\text{on}^2-p^2 )$
is obviously bounded by
\begin{align}
 |G(p{;p_\text{on}})|\le \mathcal{M}_G \frac{m_N}{|p^2|}\,,
 \label{Eq:bound_on_G}
\end{align}
with  $\mathcal{M}_G=1/\sin(2\alpha_{\mathcal{C}}) $.

Below, we list the relevant bounds on the LO and NLO potentials, which are derived in 
Appendix~\ref{Sec:LO_bounds} and Appendix~\ref{Sec:NLO_bounds}, respectively.

For the LO potentials with $l=0$ (in the case of coupled partial
waves, $l$ refers to the lowest orbital angular momentum), 
it holds for every matrix element:
\begin{align}
&\left|V_0(p',p)\right|< \mathcal{M}_{V_0,0}
V_{0,\text{max}}(p',p)\,,\quad
\left|V_0(p',p)\right|< \mathcal{M}_{V_0,0}
V_{0,\text{max}}(p,p')\,,
\label{Eq:bounds_V0_l_0_text}
\end{align}
with
\begin{align}
&V_{0,\text{max}}(p',p)
=\frac{8\pi^2 }{m_N \Lambda_V}\Big[F_{\tilde\Lambda}(|p'|-|p|)+F_{\tilde\Lambda}(|p'|)\Big]\,,
\label{Eq:V0max_text}
\end{align}
where $F_\Lambda(p)$ is the dipole form factor defined in Eq.~\eqref{Eq:nonlocal_formfactor},
$\tilde \Lambda=\Lambda$, 
except for the spin-singlet partial waves without short-range leading-order interactions, for which 
 $\tilde \Lambda=M_\pi$,
 and $\Lambda$ is the largest cutoff among all cutoffs used in the LO potential. 
Further, $\Lambda_V$ is the characteristic hard scale of the LO
potential given by a combination of $F_\pi$, $m_N$ and the
scale that governs the included contact interactions as explained in
Appendix \ref{Sec:AppD}.
We explicitly keep track of $\Lambda_V$ and discriminate it from $\Lambda_b$, because the ratio $\Lambda/\Lambda_V$
enters most of the estimates in this work. 

For the renormalizability proof in the $S$-waves carried out
  in the next section, we will perform subtractions of the potential.
The subtraction remainders of the LO potentials with $l=0$ 
\begin{align}
\Delta_p V_0(p',p)\equiv V_0(p',p)-V_0(p',0)\,,\quad
 \Delta_{p'} V_0(p',p)\equiv V_0(p',p)-V_0(0,p)\,,
 \end{align}
are bounded as
\begin{align}
&\left|\Delta_p V_0(p',p)\right|\le
\mathcal{M}_{V_0,0}\left|\frac{p}{p'}\right|^2V_{0,\text{max}}(p',p)\,
\,,\quad
\left|\Delta_{p'} V_0(p',p)\right|\le
\mathcal{M}_{V_0,0}\left|\frac{p'}{p}\right|^2V_{0,\text{max}}(p,p')\,
\,,
\label{Eq:bounds_Delta_V0_l_0_text1}
\end{align}
or, alternatively, as
\begin{align}
&\left|\Delta_p V_0(p',p)\right|\le
\mathcal{M}_{V_0,0}V_{0,\text{max}}(p',p)\,
\,,\quad
\left|\Delta_{p'} V_0(p',p)\right|\le
\mathcal{M}_{V_0,0}V_{0,\text{max}}(p,p')\,
\,.
\label{Eq:bounds_Delta_V0_l_0_text2}
\end{align}
For $l>0$, we have the inequalities
\begin{align}
&\left| V_0(p',p)\right|\le \mathcal{M}_{V_0,\tilde l}
\left|\frac{p}{p'}\right|^{\tilde l}V_{0,\text{max}}(p',p) \,
\,,\quad
\left| V_0(p',p)\right|\le \mathcal{M}_{V_0,\tilde l}
\left|\frac{p'}{p}\right|^{\tilde l}V_{0,\text{max}}(p,p') \,
\,,
\label{Eq:bounds_V0_l_text}
\end{align}
where  $\tilde l$ can be chosen to take any value compatible with $0\le \tilde l \le l$.

We split the NLO potential into two parts:
\begin{align}
 V_2(p',p)=:\hat V_2(p',p)+\tilde V_2(p',p)\,,\quad \hat V_2(p',p)=V_2(0,0)\,.
 \label{Eq:V2_tilde_text}
\end{align}
For the NLO potential
with $l=0$, it holds:
\begin{align}
 \left|\hat V_2(p',p)\right|\le \mathcal{\hat M}_{V_2,0} \frac{8\pi^2 }{m_N \Lambda_V}\frac{M_\pi^2}{\Lambda_{b}^2}\,,
 \label{Eq:bound_V_2_0_hat_text}
\end{align}
and
\begin{align}
&\left|\tilde V_2(p',p)\right| {\equiv}
\left|V_2(p',p)-V_2(0,0)\right|
\le \mathcal{M}_{V_2,0}
\left(|p|^2+|p'|^2\right)\tilde f_\text{log}(p',p) \,,
\label{Eq:bounds_V2_l_0_text}
\end{align}
with
\beqa
\tilde f_\text{log}(p',p) &=&\frac{8\pi^2}{m_N \Lambda_V \Lambda_{b}^2} f_\text{log}(p',p)\,,
\nn
 f_\text{log}(p',p) &=&\theta(|p|-M_\pi)\log\frac{|p|}{M_\pi}
+\theta(|p'|-M_\pi)\log\frac{|p'|}{M_\pi}+\log\frac{\tilde\Lambda}{M_\pi}+1\,.
\label{Eq:f_log_tilde_text}
\eeqa
For the subtraction remainders we obtain the inequalities
\begin{align}
&\left|\Delta_p\tilde V_2(p',p)\right|\le \mathcal{M}_{V_2,0}
|p|^2\tilde f_\text{log}(p',p) \,,\quad
\left|\Delta_{p'}\tilde V_2(p',p)\right|\le \mathcal{M}_{V_2,0}
|p'|^2\tilde f_\text{log}(p',p) \,.
\label{Eq:bounds_Delta_V2_l_0_text}
\end{align}
For $l>0$, $V_2 (0, 0) = 0$ so that $\tilde V_2=V_2$. The
employed bounds have the form
\begin{align}
&\left|V_2(p',p)\right|\le \mathcal{M}_{V_2,\tilde l}
\left|\frac{p}{p'}\right|^{\tilde l}|p'|^2\tilde f_\text{log}(p',p) \,,
\quad 
\left|V_2(p',p)\right|\le \mathcal{M}_{V_2,\tilde l}
\left|\frac{p'}{p}\right|^{\tilde l}|p|^2\tilde f_\text{log}(p',p) \,,
\label{Eq:bounds_V2_l_1_text}
\end{align}
where $1\le\tilde l \le l$.

\subsection{Perturbative LO amplitude}
\label{Sec:perturbarive_LO}
After these preparations, we are now in the position to verify the
power counting for the LO partial-wave amplitude for $l > 0$, which   
is given by the series
\begin{align}
 T_0=\sum_{n=0}^\infty T_{0}^{[n]}\,, \quad T_{0}^{[0]}=V_0\,,\quad T_{0}^{[n]}=V_0 K^n\,,
 \label{Eq:perturbative_sum}
\end{align}
or, more explicitly
\begin{align}
\left(T_{0}^{[n]}\right)_{j i}(p',p;p_\text{on})=&
\sum_{i_{1}, \ldots, i_n=1}^{n_\text{PW}}
\int \prod_{k=1}^n \frac{p_k^2 dp_k}{(2\pi)^3}
\left(V_0\right)_{j i_1}(p',p_1)
K_{i_1,i_2}(p_1,p_2;p_\text{on})\dots
K_{i_{n} i}(p_n,p;p_\text{on})\,,
\label{Eq:T_0_n}
\end{align}
where the subscripts of $T_0$, $V_0$ and $K$ denote the partial wave
index and $n_\text{PW} = 1,2$ is the number of channels.
We are now going to show by induction that 
\begin{align}
 \left|T_{0}^{[n]}(p',p;p_\text{on})\right|\le \mathcal{M}_{n} \frac{8\pi^2 }{m_N \Lambda_V}\,.
 \label{Eq:bound_T_0_n_0}
\end{align}

From Eq.~\eqref{Eq:bounds_V0_l_text} with $\tilde l=0$, it follows that
\begin{align}
 \mathcal{M}_{0}=2 \mathcal{M}_{V_0, 0}\,.
\end{align}
Assuming that Eq.~\eqref{Eq:bound_T_0_n_0} holds and
applying the bound~\eqref{Eq:bounds_V0_l_text} with $\tilde l=0$,
we get for $T_{0}^{[n+1]}$ (here, we suppress the 
channel indices):
\beqa
\left|T_{0}^{[n+1]}(p',p;p_\text{on})\right|&=&
\left|\int \frac{p''^2 dp''}{(2\pi)^3}
T_{0}^{[n]}(p',p'';p_\text{on})
G(p'';p_\text{on})V_0(p'',p)\right|\nonumber\\
&\le& \frac{8\pi^2 \mathcal{M}_{n} }{m_N \Lambda_V} n_\text{PW} \mathcal{M}_{V_0, 0}
\int \frac{|p''|^2 d|p''|}{(2\pi)^3}
\left|G(p'';p_\text{on})\right|V_{0,\text{max}}(p'',p)\nonumber\\
&{\equiv}&\frac{8\pi^2 \mathcal{M}_{n} }{m_N \Lambda_V} 
n_\text{PW}\mathcal{M}_{V_0, 0} I_{V_0}(p',p;p_\text{on}) \nn
&\le&
\frac{8\pi^2 \mathcal{M}_{n} }{m_N \Lambda_V} n_\text{PW}\mathcal{M}_{V_0, 0} 
\mathcal{M}_{I_{V_0}}
\frac{\tilde \Lambda}{\Lambda_V}\,.
\label{Eq:bound_T_0_n1}
\eeqa
The integral $ I_{V_0}(p',p;p_\text{on})$ is estimated in Appendix~\ref{Sec:integrals}.
Introducing the quantity 
\begin{align}
&\Sigma= n_\text{PW} \mathcal{M}_{V_0, 0} \mathcal{M}_{I_{V_0}}\frac{\tilde \Lambda}{\Lambda_V}\,,
\label{Eq:Sigma}
\end{align}
we finally arrive at the following bound
\begin{align}
&\left|T_{0}^{[n]}(p_\text{on})\right|\le \frac{8\pi^2}{m_N \Lambda_V}\mathcal{M}_{0}\,
\Sigma^n\,.
\label{Eq:bound_T_0_n}
\end{align}

For spin-triplet partial waves and for spin-singlet partial
waves featuring a short-range LO interaction, all terms $T_0^{[n]}$ are, as expected, of order $\mathcal{O}(Q^0)$ as long as the cutoff is of the order of the 
hard scale, $\Lambda\sim\Lambda_V$.
If $\Sigma<1$, the series for $T_0$ is perturbative, although 
for $\Sigma$ close to $1$, the convergence of this series might be very slow.
Obviously, the convergence gets better for smaller values of $\Lambda$ until
the cutoff artifacts become relevant.
For spin-singlet partial waves without short-range LO interactions,
we have a better convergence, $T_0^{[n]}\sim \left(M_\pi/\Lambda\right)^n$,
because the integrals converge at momenta $p\sim M_\pi$.

We do not provide here the estimates for $\Sigma$ for
individual partial waves although it is obvious that
$\Sigma$ will get smaller for larger values of $l$.
It is also straightforward to obtain a better bound for
$T_{0}^{[n]}$ for spin-triplet channels and  spin-singlet channels
with short-range interactions when $n<2l$, so that the unregularized loop
  integrals are ultraviolet convergent, by applying Eq.~\eqref{Eq:bounds_V0_l_text}
with $\tilde l=l$ for the integration regions $|p''|\ge p_\text{on}$:
\begin{align}
&\left|T_{0}^{[n]}(p_\text{on})\right|\le \frac{8\pi^2}{m_N \Lambda_V}
\mathcal{\tilde M}_{n,l} \left(\frac{p_\text{on}}{\Lambda_V}\right)^n\,.
\label{Eq:bound_T_0_n_l}
\end{align}
However, this does not affect our general conclusions about the
validity of the power
counting.

\subsection{Perturbative amplitude at NLO: Higher partial waves}
\label{Sec:perturbarive_NLO_higherPW}
In a complete analogy with the previous section, we can analyze the
NLO amplitude of order $m+n$ in $V_0$:
\begin{align}
T_2^{[m,n]}=\bar K^m V_2 K^n\,,
 \label{Eq:T_2_mn}
\end{align}
for partial waves with $l\ge 1$.
 In the following, we will show by induction that
\begin{align}
 \left|T_{2}^{[m,n]}(p',p;p_\text{on})\right|\le 
 \mathcal{M}_{T_2}^{m,n} 
 |p'||p|  \tilde f_\text{log}(p',p)\,.
 \label{Eq:bound_T_2_mn_l}
\end{align}
From Eq.~\eqref{Eq:bounds_V2_l_1_text} with $\tilde l=1$, it follows
\begin{align}
 \mathcal{M}_{T_2}^{0,0}=\mathcal{M}_{V_2, 1}\,.
\end{align}
Assuming that Eq.~\eqref{Eq:bound_T_2_mn_l} holds and
applying the bound~\eqref{Eq:bounds_V0_l_text} with $\tilde l=1$,
we obtain for $T_{2}^{[m,n+1]}$:
\beqa
   \left|T_{2}^{[m,n+1]}(p',p{;p_\text{on}})\right| &=& 
   \left|\int \frac{p''^2dp''}{(2\pi)^3}
 T_{2}^{[m,n]}(p',p''{;p_\text{on}}) G(p''{;p_\text{on}}) V_0(p'',p)\right|\nonumber \\
 &\le&  n_\text{PW} \mathcal{M}_{T_2}^{m,n} \mathcal{M}_{V_0,1}
 \int \frac{|p''|^2dp''}{(2\pi)^3}|p'||p''| \tilde f_\text{log}(p',p'')  
\left|G(p''{;p_\text{on}})\right|
\left|\frac{p}{p''}\right|V_{0,\text{max}}(p',p)\nonumber\\
&\le& n_\text{PW} \mathcal{M}_{T_2}^{m,n}\mathcal{M}_{V_0,1}|p'||p|
\int \frac{|p''|^2dp''}{(2\pi)^3}\tilde f_\text{log}(p',p'')  
\left|G(p''{;p_\text{on}})\right|V_{0,\text{max}}(p',p)\nonumber\\
&{\equiv}&  n_\text{PW} \mathcal{M}_{T_2}^{m,n}\mathcal{M}_{V_0,1}|p'||p|
I_{f_\text{log}}(p',p{;p_\text{on}}) \nn
&\le&  n_\text{PW} \mathcal{M}_{T_2}^{m,n}\mathcal{M}_{V_0,1}\mathcal{M}_{f_\text{log}}
\frac{\tilde\Lambda}{\Lambda_V} |p'||p|  \tilde f_\text{log}(p',p)\,,
\eeqa
and, analogously, for $T_{2}^{[m+1,n]}$.
The integral $I_{f_\text{log}}(p',p{;p_\text{on}})$ is estimated in Appendix~\ref{Sec:integrals}.
Introducing
\begin{align}
&\Sigma_{1}= n_\text{PW} \mathcal{M}_{V_0,1}\mathcal{M}_{f_\text{log}}\frac{\tilde\Lambda}{\Lambda_V}\,,
\label{Eq:Sigma1}
\end{align}
we obtain
\begin{align}
&\mathcal{M}_{T_2}^{m,n}=\mathcal{M}_{V_2,1}\Sigma_{1}^{m+n}\,.
\end{align}
Therefore, 
for the on-shell kinematics with $p=p'=p_\text{on}$, 
the amplitude $T_{2}^{[m,n]}(p_\text{on})$ is bounded by
\begin{align}
&\left|T_{2}^{[m,n]}(p_\text{on})\right|\le \frac{8\pi^2 \mathcal{M}_{V_2,1}}{m_N\Lambda_V} 
\Sigma_{1}^{m+n}
\frac{p_\text{on}^2}{\Lambda_{b}^2}f_\text{log}(p_\text{on},p_\text{on})\,,
\label{Eq:Power_Counting_T2_m_n_1}
\end{align}
where we have used Eq.~(\ref{Eq:f_log_tilde_text}). 
For the spin-triplet partial waves and for the spin-singlet partial
waves with a short-range LO interaction, we get
\begin{align}
 \left|T_{2}^{[m,n]}(p_\text{on})\right|\le \frac{8\pi^2 \mathcal{M}_{V_2,1}}{m_N\Lambda_V} 
\Sigma_{1}^{m+n}
\frac{p_\text{on}^2}{\Lambda_{b}^2}
\, \log{\frac{\Lambda}{M_\pi}}\,,
\label{Eq:Power_Counting_T2_m_n_triplet}
\end{align}
where we have used $p_\text{on}\le \mathcal{M}_{p_\text{on}} M_\pi$
and absorbed terms of order $1$ into the  $\log\Lambda/M_\pi$-term
in $f_\text{log}(p_\text{on},p_\text{on})$, thereby redefining
$\mathcal{M}_{V_2,1}$.
If we choose the cutoff to be of the order of the 
hard scale $\Lambda\sim\Lambda_V$,
all terms $T_{2}^{[m,n]}$ are of order $\mathcal O(Q^2\log Q)$
and do not violate power counting.
The appearance of the $\log\Lambda/M_\pi$-factor in Eq.~\eqref{Eq:Power_Counting_T2_m_n_triplet} as compared to
the bare potential $V_2$ can be traced back to the $\log p/M_\pi$ terms in $V_2$
and the fact that the integrals converge at momenta $p\sim\Lambda$.
As in the case of the leading order amplitude, one can improve
the bounds on $T_{2}^{[m,n]}$ for $n+m<2l$
using Eqs.\eqref{Eq:bounds_V2_l_1_text},~\eqref{Eq:bounds_V0_l_text}
for the integration momenta $|p''|\ge p_\text{on}$.
This does not modify our findings about the power counting for the NLO amplitude either. 

For spin-singlet partial waves without short-range LO interactions,
where the integrals converge at $|p''|\sim M_\pi$,
we have 
\begin{align}
&\left|T_{2}^{[m,n]}(p_\text{on})\right|\le \frac{8\pi^2 \mathcal{M}_{V_2,1}}{m_N\Lambda_V} 
\Sigma_{1}^{m+n}
\frac{p_\text{on}^2}{\Lambda_{b}^2}\,,
\label{Eq:Power_Counting_T2_m_n_singlet}
\end{align}
which gives
\begin{displaymath}
T_{2}^{[m,n]}\sim \frac{p_\text{on}^2}{\Lambda_{b}^2}\left(\frac{M_\pi}{\Lambda_V}\right)^{m+n}\,.
\end{displaymath}

To summarize, we have shown in this section that (i) the LO scattering
amplitude scales according to the power counting and (ii) the NLO
contribution to the amplitude in $P$- and higher partial waves does not
contain power-counting breaking  terms, i.e.~it is also compatible
with the power counting.

\section{Perturbative renormalization of the amplitude at NLO: $S$-waves}
\label{Sec:perturbarive_NLO_Swave}
In this section, we consider the renormalization of the next-to-leading-order amplitude
in the $S$-waves, i.e. in the partial waves where the power-counting breaking
terms have to be absorbed by the counter terms.

As specified in Eq.~\eqref{Eq:V2_tilde_text}, the next-to-leading order $S$-wave partial wave
potential is split into two parts  
$V_2(p',p)=\hat V_2(p',p)+\tilde V_2(p',p)$
($V_2$ is a matrix of dimension $n_\text{PW}\times
n_\text{PW}$ with $n_\text{PW}=1$ for $^1S_0$ and $n_\text{PW}=2$ for $^3S_1-{^3D}_1$
system).
Analogously, for the $T$-matrix, we define:
\begin{align}
 \hat T_2^{[m,n]}=\bar K^m\hat V_2K^n\,,\quad \tilde T_2^{[m,n]}=\bar K^m\tilde V_2K^n\,.
\end{align}
The amplitude $\hat T_2^{[m,n]}$ can be estimated in the same way as 
$T_0$ using Eq.~\eqref{Eq:bound_V_2_0_hat_text},
first obtaining the bounds for $\hat T_2^{[0,n]}$:
\begin{align}
&\left|\hat T_{2}^{[0,n]}(p_\text{on})\right|\le \mathcal{\hat M}_{V_2,0}\frac{8\pi^2}{m_N \Lambda_V} \frac{M_\pi^2}{\Lambda_{b}^2}\,
\Sigma^n\,,
\label{Eq:bound_T_2_hat_n}
\end{align}
and then symmetrically for $\hat T_2^{[m,n]}$:
\begin{align}
&\left|\hat T_{2}^{[m,n]}(p_\text{on})\right|\le \mathcal{\hat M}_{V_2,0}\frac{8\pi^2}{m_N \Lambda_V} \frac{M_\pi^2}{\Lambda_{b}^2}\Sigma^{m+n}\,.
\label{Eq:bound_T_2_hat_mn}
\end{align}

\subsection{Subtractions in the BPHZ scheme}
The amplitude $\tilde T_{2}^{[m,n]}$ contains power-counting breaking contributions
from the integration regions where all or some of the loop momenta are $p\sim\Lambda$.
It is natural to expect that such contributions can be removed by 
performing subtractions in all possible subdiagrams in the spirit of
the Bogoliubov-Parasiuk-Hepp-Zimmermann (BPHZ) renormalization procedure
\cite{Bogoliubov:1957gp,Hepp:1966eg,Zimmermann:1969jj}
(for a detailed discussion of the BPHZ scheme see 
Refs.~\cite{Collins:1984xc,Zavyalov:1990kv,Smirnov:1991jn}).
In the following, we will show that removing all power-counting breaking terms 
in $\tilde T_{2}^{[m,n]}$ can be achieved by introducing the 
appropriate momentum independent counter terms $\delta V^{(2)}_{C_S}$ and
$\delta V^{(2)}_{C_T}$, or, equivalently, $\delta V^{(2)}_{^1S_0}$ and $\delta V^{(2)}_{^3S_1}$.

In the present work, we consider for simplicity subtractions at
zero momenta and introduce the 
linear operation for the overall subtraction $\mathds{T}$ that replaces an operator $X$ with matrix elements
$X_{ji}(p',p{;p_\text{on}})$ with its value at $p=p'=p_\text{on}=0$:
\begin{align}
 \mathds{T}(X)=X_{11}(0,0,0) V_\text{ct}\,,
\end{align}
where $V_\text{ct}$ is the contact operator
$ \left(V_\text{ct}\right)_{ji}(p',p)=\delta_{1,i}\delta_{1,j}$.
For convenience, we also introduce the operation $\mathds{\bar T}=\mathds{1-T}$:
\begin{align}
 \mathds{\bar T}(X)=X- \mathds{T}(X).
\end{align}

Since the leading order amplitude does not violate power counting (see Sec.~\ref{Sec:perturbarive_LO}), the subgraphs leading to power-counting violating terms must necessarily include the vertex $\tilde V_2$
and have the form $\bar K^{m_1}\tilde V_2 K^{n_1}=\tilde T_2^{[m_1,n_1]}$ for some $m_1$, $n_1$.
We denote the diagram corresponding to the amplitude $\tilde T_2^{[m,n]}$
as $(m,n)$ and
define the subtraction operation $\mathds{T}^{m_1,n_1}$ for a subgraph $(m_1,n_1)$, $m_1\le m$, $n_1\le n$:
\begin{align}
\mathds{T}^{m_1,n_1}(\tilde T_2^{[m,n]})=\bar{K}^{m-m_1}\mathds{T}(\tilde T_2^{[m_1,n_1]}) K^{n-n_1}\,.
\label{Eq:subtraction_m1n1}
\end{align}
A composition of two subtractions is defined analogously:
\begin{align}
\mathds{T}^{m_2,n_2}\big(\mathds{T}^{m_1,n_1}( \tilde T_2^{[m,n]}) \big)=
\bar{K}^{m-m_2}\mathds{T}\big(\bar{K}^{m_2-m_1}\mathds{T}(\tilde
  T_2^{[m_1,n_1]}) K^{n_2-n_1} \big) K^{n-n_2}\,,
\label{Eq:two_subtractions}
\end{align}
where $m_1\le m_2\le m$, $n_1\le n_2\le n$ 
and either $m_1<m_2$ or $n_1<n_2$.

The BPHZ $\mathds{R}$-operation that leads to a 
renormalized expression for $\tilde T_2^{[m,n]}$ is given 
by 
\begin{align}
\mathds{R}(\tilde T_2^{[m,n]})=\tilde T_2^{[m,n]}+\sum_{U_k\in \mathcal{F}^{m,n}}
 \bigg(\prod_{(m_i,n_i)\in U_k} -\mathds{T}^{m_i,n_i}\bigg) \tilde T_2^{[m,n]}\,,
\label{Eq:R_operation1}
\end{align}
where $\mathcal{F}^{m,n}$ represents the set of all forests, i.e, 
the set of all possible distinct sequences of nested subdiagrams $(m_i,n_i)$:
\begin{align}
&U_k=((m_{k;1},n_{k;1}),(m_{k;2},n_{k;2}),\dots)\,,\nonumber\\
& m\ge m_{k;i+1}\ge m_{k;i}\ge 0\,, \quad n\ge n_{k;i+1}\ge n_{k;i}\ge 0\,,
\quad n+m>0\,,
\label{Eq:R_operation2}
\end{align}
where the last inequality can be omitted because $\tilde T_2^{[0,0]}(0,0,0)=\tilde V_2(0,0)=0$
and, therefore, $\mathds{T}^{0,0}( \tilde T_2^{[m,n]})=0 $.
For instance, the set $\mathcal{F}^{1,1}$ contains the following 
forests:
\begin{align}
 U_1=((0,1))\,,\quad U_2=((0,1),(1,1))\,,\quad 
 U_3=((1,0))\,,\quad U_4=((1,0),(1,1))\,,\quad 
 U_5=((1,1))\,.
\end{align}
When proving the power counting for $\mathds{R}(\tilde T_2^{[m,n]})$,
we will also use an equivalent representation of the $\mathds{R}$-operation 
\cite{Anikin:1973ra,Smirnov:1991jn,Zavyalov:1990kv,Bogolyubov:1980nc}:
\begin{align}
\mathds{R}\left(\tilde T_2^{[m,n]}\right)=\vdots \bigg(\prod_{m_i=0}^m\prod_{n_j=0}^n \mathds{\bar T}^{m_i,n_j}\bigg) \tilde T_2^{[m,n]}\,\vdots
=\vdots \bigg[\prod_{m_i=0}^m\prod_{n_j=0}^n \left(\mathds{1}-\mathds{T}^{m_i,n_j}\right)\bigg] \tilde T_2^{[m,n]}\,\vdots\,,
\label{eq:BPHZ_with_dots}
\end{align}
where the three-dot-product means that overlapping subdiagrams 
(i.e.~$(m_1,n_1)$ and $(m_2,n_2)$ such that $m_2>m_1$ and $n_2<n_1$
or $n_2>n_1$ and $m_2<n_1$) should be dropped from the product after expanding the brackets.
The ordering of $\mathds{T}^{m_i,n_j}$ in any product is such that $\mathds{T}$
corresponding to smaller subdiagrams (with smaller $m_i$ or $n_j$)
are applied first, e.g. $\mathds{T}^{4,5}\mathds{T}^{3,2}\mathds{T}^{1,2}$.

The presence of the overlapping diagrams prevents one from directly applying 
the recurrent procedure we used in the case of higher partial waves
except for the cases of $\tilde T_2^{[m,0]}$ and $\tilde T_2^{[0,n]}$.
In particular, the residues 
$\Delta_p\mathds{R}(\tilde T_2^{[m,n]})$ and
$\Delta_{p'}\mathds{R}(\tilde T_2^{[m,n]})$
do not satisfy the same inequalities as $\tilde V_2$, only some parts of them do. As we will see later in this section, the correct way to estimate 
$\mathds{R}(\tilde T_2^{[m,n]})$ is not the induction from $\mathds{R}(\tilde T_2^{[m,n-1]})$ or from
$\mathds{R}(\tilde T_2^{[m-1,n]})$, but rather from some combination of the two.

In an analysis of the renormalized amplitude in the 
BPHZ scheme, one typically performs a decomposition of the integration
region into sectors, usually, in the parametric representation~\cite{Hepp:1966eg,Speer:1975dc} (see also Refs.~\cite{Heinrich:2008si,Smirnov:2008aw} for more recent applications). 
In our case, we would like to keep a rather general form of the
interaction and the regulator,
and it appears convenient to decompose the whole integration region into sectors 
in momentum space as described in the next subsection.
For the examples of splitting the integration in momentum space into
subregions in the analysis of Feynman integrals we refer the reader to 
Refs.~\cite{Weinberg:1959nj,Hahn1968,Caswell:1981xt}.

\subsection{Sector decomposition}
\label{Sec:Sector_Decomposition}
In this subsection, we describe the sector decomposition
of the whole $(m+n)$-dimensional integration region
that allows one to prove that the BPHZ procedure indeed leads to the renormalized
amplitude satisfying the power counting.

We denote the momenta of $\bar K^m \tilde V_2 K^n$
corresponding to the loops involving $K$ as $p_1$, $p_2,\, \dots ,\, p_n$ ($p_{n+1}\equiv p$)
starting with the loop closest to $V_2$.
Analogously, we denote the momenta corresponding to the loops with
$\bar K$ as $p'_1$, $p'_2, \, \dots , \, p'_m$ ($p'_{m+1}\equiv p'$).
This means:
\begin{align}
\tilde T_2^{[m,n]}(p',p{;p_\text{on}}) 
 =\int\prod_{i=1}^n\prod_{j=1}^m\frac{ p_i^2dp_i}{(2\pi)^3}\frac{ p'^2_jdp'_j}{(2\pi)^3}\bar K(p',p'_m)\dots \bar K(p'_2,p'_1)
 \tilde V_2(p'_1,p_1)K(p_1,p_2)\dots K(p_n,p)\,.
\end{align}
We represent $\tilde T_2^{[m,n]}$ as a sum
\begin{align}
 \tilde T_2^{[m,n]}=\sum_i \tilde T_2^{\mathcal{P}_i^{m,n}}\,,
\end{align}
where the sum runs over all paths $\mathcal{P}_i^{m,n}$ that have the form of sequences of momenta
\begin{align}
  \mathcal{P}_1^{m,n}=(p_1,\dots,p_n,p'_1,\dots p'_m)\,,
   \quad
  \mathcal{P}_2^{m,n}=(p_1,\dots,p_{n-1},p'_1,p_n, p_2', \dots p'_m)\,,
  \quad \ldots\,,
\label{Eq:sequences}
\end{align}
with all ${n+m\choose n}$ possible permutations of $p_i$ and $p'_j$ in  
Eq.~\eqref{Eq:sequences} that do not change the order 
within $p_i$ and $p'_j$. I.e., for $j_1>j_2$, 
$p_{j_1}$ appears to the right of $p_{j_2}$, and analogously 
for $p'$, e.g.~$(p_1,p_2,p'_1,p_3,p_4,p'_2,p'_3,p'_4,p'_5)$.
To each path $\mathcal{P}_i$, we assign 
different $(n+m)$-dimensional integration sectors $\mathcal{D}_i^{m,n}=\mathcal{D}(\mathcal{P}_i^{m,n})$.
The choice of the integration sectors for a path $\mathcal{P}_i$ is as follows:
if $p_k'$ in $\mathcal{P}_i$ is immediately preceded  by $r$ variables $p_j,\dots ,p_{j+r-1}$, then the following $r$ inequalities must hold:
\begin{align}
 |p_k'|\le |p_j|\,, \; \dots\,, \; |p_k'|\le |p_{j+r-1}|\,, \label{Eq:sector_definition1}
\end{align}
and, conversely, if $p_k$ is immediately preceded by $r$ variables $p'_j,\dots, p'_{j+r-1}$, then 
\begin{align}
 |p_k|\le |p'_j|\,,\; \dots\,, \;  |p_k|\le |p'_{j+r-1}|\,. \label{Eq:sector_definition2}
\end{align}
If $p_k$ ($p'_k$) is preceded by $p_{k-1}$ ($p'_{k-1}$) or is the first momentum in the path,
then there are no constraints imposed on such a momentum.
The motivation for this choice of sectors will become clear during the proof of the renormalizability of $\tilde T_2^{[m,n]}$ and the analysis of the overlapping diagrams. 

We denote the set of all $\mathcal{P}_i^{m,n}$ as
$\mathcal{S}^{m,n}\equiv \{\mathcal{P}_i^{m,n}\}$. The corresponding sectors $\{\mathcal{D}_i^{m,n}\}$
do not overlap (except along the boundaries), and their union
covers the whole integration region.
We will show this below by induction.

The case of $\mathcal{S}^{0,n}$ is trivial.
Assume that the statement holds for $\mathcal{S}^{m,n}$.
The set $\mathcal{S}^{m,n}$ can be split into the following subsets:
\begin{align}
& \mathcal{S}^{m,n}= \bigcup_{k=0,n} \mathcal{S}^{m,n}_{k}\,,\quad
\mathcal{S}^{m,n}_{k}=\{\mathcal{P}^{m,n}_{i_k}\}\,,
\end{align}
where $i_k$ runs over the sequences that end with exactly $k$ momenta $p_j$: $p_{n-k+1},\dots,p_n$.
Then, the set $\mathcal{S}^{m+1,n}$ is given by
\begin{align}
\mathcal{S}^{m+1,n}= \{\mathcal{P}^{m+1,n}_i\} = \bigcup_{k=0}^n
  \mathcal{\tilde S}^{m+1,n}_{k}\,,\quad \text{where} \; \; \; 
 \mathcal{\tilde S}^{m+1,n}_{k}=\bigcup_{j=0}^k\{\mathcal{P}^{m+1,n}_{i_k,\,j}\}\,,
\label{Eq:proof_sectors_m_p_prime}
 \end{align}
where the sequence $\mathcal{P}^{m+1,n}_{i_k,\,j}$ is obtained
from $\mathcal{P}^{m,n}_{i_k}$ by inserting the momentum $p'_{m+1}$
into the $(k-j+1)$-th position from the right, i.e. into the $(n+m+1-k+j)$-th position.
The union over $j$ of the sectors corresponding to $\mathcal{P}^{m+1,n}_{i_k,\,j}$ is determined by the sector $\mathcal{D}\left(\mathcal{P}^{m+1,n}_{i_k}\right)$ and arbitrary $p'_{m+1}$:
\begin{align}
\bigcup_{j=0}^k\mathcal{D}\left(\mathcal{P}^{m+1,n}_{i_k,\,j}\right)=\mathcal{D}\left(\mathcal{P}^{m,n}_{i_k}\right)
\times \left\{p'_{m+1}\in\mathcal{C}\right\}\,,
\label{Eq:union_over_j}
\end{align}
which can be shown as follows.

The sector $\mathcal{D}\left(\mathcal{P}^{m+1,n}_{i_k,\,0}\right)$ is determined by
(here $k>0$, the case $k=0$ is trivial)
\begin{align}
\mathcal{D}\left(\mathcal{P}^{m+1,n}_{i_k,\,0}\right)=\mathcal{D}\left(\mathcal{P}^{m,n}_{i_k}\right)
\times \left\{p'_{m+1}\Big||p'_{m+1}|\le|p_{n}|,
\; \dots\,, \; |p'_{m+1}|\le|p_{n-k+1}|\right\}\,.
\end{align}
It is easy to prove by induction that for $0\le l<k$,
\begin{align}
 \bigcup_{j=0}^{l}\mathcal{D}\left(\mathcal{P}^{m+1,n}_{i_k,\,j}\right)=\mathcal{D}\left(\mathcal{P}^{m,n }_{i_k}\right)
\times \left\{p'_{m+1}\Big||p'_{m+1}|\le|p_{n-l}|,
\; \dots\,,\; |p'_{m+1}|\le|p_{n-k+1}|\right\}\,.
\label{Eq:union_l}
\end{align}
Indeed, assuming Eq.~(\ref{Eq:union_l}) is fulfilled, we obtain 
\beqa
 \bigcup_{j=0}^{l+1}\mathcal{D}\left(\mathcal{P}^{m+1,n}_{i_k,\,j}\right)&=&
\bigcup_{j=0}^{l}\mathcal{D}\left(\mathcal{P}^{m+1,n}_{i_k,\,j}\right)\cup\mathcal{D}\left(\mathcal{P}^{m+1,n}_{i_k,\,l+1}\right)
\nonumber\\&=&
\mathcal{D}\left(\mathcal{P}^{m,n }_{i_k}\right)
\times \left\{p'_{m+1}\Big||p'_{m+1}|\le|p_{n-l}|,\; 
\dots\,, \; |p'_{m+1}|\le|p_{n-k+1}|\right\}
\nonumber\\&& {} \cup
\mathcal{D}\left(\mathcal{P}^{m,n }_{i_k}\right)
\times \left\{p'_{m+1}\Big||p'_{m+1}|\le|p_{n-l-1}|,\; 
\dots\,, \; |p'_{m+1}|\le|p_{n-k+1}|,|p'_{m+1}|\ge|p_{n-l}|\right\}
\nonumber\\&=&\mathcal{D}\left(\mathcal{P}^{m,n }_{i_k}\right)
\times \left\{p'_{m+1}\Big||p'_{m+1}|\le|p_{n-l-1}|,
\dots\,,|p'_{m+1}|\le|p_{n-k+1}|\right\}\,.
\eeqa
Therefore,
\begin{align}
 \bigcup_{j=0}^{k-1}\mathcal{D}\left(\mathcal{P}^{m+1,n}_{i_k,\,j}\right)=\mathcal{D}\left(\mathcal{P}^{m,n }_{i_k}\right)
\times \left\{p'_{m+1}\Big||p'_{m+1}|\le|p_{n-k+1}|\right\}\,.
\label{Eq:union_k_minus1}
\end{align}
On the other hand, 
\begin{align}
\mathcal{D}\left(\mathcal{P}^{m+1,n}_{i_k,\,k}\right)=\mathcal{D}\left(\mathcal{P}^{m,n }_{i_k}\right)
\times \left\{p'_{m+1}\Big||p'_{m+1}|\ge|p_{n-k+1}|\right\}\,,
\label{Eq:D_k}
\end{align}
since in $\mathcal{P}^{m+1,n}_{i_k,\,k}$, the momentum $p_{n-k+1}$ is preceded by $p'_{m+1}$, $p'_{m}$
and other $p'_r$ that stay in $\mathcal{P}^{m,n }_{i_k}$ in front of $p'_{m}$.
Equations~\eqref{Eq:union_k_minus1},~\eqref{Eq:D_k} yield Eq.~\eqref{Eq:union_over_j}, which together with Eq.~\eqref{Eq:proof_sectors_m_p_prime} proves the completeness of the partitions $\mathcal{S}^{m+1,n}$ and completes the induction.

The subtraction operation $\mathds{T}^{m_1,n_1}$  for 
$\tilde T_2^{\mathcal{P}_i^{m,n}}$  is defined as in Eq.~\eqref{Eq:subtraction_m1n1},
where the integration region in 
\begin{align}
\bar{K}^{m-m_1}V_\text{ct}K^{n-n_1} 
\end{align}
is constrained by those inequalities from
Eqs.~\eqref{Eq:sector_definition1},~\eqref{Eq:sector_definition2}
that relate the momenta $\{p'_{m_1+1},\dots, p'_m,p_{n_1+1},\dots,p_n\}$
among each other, whereas the integration region in 
\begin{align}
\tilde T_2^{[m_1,n_1]}(p'_{m_1+1}=0,p_{n_1+1}=0,p_\text{on}=0) 
\label{Eq:T2_m1n1}
\end{align}
is constrained by the inequalities from
Eqs.~\eqref{Eq:sector_definition1},~\eqref{Eq:sector_definition2}
that relate the momenta $\{p'_1,\dots, p'_{m_1+1},p_1,\dots,p_{n_1+1}\}$
among each other.
Since in Eq.~\eqref{Eq:T2_m1n1}, $p'_{m_1+1}=0$ and $p_{n_1+1}=0$,
any inequality involving $p'_{m_1+1}$ or $p_{n_1+1}$ is
either satisfied automatically 
(if $|p_i|\ge |p'_{m_1+1}|$ or $|p'_j|\ge |p_{n_1+1}|$) or
makes the whole integral vanish
(if $|p_i|\le |p'_{m_1+1}|$ or $|p'_j|\le |p_{n_1+1}|$).
Analogously, for the subsequent subtractions (see Eq.~\eqref{Eq:two_subtractions}),
only the momenta entering into the remaining integration in
\begin{align}
\bar{K}^{m-m_2}V_\text{ct}K^{n-n_2} 
\end{align}
are constrained.
That such partitions of the integration regions in the subtracted
amplitudes cover the whole space and do not overlap will be seen
in the next subsection.

\subsection{Overlapping diagrams}
Now, we will show how the paths $\mathcal{P}_i^{m,n}$
and the corresponding sectors $\mathcal{D}\left(\mathcal{P}_i^{m,n}\right)$
help disentangling overlapping diagrams.

First, we associate with the sequence  $\mathcal{P}_i^{m,n}$ the 
maximal forest $U_\text{max}(\mathcal{P}^{m,n}_i)$
(a maximal forest is a forest that is not contained in any other forest)
as follows: 
\beq
\label{EqUmax}
U_\text{max}(\mathcal{P}^{m,n}_i) =
                ((m_{i;1},n_{i;1}), \, (m_{i;2},n_{i;2}),\, \dots , \,
                (m_{i;m+n},n_{i;m+n}))\,,\quad \text{with} \; \; 
 \quad m_{i;m+n}=m\,, \quad n_{i;m+n}=n\,,
\eeq
where $m_{i;k}=m_{i;k-1}+1$, $n_{i;k}=n_{i;k-1}$ if 
the $k$-th element of $\mathcal{P}^{m,n}_i$ is $p_r'$,
$(\mathcal{P}^{m,n}_i)_k=p_r'$ (for some $r$),
and $m_{i;k}=m_{i;k-1}$, $n_{i;k}=n_{i;k-1}+1$ if
$(\mathcal{P}^{m,n}_i)_k=p_s$ (for some $s$).
By definition, $m_{i;0}=n_{i;0}=0$.
We denote by  $\mathcal{F}^{m,n}_i$ the set of forests that consists of all 
subsequences of $U_\text{max}(\mathcal{P}^{m,n}_i)$.
All subdiagrams of the diagram $(m,n)$ that belong to  $U_\text{max}(\mathcal{P}^{m,n}_i)$ are non-overlapping.

Assume that a subdiagram $(p,q)$ of the diagram $(m,n)$ does not belong to any forest in $\mathcal{F}(\mathcal{P}^{n,m}_i)$, i.e.
it does not belong to the maximal forest
$U_\text{max}(\mathcal{P}^{m,n}_i)=((m_{i;1},n_{i;1}), \,
  \ldots ,\, (m_{i;k},n_{i;k}), \,
  \ldots ,\, (m_{i;m+n},n_{i;m+n}) )$.
Therefore, 
\begin{align}
 \nexists k: \quad  m_{i;k}=p \text{ and } n_{i;k}=q\,.
\end{align}
For some $k=k_0$, the subdiagram $(m_{i;k_0},n_{i;k_0})$
starts to overlap with the subdiagram $(p,q)$.
Two symmetric cases are possible:
\begin{align}
 p<m\,, \quad  q\le n\,, \quad m_{i;k_0-1}=p\,,\
 m_{i;k_0}=p+1\,,\quad n_{i;k_0}<q\,,
\end{align}
or
\begin{align}
 p\le m\,, \quad  q< n\,, \quad n_{i;k_0-1}=q\,,\
 n_{i;k_0}=q+1\,,\quad m_{i;k_0}<p\,.
\end{align}
Consider, for definiteness, the former case (the latter case can be treated analogously).
Since $n_{i;k_0}<q\le n$, there exists $k_1>k_0$ such that 
\begin{align}
 \text{for } k_0\le k<k_1:\quad n_{i;k}=n_{i;k_0}\,\text{ and }  n_{i;k_1}=n_{i;k_0}+1 \,.
\end{align}
From the definitions of the sectors given in the previous subsection, it follows that in $\mathcal{D}\left(\mathcal{P}_i^{m,n}\right)$, $|p_{n_{i;k_1}}|\le |p'_{m_{i;k_0}}|=|p'_{p+1}|$.

The subtraction operation $\mathds{T}^{(p,q)}$ for 
$\tilde T_2^{\mathcal{P}_i^{m,n}}$  gives
\begin{align}
 \mathds{T}^{p,q}\left(\tilde T_2^{\mathcal{P}_i^{m,n}}\right)=(\bar{K})^{m-p}\tilde T_2^{[p,q]}(p'_{p+1}=0,p_{q+1}=0,p_\text{on}=0)V_\text{ct} K^{n-q}\,,
\end{align}
where the integration region is constrained, in particular, by $|p_{n_{i,k_1}}|\le |p'_{p+1}|=0$, which makes the whole integral vanish:
\begin{align}
 \mathds{T}^{p,q}\left(\tilde T_2^{\mathcal{P}_i^{m,n}}\right)=0\,.
\label{Eq:Overlapping_subtraction}
 \end{align}
The same set of arguments leads to an analogous result for multiple subtractions
\begin{align}
 \mathds{T}^{p,q}\mathds{T}^{m_{i;b},n_{i;b}}\dots\mathds{T}^{m_{i;a},n_{i;a}}\left(\tilde T_2^{\mathcal{P}_i^{m,n}}\right)=0\,,
\label{Eq:Overlapping_subtraction2}
 \end{align}
with  $(m_{i;a},n_{i;a})\in U_\text{max}(\mathcal{P}^{m,n}_i)\,,\dots, (m_{i;b},n_{i;b})\in U_\text{max}(\mathcal{P}^{m,n}_i)$.

Since $(p,q)$ can be any subdiagram of $(m,n)$
overlapping with some subdiagram $(m_{i;x},n_{i;x})\in U_\text{max}(\mathcal{P}^{m,n}_i)$ (otherwise it would belong to $U_\text{max}(\mathcal{P}^{m,n}_i)$), we can replace the three-dot products in Eq.~\eqref{eq:BPHZ_with_dots} with ordinary products:
\beqa
\mathds{R}\left(\tilde T_2^{[m,n]}\right)&=&\vdots \prod_{m_k=0}^m\prod_{n_l=0}^n 
\left(\mathds{\bar T}^{m_k,n_l}\right)\sum_i \tilde T_2^{\mathcal{P}_i^{m,n}}\,\vdots\nonumber\\
&=&\sum_i \mathds{R}
 \left(\tilde T_2^{\mathcal{P}_i^{m,n}}\right) \nn
&=&\sum_i \prod_{(m_{i;k},n_{i;k})\in U_\text{max}(\mathcal{P}^{m,n}_i)}
\left(\mathds1-\mathds{T}^{m_{i;k},n_{i;k}}\right) 
 \tilde T_2^{\mathcal{P}_i^{m,n}}\,.
\label{Eq:R_for_paths}
\eeqa
Interchanging the sum and the products in Eq.~\eqref{Eq:R_for_paths}
is possible because in each product (after expanding the brackets)
\begin{align}
 \mathds{T}^{m_b,n_b}\dots \mathds{T}^{m_a,n_a}\sum_i \tilde T_2^{\mathcal{P}_i^{m,n}}\,,
\end{align}
the sum runs over the paths $\mathcal{P}_i^{m,n}$ that 
do not overlap with $(m_a,n_a),\dots ,(m_b,n_b)$ 
(as follows from
Eqs.~\eqref{Eq:Overlapping_subtraction},~\eqref{Eq:Overlapping_subtraction2}),
i.e.~over the sequences that contain $(m_a,n_a),\dots ,(m_b,n_b)$.
In other words, the sum turns into the sums over subsequences
that have the same endpoints $(m_a,n_a),\dots ,(m_b,n_b)$.
Therefore, one can apply the arguments from the previous subsection
concerning the completeness of the sector decomposition
to the integration regions in the subtracted amplitudes for each subsequence.

Equation~\eqref{Eq:R_for_paths} will allow us to derive a recurrence relation for 
the $\mathds{R}$-operation in the next subsection.
\subsection{Recurrence relation for the $\mathds{R}$-operation}
Let either the last two momenta of the sequence $\mathcal{P}_i^{m,n}$
be $p_{n-1}$ and $p_n$:
\begin{align}
  \mathcal{P}_i^{m,n}=(\mathcal{P}_j^{m,n-1},p_n)\,,\; \text{ and } \;
 \mathcal{P}_j^{m,n-1}=(\mathcal{P}_k^{m,n-2},p_{n-1})\,,
\end{align}
or $m=0$, $n=1$.
Then, as follows from Eq.~\eqref{Eq:R_for_paths},
$\mathds{R}\left(\tilde T_2^{\mathcal{P}_i^{m,n}}\right)$
is obtained from $\mathds{R}\left(\tilde T_2^{\mathcal{P}_j^{m,n-1}}\right)$
by performing an overall subtraction:
\begin{align}
 \mathds{R}\left(\tilde T_2^{\mathcal{P}_i^{m,n}}\right) = 
\mathds{\bar T}^{m,n}\left(\mathds{R}\left(\tilde T_2^{\mathcal{P}_j^{m,n-1}}\right)K\right) = 
\mathds{\bar T}\left(\mathds{R}\left(\tilde T_2^{\mathcal{P}_j^{m,n-1}}\right)K\right)\,.
\end{align}
Analogously, for 
\begin{align}
  \mathcal{P}_i^{m,n}=(\mathcal{P}_j^{m-1,n},p'_{m})\,,\; \text{ and
  }\; 
 \mathcal{P}_j^{m-1,n}=(\mathcal{P}_k^{m-2,n},p'_{m-1})\,,\; \text{ or
  } \; m=1\,, \quad n=0\,,
\end{align}
it holds:
\begin{align}
\mathds{R}\left(\tilde T_2^{\mathcal{P}_i^{m,n}}\right) = 
\mathds{\bar T}^{m,n}\left(\bar K\mathds{R}\left(\tilde T_2^{\mathcal{P}_j^{m-1,n}}\right)\right) = 
\mathds{\bar T}\left(\bar K\mathds{R}\left(\tilde T_2^{\mathcal{P}_j^{m-1,n}}\right)\right)\,.
\end{align}
If 
\begin{align}
  \mathcal{P}_j^{m,n}=(\mathcal{P}_i^{m,n-1},p_{n})\,,\; \text{ and
  }\; 
 \mathcal{P}_j^{m,n-1}=(\mathcal{P}_i^{m-1,n-1},p'_{m})\,,
\end{align}
then we have to keep track of the number of momenta $p'_k$ directly preceding $p'_m$ in $\mathcal{P}_j^{m,n}$
to constrain the integration region to the correct sector: $|p_{n}|\le |p'_k|$
and $|p_{n}|\le |p'_m|$
(and similarly for the symmetric case).
In order to implement these constraints, we introduce the following notation.
For an operator $X$ with matrix elements $X(p',p{;p_\text{on}})$, we define
\begin{align}
& X^>(p',p{;p_\text{on}})=X(p',p{;p_\text{on}})\theta(|p'|-|p|)\,, \nonumber\\
& X^<(p',p{;p_\text{on}})=X(p',p{;p_\text{on}})\theta(|p|-|p'|)\,.
\end{align}
Then, we represent the subtraction operation acting on $\tilde T_2^{\mathcal{P}_j^{m,n}}$
as  a sum of three terms:
\beq
\mathds{R}\Big(\tilde T_2^{\mathcal{P}_j^{m,n}}\Big)=
\mathds{R}_{p}\Big(\tilde T_2^{\mathcal{P}_j^{m,n}}\Big)
+\mathds{R}_{p'}\Big(\tilde T_2^{\mathcal{P}_j^{m,n}}\Big)+\mathds{R}_{pp'}\Big(\tilde T_2^{\mathcal{P}_j^{m,n}}\Big)\,,
\eeq
where
\beqa
\mathds{R}_{p}\Big(\tilde T_2^{\mathcal{P}_j^{m,n}}\Big) &=&
\mathds{R}_{p}\Big(\tilde T_2^{\mathcal{P}_i^{m,n-1}}\Big)K+ \mathds{R}_{pp'}\Big(\tilde T_2^{\mathcal{P}_i^{m,n-1}}\Big)^>K\,,\nonumber\\
\mathds{R}_{p'}\Big(\tilde T_2^{\mathcal{P}_j^{m,n}}\Big) &=& 0\,,\nonumber\\
\mathds{R}_{pp'}\Big(\tilde T_2^{\mathcal{P}_j^{m,n}}\Big)&=&
\mathds{\bar T}\left(\mathds{R}_{pp'}\Big(\tilde
  T_2^{\mathcal{P}_i^{m,n-1}}\Big)^<K\right)
\label{Eq:R_p_1}
\eeqa
if 
\begin{align}
  \mathcal{P}_j^{m,n}=(\mathcal{P}_i^{m,n-1},p_{n})\,,
\end{align}
and
\beqa
\mathds{R}_{p'}\Big(\tilde T_2^{\mathcal{P}_j^{m,n}}\Big)&=&
\bar{K}\mathds{R}_{p'}\Big(\tilde T_2^{\mathcal{P}_i^{m-1,n}}\Big)+ 
\bar{K}\mathds{R}_{pp'}\Big(\tilde T_2^{\mathcal{P}_i^{m-1,n}}\Big)^<\,,\nonumber\\
\mathds{R}_{p}\Big(\tilde T_2^{\mathcal{P}_j^{m,n}}\Big)&=&0\,,\nonumber\\
\mathds{R}_{pp'}\Big(\tilde T_2^{\mathcal{P}_j^{m,n}}\Big)&=&
\mathds{\bar T}\left(\bar{K}\mathds{R}_{pp'}\Big(\tilde
  T_2^{\mathcal{P}_i^{m-1,n}}\Big)^>\right)
\label{Eq:R_p_2}
\eeqa
if 
\begin{align}
  \mathcal{P}_j^{m,n}=(\mathcal{P}_i^{m-1,n},p'_{m})\,.
\end{align}

The part $\mathds{R}_{p}\Big(\tilde T_2^{\mathcal{P}_j^{m,n}}\Big)$
 collects all integration sectors of $\tilde T_2^{\mathcal{P}_j^{m,n}}$
with the last $|p_i|\le|p'_{m+1}|=|p'|$ for $i=n,\dots,n-k$ if $\left({\mathcal{P}_j^{m,n}}\right)_{m+i}=p_i$ and
$\left({\mathcal{P}_j^{m,n}}\right)_{m+n-k-1}=p'_m$ (or if $k+1=m+n$). 
As follows from Eq.~\eqref{Eq:R_p_2} these sectors are excluded from the integration if $\left({\mathcal{P}_l^{m,n}}\right)_{m+n+1}=p'_{m+1}$
in accordance with the sector decomposition defined in Sec.~\ref{Sec:Sector_Decomposition}.
Analogous arguments apply to $\mathds{R}_{p'}\Big(\tilde T_2^{\mathcal{P}_j^{m,n}}\Big)$ with the replacement $p\leftrightarrow p'$.
For the operations $\mathds{R}_{p}$ and $\mathds{R}_{p'}$, the overall subtraction $\mathds{\bar T}$
is not necessary as there is at least one integration variable $|p_i|\le 
|p'_{m+1}|=|p'|$ for $\mathds{R}_{p}$
and at least one integration variable $|p'_j|\le |p_{n+1}|=|p|$ for
$\mathds{R}_{p'}$ so that the subtraction yields zero.

Now, we can construct a recurrence relation for
the $\mathds{R}$ operation for the full $\tilde T_2^{[m,n]}$.
Since $\tilde T_2^{[m,n]}$ is a sum over all possible paths $\tilde T_2^{[m,n]}=\sum_i \tilde T_2^{\mathcal{P}_i^{m,n}}$ (or the corresponding integration sectors), and each path can be obtained from a smaller one either as $\mathcal{P}_i^{m,n}=(\mathcal{P}_j^{m,n-1},p)$
or as $\mathcal{P}_s^{m,n}=(\mathcal{P}_t^{m-1,n},p')$ (if $m\ne 0$, $n\ne 0$), 
we can combine Eq.~\eqref{Eq:R_p_1} and Eq.\eqref{Eq:R_p_2} to obtain
\beq
\mathds{R}(\tilde T_2^{[m,n]})=\mathds{R}_{p}(\tilde T_2^{[m,n]})+
\mathds{R}_{p'}(\tilde T_2^{[m,n]})+\mathds{R}_{pp'}(\tilde T_2^{[m,n]})\,,
\label{Eq:R_p_full}
  \eeq
  with
  \beqa
\mathds{R}_{p}(\tilde T_2^{[m,n]})&=&\mathds{R}_{p}(\tilde T_2^{m,n-1})K+ 
\mathds{R}_{pp'}(\tilde T_2^{m,n-1})^>K\,,\nonumber\\
\mathds{R}_{p'}(\tilde T_2^{[m,n]})&=&\bar{K}\mathds{R}_{p'}(\tilde T_2^{m-1,n})+ 
\bar{K}\mathds{R}_{pp'}(\tilde T_2^{m-1,n})^<\,,\nonumber\\
\mathds{R}_{pp'}(\tilde T_2^{[m,n]})&=&\mathds{\bar T}\left(\mathds{R}_{pp'}(\tilde T_2^{m,n-1})^<K\right)
+\mathds{\bar T}\left(\bar{K}\mathds{R}_{pp'}(\tilde
  T_2^{m-1,n})^>\right)\,,
\label{Eq:R_p_full_X}
\eeqa
subject to the initial conditions
\begin{align}
 \mathds{R}_{p}(\tilde T_2^{0,0})=\mathds{R}_{p'}(\tilde T_2^{0,0})=0\,, \quad 
\mathds{R}_{pp'}(\tilde T_2^{0,0})=\tilde V_2\,.
\end{align}

Note that for $m=0$ (and, analogously, for $n=0$ ), Eq.~\eqref{Eq:R_p_full} simplifies to
\begin{align}
\mathds{R}(\tilde T_2^{0,n})=\mathds{\bar T}\left(\mathds{R}(\tilde T_2^{0,n-1}) K\right)\,,
 \label{eq:R_p_4}
\end{align}
which is a consequence of the absence of overlapping subdiagrams.

\subsection{Power counting for the renormalized perturbative amplitude}
\label{Sec:Power_Counting_Swave}
We are now going to prove by induction the following inequalities for
the  $\mathds{R}_{p}(\tilde T_2^{[m,n]})$, $\mathds{R}_{p}(\tilde T_2^{[m,n]})$
and $\mathds{R}_{pp'}(\tilde T_2^{[m,n]})$ operations:
\beqa
\left|\mathds{R}_{p}(\tilde T_2^{[m,n]})(p',p{;p_\text{on}})\right|
&\le & \mathcal{M}^{m,n}
\left(|p'|^2+p_\text{on}^2\right)
\tilde f_\text{log}(p',p)\,,\nonumber\\
\left|\mathds{R}_{p'}(\tilde T_2^{[m,n]})(p',p{;p_\text{on}})\right|&\le& \mathcal{M}^{m,n}
\left(|p|^2+p_\text{on}^2\right)
\tilde f_\text{log}(p',p)\,,\nonumber\\
 \left|\mathds{R}_{pp'}(\tilde T_2^{[m,n]})(p',p{;p_\text{on}})\right|&\le& \mathcal{M}^{m,n}
\left(|p|^2+|p'|^2+p_\text{on}^2\right)
\tilde f_\text{log}(p',p)\,,\nonumber\\
 \left|\Delta_{p'}\mathds{R}_{pp'}(\tilde T_2^{[m,n]})(p',p{;p_\text{on}})\right|&\le& \mathcal{M}^{m,n}
\left(|p'|^2+p_\text{on}^2\right)
\tilde f_\text{log}(p',p)\,,\nonumber\\
 \left|\Delta_{p}\mathds{R}_{pp'}(\tilde T_2^{[m,n]})(p',p{;p_\text{on}})\right|&\le& \mathcal{M}^{m,n}
\left(|p|^2+p_\text{on}^2\right)
\tilde f_\text{log}(p',p)\,,\nonumber\\
\left|\Delta_{p_\text{on}}\mathds{R}_{pp'}(\tilde T_2^{[m,n]})(p',p{;p_\text{on}})\right|&\le& \mathcal{M}^{m,n}
p_\text{on}^2 \tilde f_\text{log}(p',p)\,,
\label{Eq:inequalities_R}
\eeqa
where the remainder $\Delta_{p_\text{on}}$ for an operator $X$ with matrix elements $X(p',p;p_\text{on})$ is defined as
\begin{align}
 \Delta_{p_\text{on}}X(p',p;p_\text{on})=X(p',p;p_\text{on})-X(p',p;0)\,.
\end{align}
For $m=n=0$, the above inequalities hold with $\mathcal{M}^{0,0}= \mathcal{M}_{V_2,0}$.
We set by definition $\mathcal{M}^{-1,n}=\mathcal{M}^{m,-1}=0$ and assume
that these inequalities hold also for 
$\mathds{R}_{p}(\tilde T_2^{m,n-1})$, $\mathds{R}_{p'}(\tilde T_2^{m,n-1})$
$\mathds{R}_{pp'}(\tilde T_2^{m,n-1})$ and
$\mathds{R}_{p}(\tilde T_2^{m-1,n})$, $\mathds{R}_{p'}(\tilde T_2^{m-1,n})$,
 $\mathds{R}_{pp'}(\tilde T_2^{m-1,n})$.
Under these assumptions, we will derive Eq.~\eqref{Eq:inequalities_R}.

First, we consider $\mathds{R}_{p}(\tilde T_2^{[m,n]})(p',p{;p_\text{on}})$.
Equations~\eqref{Eq:bounds_V0_l_0_text},~\eqref{Eq:bound_on_G} together with the inequality
\beqa
\left|\mathds{R}_{pp'}(\tilde T_2^{m,n-1})^>(p',p''{;p_\text{on}})\right|
 &=&\left|\mathds{R}_{pp'}(\tilde T_2^{m,n-1})(p',p''{;p_\text{on}})\right|
\theta(|p'|-|p''|)\nonumber\\
&\le& \left(|p'|^2+|p''|^2+p_\text{on}^2\right)
\tilde f_\text{log}(p',p'')\theta(|p'|-|p''|) \nn
&\le& 2\left(|p'|^2+p_\text{on}^2\right)\tilde f_\text{log}(p',p'')
\eeqa
give
\beqa
\left|\mathds{R}_{p}(\tilde T_2^{[m,n]})(p',p{;p_\text{on}})\right|
&=&\bigg|\int \frac{p''^2dp''}{(2\pi)^3}\mathds{R}_{p}(\tilde T_2^{m,n-1})(p',p''{;p_\text{on}})
G(p''{;p_\text{on}}) V(p'',p)\nonumber\\
&&{}+\int \frac{p''^2dp''}{(2\pi)^3}\mathds{R}_{pp'}(\tilde T_2^{m,n-1})^>(p',p''{;p_\text{on}})
G(p''{;p_\text{on}}) V(p'',p)\bigg|\nonumber\\
&\le& 3 \mathcal{M}^{m,n-1}{n_\text{PW}}\mathcal{M}_{V_0,0}\left(|p'|^2+p_\text{on}^2\right)
\int \frac{|p''|^2 d|p''|}{(2\pi)^3} \tilde f_\text{log}(p',p'') 
\left|G(p''{;p_\text{on}}) \right|
 V_{0,\text{max}}(p'',p)\nonumber\\
&=&3
\mathcal{M}^{m,n-1}{n_\text{PW}}\mathcal{M}_{V_0,0}\left(|p'|^2+p_\text{on}^2\right)I_{f_\text{log}}(p',p{;p_\text{on}}) \nn
&\le& 3 \mathcal{M}^{m,n-1}{n_\text{PW}}\mathcal{M}_{V_0,0}
\mathcal{M}_{f_\text{log}}\frac{{\Lambda}}{\Lambda_V}\left(|p'|^2+p_\text{on}^2\right)\tilde f_\text{log}(p',p)\,.
\label{Eq:R_p}
\eeqa
In a completely analogous way we obtain:
\begin{align}
&\left|\mathds{R}_{p'}(\tilde T_2^{[m,n]})(p',p{;p_\text{on}})\right|
\le 3 \mathcal{M}^{m-1,n}\mathcal{M}_{f_\text{log}}\frac{{\Lambda}}{\Lambda_V}\left(|p|^2+p_\text{on}^2\right)\tilde f_\text{log}(p',p)\,.
\end{align}

Secondly, we analyze the term $\mathds{R}_{pp'}(\tilde T_2^{[m,n]})$.
It is sufficient to consider the first term in the definition of $\mathds{R}_{pp'}(\tilde T_2^{[m,n]})$
in Eq.~\eqref{Eq:R_p_full_X} (the second term gives a similar bound from 
the symmetry arguments):
\beqa
{\bar T}\left(\mathds{R}_{pp'}(\tilde T_2^{m,n-1})^<K\right)(p',p{;p_\text{on}})
&=&\int \frac{p''^2dp''}{(2\pi)^3}\mathds{R}_{pp'}(\tilde T_2^{m,n-1})^<(p',p''{;p_\text{on}})
G(p''{;p_\text{on}}) V(p'',p) \\
&-&\int \frac{p''^2dp''}{(2\pi)^3}\mathds{R}_{pp'}(\tilde
T_2^{m,n-1})^<(0,p'',0)G(p'',0) V(p'',0) \nn
&=&\delta_1+\tilde\delta_2+\tilde\delta_3+\tilde\delta_4\,,\nonumber
\label{Eq:R_p_pprime}
\eeqa
with
\begin{align}
 \delta_1 &= \int \frac{p''^2dp''}{(2\pi)^3} 
\Delta_{p'}\mathds{R}_{pp'}\big(T_{2}^{m,n-1}\big)^<(p',p''{;p_\text{on}}) G(p''{;p_\text{on}}) V(p'',p)\,,\nonumber\\
  \tilde\delta_2 &= \int \frac{p''^2dp''}{(2\pi)^3} 
\Delta_{p_\text{on}}\mathds{R}_{pp'}\big(T_{2}^{m,n-1}\big)(0,p'',p_\text{on})
G(p''{;p_\text{on}})V(p'',p)\,,  \nonumber\\
  \tilde\delta_3 &= \int \frac{p''^2dp''}{(2\pi)^3}
 \mathds{R}_{pp'}\big(T_{2}^{m,n-1}\big)(0,p'',0)\Delta_{p_\text{on}} G(p''{;p_\text{on}}) V(p'',p)\,, \nonumber\\
  \tilde\delta_4 &= \int \frac{p''^2dp''}{(2\pi)^3}
 \mathds{R}_{pp'}\big(T_{2}^{m,n-1}\big)(0,p'',0) G(p'',0) \Delta_p V(p'',p)\,. 
\end{align}
Further, we obtain
\beqa
\Delta_{p'}\mathds{\bar T}\left(\mathds{R}_{pp'}(\tilde T_2^{m,n-1})^<K\right)&=&
\Delta_{p'}\left(\mathds{R}_{pp'}(\tilde T_2^{m,n-1})^<K\right) =\delta_1\,,\nonumber\\
\Delta_{p_\text{on}}\mathds{\bar T}\left(\mathds{R}_{pp'}(\tilde T_2^{m,n-1})^<K\right)&=&
\Delta_{p_\text{on}}\left(\mathds{R}_{pp'}(\tilde T_2^{m,n-1})^<K\right) =\delta_2+\delta_3\,,\nonumber\\
\Delta_p\mathds{\bar T}\left(\mathds{R}_{pp'}(\tilde T_2^{m,n-1})^<K\right)&=&
\Delta_p\left(\mathds{R}_{pp'}(\tilde T_2^{m,n-1})^<K\right) =\delta_4\,,
\label{Eq:Delta_R_p_pprime}
\eeqa
with
\begin{align}
 \delta_2 &= \int \frac{p''^2dp''}{(2\pi)^3} 
\Delta_{p_\text{on}}\mathds{R}_{pp'}\big(T_{2}^{m,n-1}\big)(p,p'',p_\text{on})^<
G(p''{;p_\text{on}})V(p'',p)\,,  \nonumber\\
\delta_3 &= \int \frac{p''^2dp''}{(2\pi)^3}
 \mathds{R}_{pp'}\big(T_{2}^{m,n-1}\big)^<(p,p'',0) \Delta_{p_\text{on}}G(p''{;p_\text{on}}) V(p'',p)\,, \nonumber\\
  \delta_4 &= \int \frac{p''^2dp''}{(2\pi)^3}
 \mathds{R}_{pp'}\big(T_{2}^{m,n-1}\big)^<(p',p''{;p_\text{on}}) G(p''{;p_\text{on}}) \Delta_p V(p'',p)\,. 
\end{align}
As the next step, we estimate the integrals $\delta_1$, $\delta_2$, $\delta_3$, $\delta_4$.
Taking into account that 
\beqa
\left|\Delta_{p'}\mathds{R}_{pp'}\big(T_{2}^{m,n-1}\big)^<(p',p''{;p_\text{on}})\right|&=&
 \left|\Delta_{p'}\mathds{R}_{pp'}\big(T_{2}^{m,n-1}\big)(p',p''{;p_\text{on}})\theta(|p''|-|p'|)
-\mathds{R}_{pp'}\big(T_{2}^{m,n-1}\big)(0,p'',p_\text{on})\theta(|p'|-|p''|)\right|\nonumber\\
&\le& \left|\Delta_{p'}\mathds{R}_{pp'}\big(T_{2}^{m,n-1}\big)(p',p''{;p_\text{on}})\right|\theta(|p''|-|p'|)
+\left|\mathds{R}_{pp'}\big(T_{2}^{m,n-1}\big)(0,p'',p_\text{on})\right|\theta(|p'|-|p''|)\nonumber\\
&\le& 2 \mathcal{M}^{m,n-1}\left(|p'|^2+p_\text{on}^2\right)\tilde f_\text{log}(p',p'') 
\,,\nonumber
\eeqa
we obtain
\beqa
 |\delta_1|&\le& 2{\mathcal{M}^{m,n-1}}
\left(|p'|^2+p_\text{on}^2\right)
\int \frac{|p''|^2d|p''|}{(2\pi)^3}\tilde f_\text{log}(p',p'')  
\left|G(p''{;p_\text{on}}) V(p'',p)\right|\nonumber\\
&\le& 2{\mathcal{M}^{m,n-1}}{n_\text{PW}}\mathcal{M}_{V_0,0}
\left(|p'|^2+p_\text{on}^2\right)
\int \frac{|p''|^2d|p''|}{(2\pi)^3}\tilde f_\text{log}(p',p'')  
\left|G(p''{;p_\text{on}})\right| V_{0,\text{max}}(p'',p)\nonumber\\
&=&2{\mathcal{M}^{m,n-1}}{n_\text{PW}}\mathcal{M}_{V_0,0}\left(|p'|^2+p_\text{on}^2\right)I_{f_\text{log}}(p',p{;p_\text{on}}) \nn
&\le& 2{\mathcal{M}^{m,n-1}}{n_\text{PW}}\mathcal{M}_{V_0,0}\mathcal{M}_{f_\text{log}}\frac{{\Lambda}}{\Lambda_V}\left(|p'|^2+p_\text{on}^2\right)\tilde f_\text{log}(p',p)\,.
\eeqa
Analogously, the estimates for $\delta_2$ and $\delta_3$ (and for $\tilde\delta_2$ and $\tilde\delta_3$) are given by
\beqa
 |\delta_2| &\le& {\mathcal{M}^{m,n-1}}p_\text{on}^2
\int \frac{|p''|^2d|p''|}{(2\pi)^3}\tilde f_\text{log}(p',p'')  
\left|G(p''{;p_\text{on}}) V(p'',p)\right|\nn
&\le& {\mathcal{M}^{m,n-1}}{n_\text{PW}}\mathcal{M}_{V_0,0}\mathcal{M}_{f_\text{log}}\frac{{\Lambda}}{\Lambda_V}p_\text{on}^2  \tilde f_\text{log}(p',p)\,,
\eeqa
and
\beqa
 |\delta_3|&\le& {\mathcal{M}^{m,n-1}}
\int \frac{|p''|^2d|p''|}{(2\pi)^3}2|p''|^2\tilde f_\text{log}(p',p'')  
\frac{p_\text{on}^2}{|p''|^2}\left|G(p''{;p_\text{on}}) V(p'',p)\right|\nn
&\le& 2{\mathcal{M}^{m,n-1}}{n_\text{PW}}\mathcal{M}_{V_0,0}\mathcal{M}_{f_\text{log}}\frac{{\Lambda}}{\Lambda_V} p_\text{on}^2  \tilde f_\text{log}(p',p)\,,
\eeqa
where we have used that $\Delta_{p_\text{on}} G(p''{;p_\text{on}})=\frac{p_\text{on}^2}{p''^2}G(p''{;p_\text{on}})$.

For the particular kinematics $p'=0$, the same bounds hold
for $\tilde \delta_2$ and $\tilde \delta_3$:
\begin{align}
&|\tilde \delta_2|\le {\mathcal{M}^{m,n-1}}{n_\text{PW}}\mathcal{M}_{V_0,0}\mathcal{M}_{f_\text{log}}\frac{{\Lambda}}{\Lambda_V}p_\text{on}^2  \tilde f_\text{log}(p',p)\,,\nonumber\\
 &|\tilde \delta_3|\le 
 2{\mathcal{M}^{m,n-1}}{n_\text{PW}}\mathcal{M}_{V_0,0}\mathcal{M}_{f_\text{log}}\frac{{\Lambda}}{\Lambda_V} p_\text{on}^2  \tilde f_\text{log}(p',p)\,,
\end{align}
since $\tilde f_\text{log}(0,p)\le\tilde f_\text{log}(p',p)$.

Finally, using Eqs.~\eqref{Eq:bounds_Delta_V0_l_0_text1}~\eqref{Eq:bounds_Delta_V0_l_0_text2}, we obtain the following bound for $\delta_4$:
\beqa
 |\delta_4|&\le& {\mathcal{M}^{m,n-1}}
\int \frac{|p''|^2dp''}{(2\pi)^3}(|p|''^2+p_\text{on}^2)f_\text{log}(p',p'')  
\left|G(p''{;p_\text{on}})
\Delta_{p}V(p'', p)\right|\nonumber\\
&\le&{\mathcal{M}^{m,n-1}}{n_\text{PW}}\mathcal{M}_{V_0,0}\left(|p|^2+p_\text{on}^2\right)
\int \frac{|p''|^2dp''}{(2\pi)^3}f_\text{log}(p',p'')  
\left|G(p''{;p_\text{on}})\right|V_{0,\text{max}}(p'',p)\nonumber\\
&\le& {\mathcal{M}^{m,n-1}}{n_\text{PW}}\mathcal{M}_{V_0,0}\mathcal{M}_{f_\text{log}}\frac{{\Lambda}}{\Lambda_V} \left(|p|^2+p_\text{on}^2\right)  \tilde f_\text{log}(p',p)\,,
\eeqa
and for $\tilde\delta_4$:
\begin{align}
 &|\tilde\delta_4|<{\mathcal{M}^{m,n-1}}{n_\text{PW}}\mathcal{M}_{V_0,0}
\mathcal{M}_{f_\text{log}}\frac{{\Lambda}}{\Lambda_V} |p|^2  \tilde f_\text{log}(p',p)\,.
\end{align}

Substituting the estimates for $\delta_1$, $\delta_2$, $\delta_3$, $\delta_4$, 
$\tilde\delta_2$, $\tilde\delta_3$, $\tilde\delta_4$ into Eqs.~(\ref{Eq:R_p_pprime},~\ref{Eq:Delta_R_p_pprime})
and adding symmetrically the second term in the definition of
$\mathds{R}_{pp'}(\tilde T_2^{[m,n]})$ in Eq.~\eqref{Eq:R_p_full_X}, we obtain
\beqa
 \left|\mathds{R}_{pp'}(\tilde T_2^{[m,n]})(p',p{;p_\text{on}})\right|
 &\le& 5 {n_\text{PW}}\mathcal{M}_{V_0,0}\mathcal{M}_{f_\text{log}}\frac{{\Lambda}}{\Lambda_V}
 (|p|^2+|p'|^2+p_\text{on}^2)
 \tilde f_\text{log}(p',p)
 \left({\mathcal{M}^{m,n-1}}+{\mathcal{M}^{m-1,n}}\right)\,,\nonumber\\
 \left|\Delta_{p'}\mathds{R}_{pp'}(\tilde T_2^{[m,n]})(p',p{;p_\text{on}})\right|
  &\le& 2 {n_\text{PW}}\mathcal{M}_{V_0,0}\mathcal{M}_{f_\text{log}}\frac{{\Lambda}}{\Lambda_V}
 (|p'|^2+p_\text{on}^2)
 \tilde f_\text{log}(p',p)
 \left({\mathcal{M}^{m,n-1}}+{\mathcal{M}^{m-1,n}}\right)\,,\nonumber\\
\left|\Delta_{p_\text{on}}\mathds{R}_{pp'}(\tilde T_2^{[m,n]})(p',p{;p_\text{on}})\right|
&\le& 3 {n_\text{PW}}\mathcal{M}_{V_0,0}\mathcal{M}_{f_\text{log}}\frac{{\Lambda}}{\Lambda_V}
 p_\text{on}^2
 \tilde f_\text{log}(p',p)
 \left({\mathcal{M}^{m,n-1}}+{\mathcal{M}^{m-1,n}}\right)\,,\nonumber\\
\left|\Delta_p\mathds{R}_{pp'}(\tilde T_2^{[m,n]})(p',p{;p_\text{on}})\right|
&\le& 2 {n_\text{PW}}\mathcal{M}_{V_0,0}\mathcal{M}_{f_\text{log}}\frac{{\Lambda}}{\Lambda_V}
 (|p|^2+p_\text{on}^2)
 \tilde f_\text{log}(p',p)
 \left({\mathcal{M}^{m,n-1}}+{\mathcal{M}^{m-1,n}}\right)\,.
\eeqa
Taking the maximum of all numerical prefactors, we finally arrive
at Eq.~\eqref{Eq:inequalities_R} with 
\begin{align}
 \mathcal{M}^{m,n}=
  \mathcal{M}_{\text{max}}\frac{{\Lambda}}{\Lambda_V}\left({\mathcal{M}^{m,n-1}}+{\mathcal{M}^{m-1,n}}\right),\quad
  \text{with} \; \;
\mathcal{M}_{\text{max}}=5 {n_\text{PW}}\mathcal{M}_{V_0,0}\mathcal{M}_{f_\text{log}}\,.
\end{align}
The above equation can be solved as
\begin{align}
\mathcal{M}^{m,n}\le \mathcal{M}^{0,0}
\Sigma_{2,0}^{m+n},\quad
  \text{with} \; \;
\Sigma_{2,0}=2\mathcal{M}_{\text{max}}\frac{{\Lambda}}{\Lambda_V}\,.
\end{align}

Now, we can combine the terms $\mathds{R}_{p}(\tilde T_2^{[m,n]})$,
$\mathds{R}_{p'}(\tilde T_2^{[m,n]})$ and $\mathds{R}_{pp'}(\tilde T_2^{[m,n]})$
to obtain $\mathds{R}(\tilde T_2^{[m,n]})$ (see Eq.~\eqref{Eq:R_p_full})
for the on-shell kinematics ($p=p'=p_\text{on}$):
\begin{align}
&\left|\mathds{R}(\tilde T_2^{[m,n]})(p_\text{on})\right|\le \frac{8\pi^2 \mathcal{M}_{T_2}}{m_N\Lambda_V} 
\Sigma_{2,0}^{m+n}
\frac{p_\text{on}^2}{\Lambda_{b}^2}
\, \log{\frac{\Lambda}{M_\pi}}\,,
\label{Eq:Power_Counting_T2_m_n}
\end{align}
where we have used that $p_\text{on}\le \mathcal{M}_{p_\text{on}} M_\pi$
and absorbed terms of order $1$ into the $\log\Lambda/M_\pi$ term
in $\tilde f_\text{log}(p_\text{on},p_\text{on})$.
Equation~\eqref{Eq:Power_Counting_T2_m_n} completes the proof of the power
counting for $\mathds{R}\left(\tilde T_2^{[m,n]}\right)$.

Note that apart from the functional dependence of $\mathds{R}(\tilde T_2^{[m,n]})(p_\text{on})$ on $\Lambda$ and
$p_\text{on}$, we have determined the dependence of the numerical coefficient in Eq.~\eqref{Eq:Power_Counting_T2_m_n}
on $m$ and $n$, which turned out to be exponential and can be absorbed
into the definition of $\Sigma_{2,0}$.
This is important because if this coefficient grew faster (say, as
factorial), the series in $m$ and $n$ would never converge.

To summarize, we have explicitly demonstrated in this section that all
power-counting breaking contribution to the NN scattering amplitude
in the $^1S_0$ and $^3S_1$-$^3D_1$ channels at NLO in the chiral
expansion can be absorbed into the
renormalization of the LECs $C_S$ and $C_T$. 

\section{Numerical results}
\label{Sec:results}
Having completed the formal renormalizability proof (in the EFT
  sense) of the LO and NLO scattering amplitude in the previous two
  sections, we are now in the position to provide numerical results for
  NN scattering within the proposed scheme.\footnote{All numerical and
    analytical calculations presented here have been performed in \emph{Mathematica} \cite{mathematica12.0}.} From the practical point
  of view, the most interesting novel feature of our scheme is the
  \emph{explicit} subtraction of the leading regulator-dependent term
  $\delta_\Lambda V_0$ (see Sec.~\ref{Sec:formalism} for details) 
  at NLO in contrast to the conventional approaches, in which the
  residual cutoff dependence in the calculated phase shifts
  is implicitly removed by adjusting the
  corresponding LECs. In the following, we will provide numerical
  evidence that the explicit subtraction of the leading regulator
  artifacts indeed allows one to significantly reduce the residual
  cutoff dependence of the amplitude at NLO (especially for soft cutoff choices). 
  We restrict ourselves to partial waves,
  for which
our current scheme is applicable, i.e.~where 
the series of  iterations of the leading order potential
can be regarded as convergent.
This excludes the $^1S_0$, $^3S_1$ and $^3P_0$
partial waves where non-perturbative effects are strong.
Therefore, we present our results for the remaining
$P$-waves: $^1P_1$, $^3P_1$, $^3P_2-{^3F}_2$ and
all $D$-waves: $^1D_2$, $^3D_2$, $^3D_3-{^3G}_3$.
We have checked that a few iterations of 
the leading order potential are sufficient to 
reproduce the whole series for these channels with high accuracy.
Higher partial waves ($F$-waves and higher) are strictly perturbative
(i.e.~no iterations of the leading order potential is necessary),
and our results reproduce, to a large extent, the purely perturbative
calculations of Ref.~\cite{Kaiser:1997mw}.

We use the following values for the numerical constants that
appear in the calculation:
the pion decay constant $F_\pi=92.1$ MeV,
the isospin average nucleon and pion masses
$m_N=938.9$~MeV, $M_\pi=138.04$~MeV.
For the nucleon axial coupling constant, we adopt the
effective value $g_A = 1.29$ following Ref.~\cite{Epelbaum:2004fk}
to take into account the Goldberger-Treiman discrepancy.
This gives for to the pion-nucleon coupling constant
$g_{\pi N} = g_A m_N/F_\pi$ the value of $g_{\pi N}^2/(4 \pi)
= 13.67$, which is consistent with the determination based on
the Goldberger-Miyazawa-Oehme sum rule $g_{\pi N}^2/(4 \pi)
= 13.69 \pm 0.20$ \cite{Baru:2010xn,Baru:2011bw},
see also \cite{Reinert:2020mcu} for a recent determination of the pion-nucleon
coupling constants from NN scattering. 

We include no contact interactions at leading order in the discussed
partial waves.
For the LO one-pion-exchange potential,
we consider two different regulators: the local regulator of the dipole form
and the Gaussian non-local regulator, see Eqs.~\eqref{Eq:V_1pi_local},~\eqref{Eq:local_formfactor_1pi},~\eqref{Eq:V_1pi_nonlocal},~\eqref{Eq:F_exp}.
The dipole form of the local regulator is chosen to demonstrate that
even such a slowly decreasing with momenta regulator is sufficient 
to regularize (and renormalize) the scattering amplitude.
The results obtained with the Gaussian local regulator (see Eq.~\eqref{Eq:1pi_exchange_local_gaussian_regulator})
are very similar.
The choice of the Gaussian regulator for the non-local case is motivated
by the fact that a power-law form will not be sufficient to regularize (and renormalize)
the amplitude at all (potentially high) orders
as can be seen from the bounds in Appendix~\ref{Sec:nonlocal_formfactor_bounds}, e.g. in Eq.~\eqref{Eq:Delta_p_non_local_formfactor}, where the power of the form factor
is reduced after each subtraction.
In all the cases that we consider in this section, there is no
qualitative difference among various forms of the non-local regulators.
The values of the cutoff are varied in the wide range:
from the extremely small value $\Lambda_{1\pi}=300$~MeV to $\Lambda_{1\pi}=800$~MeV, i.e.
$\Lambda_{1\pi}\sim\Lambda_{b}$.

We have also introduced a regulator for the $2$-pion exchange potential:
the local regulator of power $2$, $F_{\Lambda_{2\pi}, 2}(q)=\left[\Lambda_{2\pi}^2/(\Lambda_{2\pi}^2+q^2)\right]^2$,
when the one-pion-exchange potential is regularized with the local form factor
and $F_{\Lambda_{2\pi}, \text{exp}}(p',p)=\exp\left[-(p^2+p'\,^2)/\Lambda_{2\pi}^2\right]$
when the one-pion-exchange potential is regularized with the non-local form factor. For definiteness, we have chosen the intermediate value of  
$\Lambda_{2\pi}=600$~MeV.

Regularizing the two-pion-exchange potential is not necessary (from the mathematical point of view), but it improves convergence in some of the 
$D$-waves by removing some short-range pieces of the potential.
We have used this possibility to demonstrate the flexibility of the scheme.

\begin{figure}[tb]
\includegraphics[width=\textwidth]{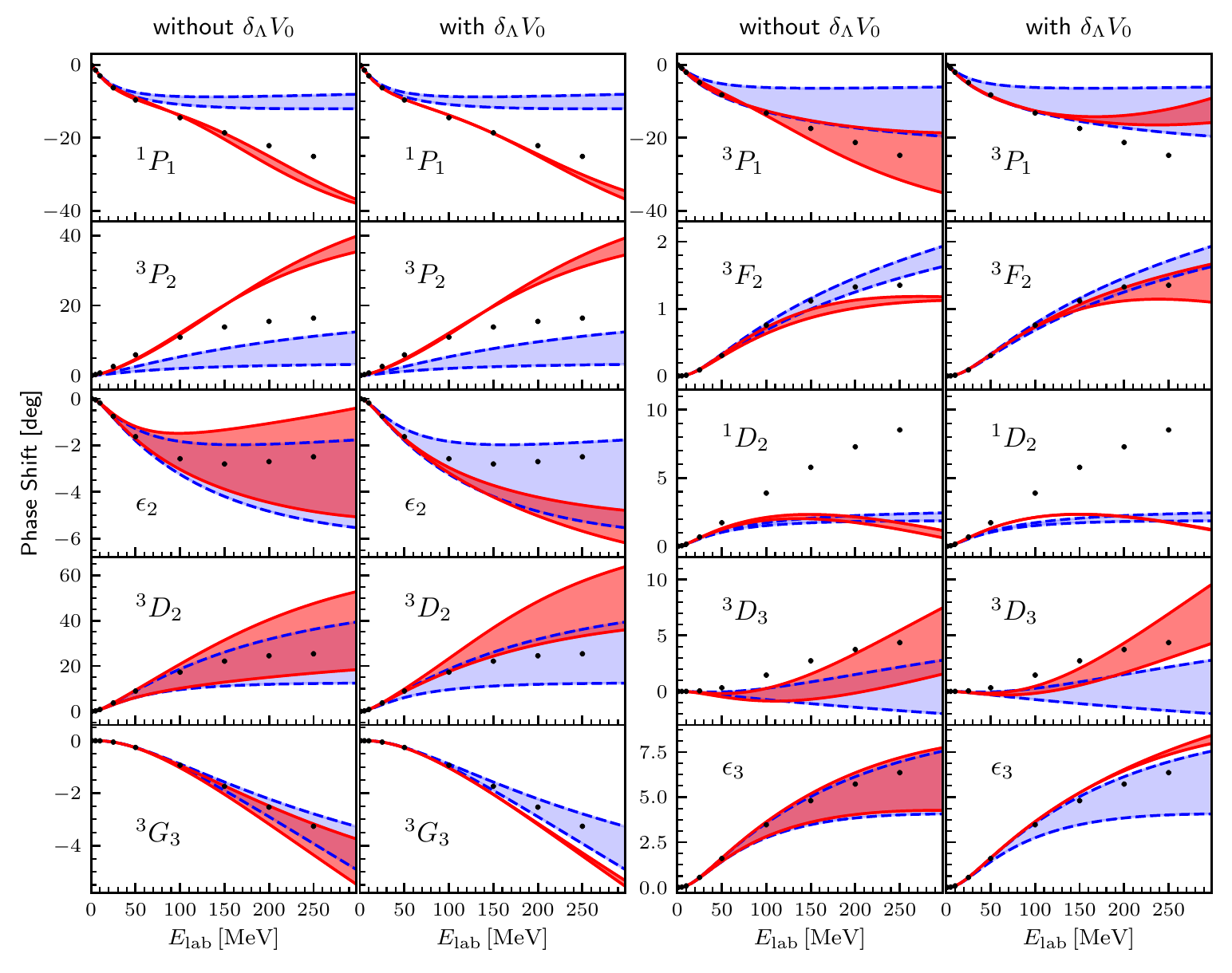}
\caption{The results of the leading-order (blue dashed lines)
and next-to-leading-order (red solid lines)  calculations with the local regulator for selected $P$- and $D$-wave phase shifts. The bands indicate the variation of the one-pion-exchange cutoff within the range $\Lambda_{1\pi}\in(300,800)$~MeV. 
The right columns correspond to the NLO potential with the regulator
correction $\delta_\Lambda V_0$, while the results in the left columns
are obtained without this term. The plots were created using Matplotlib \cite{Hunter:2007}.
} \label{Fig:plots_local_cutoff}
\end{figure}

\begin{figure}[tb]
\includegraphics[width=\textwidth]{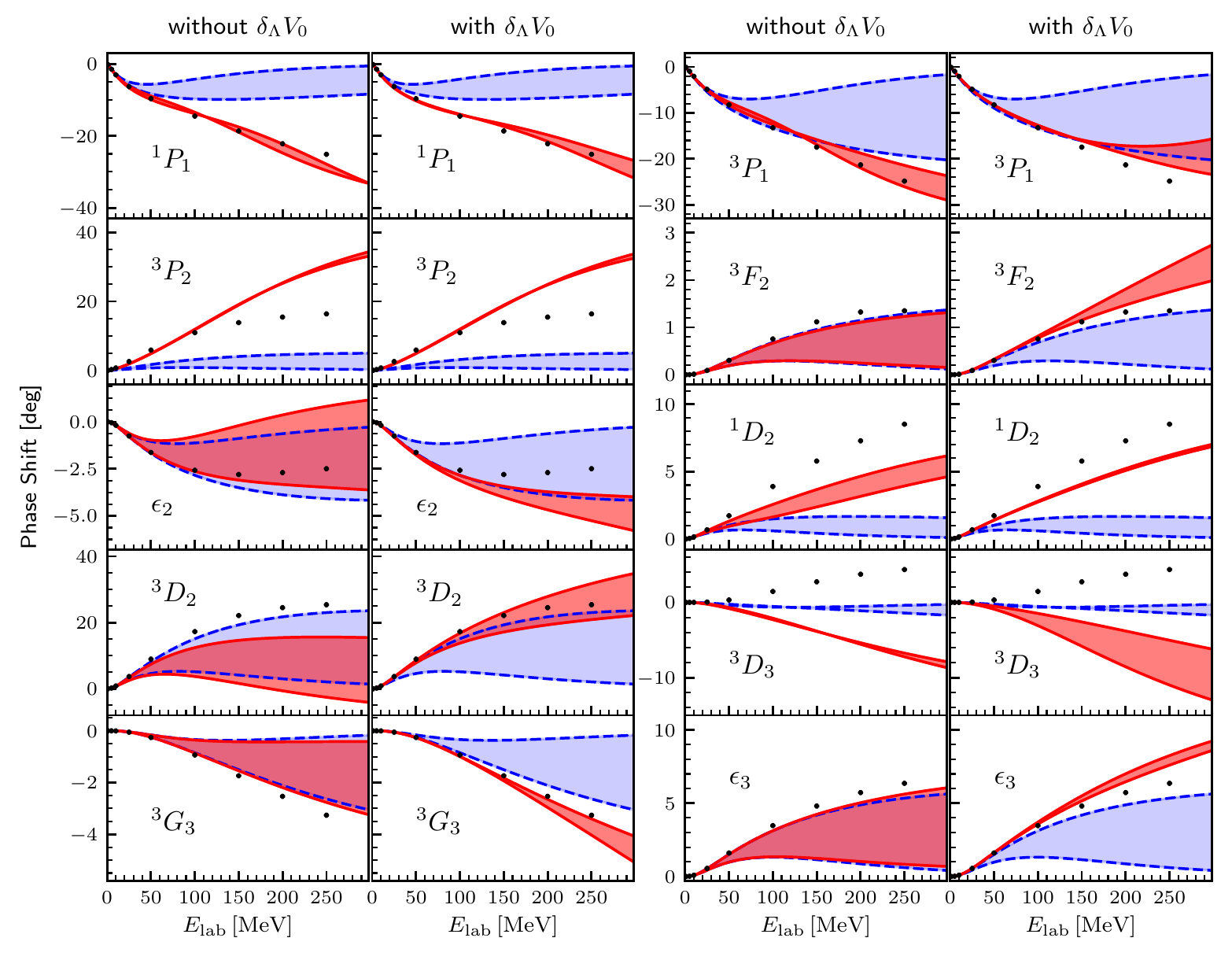}
\caption{The results of the leading-order (blue dashed lines)
and next-to-leading-order (red solid lines)  calculations with the non-local regulator for selected $P$- and $D$-wave phase shifts. The bands indicate the variation of the one-pion-exchange cutoff within the range $\Lambda_{1\pi}\in(300,800)$~MeV. 
The right columns correspond to the NLO potential with the regulator
correction $\delta_\Lambda V_0$, while the results in the left columns
are obtained without  this term.} \label{Fig:plots_nonlocal_cutoff}
\end{figure}

The results for the phase shifts
of the LO and NLO calculations
are presented in Figs.~\ref{Fig:plots_local_cutoff},~\ref{Fig:plots_nonlocal_cutoff}.
The bands correspond to the variation of the cutoff.
Two versions of the NLO interactions are shown: the one that contains 
the regulator correction $\delta_\Lambda V_0$ and the one that does not.
Our calculations are compared with the results of the Nijmegen partial wave analysis~\cite{Stoks:1993tb}.
For $P$-waves, we have fitted the leading contact interactions
appearing at next-to-leading order to the Nijmegen phase shifts up to
$E_\text{lab}=150$~MeV.
The results for $D$-waves do not involve free parameters.

As one can see from the figures, the convergence pattern is 
reasonable in all studied partial waves if one considers the energy
region $E_\text{lab}\lesssim 150$~MeV or $p_\text{on}\lesssim 260$~MeV.
If one aims at constructing an efficient scheme at higher energies
(where the expansion parameter gets larger),
the natural procedure would be to promote the $P$-wave contact interactions
to leading order (e.g.~in the $^1P_1$ and $^3P_2$ waves) and
some $D$-wave contact terms to next-to-leading order (e.g.~in the
$^3D_2$ wave) in order to prevent possible 
issues with a large violation of unitarity within the
schemes that rely on a perturbative treatment of subleading terms.

Looking at the cutoff dependence of our NLO results, one indeed 
observes that the corresponding bands, in most cases, become narrower upon explicitly
including the 
$\delta_\Lambda V_0$ correction. 
One noticeable exception is the $^3F_2$ partial wave with the local
regulator. However, a comparison to the results obtained using a nonlocal
regulator suggests that the small band in calculations without
the $\delta_\Lambda V_0$ correction is accidental. Notice further that
the phase shift in the  $^3F_2$ partial wave is several times smaller
than in the other $F$-waves, and the resulting band width is still
considerably smaller compared to the leading short-range contribution
in that channel \cite{Reinert:2017usi}. 
Another exception is the $^3D_3$ partial wave with the non-local
regulator. In this case, we again observe large cancellations 
for the NLO amplitude without $\delta_\Lambda V_0$. Notice that the boundaries of the 
shown band correspond to the limiting values of the cutoffs ($300$ and $800$~MeV), whereas for the intermediate values, the curves go slightly outside of the
band (and are not shown). This partial wave is known to suffer from a
very slow convergence of the EFT expansion due to the negligibly small
contribution of the unregularized one-pion-exchange potential. 
Finally, one observes that the  bands 
for the NLO results with $\delta_\Lambda V_0$ become particularly
narrow for the $^1D_2$, $^3G_3$, $\epsilon_3$ partial waves,
which is an indication that the LO interaction in those partial waves is
purely perturbative.

In general, one expects that the cutoff dependence will be further reduced
upon inclusion of higher-order corrections in $\delta_\Lambda V_0$.

\newpage
\section{Summary and outlook} \label{Sec:summary}
We have analyzed the nucleon-nucleon interaction in chiral effective theory with a finite cutoff,
i.e., a cutoff of the order of the hard scale.
We have formulated a scheme based on the chiral effective Lagrangian
that allows one to take into account the leading-order interaction
non-perturbatively and to perform a perturbative expansion in terms of
the subleading interactions. The dependence of the theory on the
cutoff is 
also eliminated perturbatively, order by order, without changing a non-perturbative regime.
This scheme ensures that the basic principles of EFT are met on the
formal level: the constructed $S$-matrix is most general and satisfies
analyticity, perturbative unitarity and a power counting, and the
symmetries of  
the underlying theory are maintained.

The main question we addressed in this study is whether the power
counting is still satisfied for the renormalized amplitude within our scheme.
Renormalization of the scattering amplitude is necessary not only because of divergent loop integrals,
but also due to the presence of positive powers of the (finite) cutoff 
that violate the naive power counting. We have shown that in the $S$-wave nucleon-nucleon 
amplitudes, one can absorb the power-counting breaking terms at all
orders in the LO interaction by a renormalization of the leading order
contact interactions as long as the cutoff is chosen to be of the
order of the hard scale. 
In order to prove this statement, we have implemented the BPHZ subtraction procedure. 
A suitable sector decomposition of the multiple momentum integrals,
combined with the rotation of the integration contour into the complex plain,
allowed us to impose the relevant bounds on the considered amplitudes.
For higher partial waves, we have demonstrated that no power-counting
breaking terms appear at any order in the LO interaction.  
These results have been obtained for a rather general class of local and non-local regulators.
The renormalizability has been proven at NLO to all orders in the LO
interaction (but not for the purely non-perturbative case). 

In order to illustrate how this scheme works in practice,
we have calculated the next-to-leading-order nucleon-nucleon amplitude for 
the $P$- and $D$-waves, where the leading order interaction can be regarded 
as being sufficiently perturbative, i.e.~one can numerically reproduce the full
result with a finite number of iterations. In the $P$-waves, the contact terms quadratic in
momenta have been fitted to the result of the Nijmegen
partial-wave analysis, 
whereas the $D$-wave amplitudes are parameter-free.
The resulting phase shifts are shown in
Figs.~\ref{Fig:plots_local_cutoff} and \ref{Fig:plots_nonlocal_cutoff}
up to the laboratory energy of $E_\text{lab}=300$~MeV.
We have also analyzed perturbative corrections due to the finite-cutoff
artifacts.
As expected, we found that their explicit
inclusion allows one to
significantly reduce the residual cutoff dependence in most cases, especially when the local form
of the regulator is used. 

As the next step, based on the findings of the present paper, we are
going to consider the situation when the leading order interaction is
essentially non-perturbative and the LO amplitude cannot be reproduced by a finite
number of iterations. 
It is also important to extend the scheme to higher chiral orders and
to reactions involving electroweak currents. Last but not least, it
would  also be
interesting to apply our method to few-nucleon systems, in particular,
regulated on the lattice.

\section*{Acknowledgments} 
We would like to thank Jambul Gegelia for helpful discussions and for useful comments 
on the manuscript.
This work was supported by DFG  (Grant No. 426661267), by DFG and NSFC through funds provided to the
Sino-German CRC 110 ``Symmetries and the Emergence of Structure in QCD'' (NSFC
Grant No.~12070131001, Project-ID 196253076 - TRR 110) and by BMBF (Grant No.~05P18PCFP1).

\newpage
\appendix
\section{Bounds for real and complex momenta}
The components of the initial and final nucleon c.m. momenta $p$ and $p'$
are given by
\begin{align}
\vec p = p\begin{pmatrix} 0 \\ 0 \\ 1 \end{pmatrix}\,,\quad
\vec p\,' = p'\begin{pmatrix} \sin \theta \cos \phi  \\ \sin \theta \sin \phi \\ \cos \theta \end{pmatrix}\,,
\end{align}
where $p$ is either $p=p_\text{on}$ or lies on the complex contour $p\in \mathcal{C}$:
$p=|p|\exp(-i\alpha_{\mathcal{C}})$, and  
$p'$ is either $p'=p_\text{on}$ or 
$p'=|p'|\exp(-i\alpha_{\mathcal{C}})$.

The following obvious inequalities for $\vec q = \vec p\,'-\vec p$ and $\vec k = ( \vec p\,'+\vec p)/2$ hold:
\begin{align}
&|q_i|\le |p|+|p'|\,,\quad |q_i q_j|\le 2(|p|^2+|p'|^2)\,, \quad |q^2|\le 2(|p|^2+|p'|^2)\,,
\label{Eq:bounds_qi}
\end{align}
and
\begin{align}
&|k_i|\le \frac{1}{2}(|p|+|p'|)\,,\quad |k_i k_j|\le \frac{1}{2}(|p|^2+|p'|^2)\,, \quad |k_i q_j|\le |p|^2+|p'|^2\,, \quad  |k^2|\le \frac{1}{2}(|p|^2+|p'|^2)\,.
\label{Eq:bounds_ki}
\end{align}

\section{Bounds on the typical denominator}
Below, we obtain bounds for the typical denominator 
\begin{align}
&q^2+\mu^2=p'^2+p^2-2p p' x+\mu^2\,,
\end{align}
with $\mu\ge M_\pi$.

It is convenient to provide unified bounds for all considered momenta
$p$ and $p'$ regardless of
whether they are real and on shell or lie on the complex contour.
The following estimates will be proven:
\begin{align}
&\big|q^2+\mu^2\big|\ge \mathcal{M}_{\text{den},1}\left(\sum_{i=1}^3|q_i|^2+\mu^2\right)
\ge \mathcal{M}_{\text{den},1}\big(|q|^2+\mu^2\big)\,,
\label{Eq:estimates_denominator1}\\
&\big|q^2+\mu^2\big|\ge \mathcal{M}_{\text{den},2}\big(|p|^2+|p'|^2-2|p||p'|x+\mu^2\big)\ge \mathcal{M}_{\text{den},2}\big[(|p|'-|p|)^2+\mu^2\big]\,.
\label{Eq:estimates_denominator2}
\end{align}
When both momenta $p$ and $p'$ are on shell,
$\mathcal{M}_{\text{den},1}=\mathcal{M}_{\text{den},2}=1$.
When both momenta $p$ and $p'$ lie on the complex contour $p, \,p'\in\mathcal{C}$,
$\mathcal{M}_{\text{den},1}=\mathcal{M}_{\text{den},2}=\cos(\alpha_{\mathcal{C}})$:
\begin{align}
&|q^2+\mu^2|=|e^{-2i\alpha_{\mathcal{C}}}|q^2|+\mu^2|=|e^{-i\alpha_{\mathcal{C}}}|q^2|+e^{i\alpha_{\mathcal{C}}}\mu^2|>\Re(e^{-i\alpha_{\mathcal{C}}}|q^2|+e^{i\alpha_{\mathcal{C}}}\mu^2)=\cos(\alpha_{\mathcal{C}})(|q^2|+\mu^2)\,.
\end{align}
Now, we prove Eq.~\eqref{Eq:estimates_denominator1} for the configuration when
$p=p_\text{on}$ and $p'\in\mathcal{C}$.

The real part of $q^2$ satisfies:
\beqa
\Re(q^2)&=& |p'|^2\cos(2\alpha_{\mathcal{C}})+p_\text{on}^2-2|p'|p_\text{on}x\cos(\alpha_{\mathcal{C}})
 > -\mathcal{M}_{\alpha_{\mathcal{C}},1} M_\pi^2 \,,\nonumber\\
\mathcal{M}_{\alpha_{\mathcal{C}},1}&=&\frac{1}{4}[1-\tan^2(\alpha_{\mathcal{C}})]\approx\frac{1}{4}\,,
\label{Eq:estimates_re_q2}
\eeqa
where we have used that $p_\text{on}<\frac{M_\pi}{\tan(2\alpha_{\mathcal{C}})}$\,.
It follows then:
\beqa
 |q^2+\mu^2|&>&\Re(q^2+\mu^2)>M_\mu\mu^2\,,\nonumber\\
M_\mu&=&1-\mathcal{M}_{\alpha_{\mathcal{C}},1}\frac{M^2}{\mu^2}\,.
\label{Eq:estimates_denominator3}
\eeqa
On the other hand,
\beqa
 |q^2+\mu^2|&=&||p'|^2-2|p'|p_\text{on}x e^{i\alpha_{\mathcal{C}}}+e^{2i\alpha_{\mathcal{C}}}(p_\text{on}^2+\mu^2)|\nn
&>&\Re(|p'|^2-2|p'|p_\text{on}x e^{i\alpha_{\mathcal{C}}}+e^{2i\alpha_{\mathcal{C}}}(p_\text{on}^2+\mu^2))\nonumber\\
&=&|p'|^2-2|p'|p_\text{on}x \cos(\alpha_{\mathcal{C}})+\cos(2\alpha_{\mathcal{C}})(p_\text{on}^2+\mu^2)\nn
&>&|p'|^2+p_\text{on}^2-2|p'|p_\text{on}x\cos(\alpha_{\mathcal{C}})+\mathcal{M}_{\alpha_{\mathcal{C}},2} M_\pi^2\nonumber\\
&>&|p'|^2+p_\text{on}^2-2|p'|p_\text{on}x\cos(\alpha_{\mathcal{C}})\,,\label{Eq:estimates_denominator4}\\
\mathcal{M}_{\alpha_{\mathcal{C}},2}&=&\cos(2\alpha_{\mathcal{C}})+\cot^2(2\alpha_{\mathcal{C}})
[\cos(2\alpha_{\mathcal{C}})-1]=1-\frac{1}{2\cos^2(\alpha_{\mathcal{C}})}\approx\frac{1}{2}\,.\nonumber
\eeqa
We can rewrite Eq.~\eqref{Eq:estimates_denominator4} as
\begin{align}
& |q^2+\mu^2|>|p'|^2(1-x^2)+|p_\text{on}-p'x|^2=\sum_i|q_i|^2\,.
\label{Eq:estimates_denominator5}
\end{align}
Combining Eq.~\eqref{Eq:estimates_denominator3} and Eq.~\eqref{Eq:estimates_denominator5},
we obtain the bound~\eqref{Eq:estimates_denominator1}:
\beq
|q^2+\mu^2|\ge \mathcal{M}_{\text{den},1}\left(\sum_i|q_i|^2+\mu^2\right)\,, \quad  
\mathcal{M}_{\text{den},1}=\frac{1-\mathcal{M}_{\alpha_{\mathcal{C}},1}}{2-\mathcal{M}_{\alpha_{\mathcal{C}},1}}<\frac{M_\mu}{M_\mu+1}\,.
\label{Eq:estimates_denominator6}
\eeq

From Eq.~\eqref{Eq:estimates_denominator4}, we can also deduce
\begin{align}
& |q^2+\mu^2|>\cos(\alpha_{\mathcal{C}})\left(|p'|^2+p_\text{on}^2-2p_\text{on} |p'|x\right)\,.
\label{Eq:estimates_denominator7}
\end{align}
Combining Eq.~\eqref{Eq:estimates_denominator3} and Eq.~\eqref{Eq:estimates_denominator7},
we obtain the bound~\eqref{Eq:estimates_denominator2}:
\beqa
|q^2+\mu^2|&\ge& \mathcal{M}_{\text{den},2}(|p'|^2+p_\text{on}^2-2|p'|p_\text{on}x+\mu^2)\,, \nonumber\\
\mathcal{M}_{\text{den},2}&=&\cos(\alpha_{\mathcal{C}})\frac{1-\mathcal{M}_{\alpha_{\mathcal{C}},1}}{1+\cos(\alpha_{\mathcal{C}})-\mathcal{M}_{\alpha_{\mathcal{C}},1}}<
\frac{M_\mu \cos(\alpha_{\mathcal{C}})}{M_\mu+\cos(\alpha_{\mathcal{C}})}\,.
\label{Eq:estimates_denominator8}
\eeqa
Equations.~\eqref{Eq:estimates_denominator6},~\eqref{Eq:estimates_denominator8} 
can be combined with the cases $p=p'=p_\text{on}$ and $p\,,p'\in\mathcal{C}$
to arrive at Eqs.~\eqref{Eq:estimates_denominator1},~\eqref{Eq:estimates_denominator2}.
It is important to emphasize that $\mathcal{M}_{\text{den},1}$ and $\mathcal{M}_{\text{den},2}$ do not depend on $\mu$.

From Eqs.~\eqref{Eq:estimates_denominator1},~\eqref{Eq:estimates_denominator2}, it follows:
\begin{align}
& \left|\frac{q_i q_j}{q^2+\mu^2}\right|<\mathcal{M}_{\text{den},1}^{-1}\,,\quad
\left|\frac{q^2}{q^2+\mu^2}\right|<\mathcal{M}_{\text{den},1}^{-1}\,,
\quad i,j=1,2,3\,,
\label{Eq:bounds_qi_qj_over_q2}
\end{align}
and
\beqa
  \left|\frac{( \vec{k} \times \vec{q})_i}{q^2+\mu^2}\right|
 &=&\left|\frac{( \vec{p} \times \vec{p}\,')_i}{q^2+\mu^2}\right|
 \le\mathcal{M}_{\text{den},2}^{-1}
\frac{|p| |p'|\sqrt{1-x^2}}{|p|^2+|p'|^2-2|p||p'|x+\mu^2}
=\mathcal{M}_{\text{den},2}^{-1}
\frac{|p| |p'|\sqrt{1-x^2}}{\left(|p|-|p'|\right)^2+2|p||p'|(1-x)+\mu^2}
\nonumber\\
&\le&\mathcal{M}_{\text{den},2}^{-1}
\frac{|p| |p'|\sqrt{1-x^2}}{|p||p'|(1-x^2)+\mu^2}
\le\mathcal{M}_{\text{den},2}^{-1}\left(1-x^2\right)^{-1/2}
\,,\quad i=1,2,3\,.
\label{Eq:bounds_qi_kj_over_q2}
\eeqa

\section{Bounds for subtractions}
\label{Sec:subtraction_remainders}
In this section, we provide bounds for the remainders $\Delta_p^{(n)}$, 
$\Delta_{p'}^{(n)}$ for various functions relevant for our study.

We define the subtraction remainders for a function $f(p',p)$ as:
\beqa
\Delta_p^{(n)} f(p',p)&=& 
f(p',p)-\sum_{i=0}^{n}\frac{\partial^i f(p',p)}{i!(\partial p)^i}\bigg|_{p=0}p^i\,,\nonumber\\
\Delta_{p'}^{(n)} f(p',p)&=& 
f(p',p)-\sum_{i=0}^{n}\frac{\partial^i f(p',p)}{i!(\partial p')^i}\bigg|_{p'=0}(p')^i\,,\nonumber\\
\Delta_p f(p',p)&\equiv& \Delta_p^{(0)} f(p',p)\,,\nn
\Delta_{p'} f(p',p)&\equiv& \Delta_{p'}^{(0)} f(p',p)\,.
\label{Eq:remainders}
\eeqa
For a product of two functions $f_1(p)f_2(p)$, it is convenient to utilize
the identities:
\beqa
\Delta_p \left (f_1(p)f_2(p)\right)&=&f_1(p)\Delta_p f_2(p)+\Delta_p f_1(p) f_2(0)\,,\nonumber\\
\Delta^{(n)}_p \left (f_1(p)f_2(p)\right)&=&f_1(p)\Delta^{(n)}_p f_2(p)
+\sum_{i=0}^n\Delta^{(n-i)}_p f_1(p)\frac{f_2^{(i)}(0)}{i!} p^i
\label{Eq:subtractions_product1}\\
&=&f_1(p)\Delta^{(n)}_p f_2(p)
+\Delta^{(n)}_p f_1(p) f_2(p)\nonumber\\
&+&\sum_{i=1}^n\Delta^{(n-i)}_p f_1(p) \Delta^{(i-1)}_p f_2(p)
-\sum_{i=0}^n\Delta^{(n-i)}_p f_1(p) \Delta^{(i)}_p f_2(p)
\label{Eq:subtractions_product2} \,.
\eeqa

\subsection{Polynomials in momenta}
Consider a homogeneous polynomial $Q_m(p',p)$:
\begin{align}
& Q_m(p',p)=\sum_{\alpha,\beta}
\mathcal{M}_{\alpha\beta}p^\alpha p'^\beta\,, \quad \alpha+\beta=m\,,
\label{Eq:polynomial_Qm}
\end{align}
e.g., $q^2$, $q_i q_j q^2$, $\vec q\times\vec k$, etc.
Since $Q_m(p',p)$ is a polynomial in $p$ and $p'$, the following inequalities
for the residues hold:
\beq
\left|\Delta_p^{(n)}Q_m(p',p)\right|\; \le \;  \mathcal{M}_Q^{m,n}|p'|^{m-n-1}|p|^{n+1} 
\; =\; \mathcal{M}_Q^{m,n}\left|\frac{p}{p'}\right|^{n+1}|p'|^m\,, \quad
\text{ if }|p'|>|p|\,,
\label{Eq:residue_Qm}
\eeq
and symmetrically for $\Delta_{p'}^{(n)}Q_m(p',p)$.

The derivatives of $Q_m(p',p)$ are bounded by (including the case $n=0$)
\begin{align}
p^n\frac{\partial^n Q_m(p',p)}{(\partial p)^n}\bigg|_{p=0}\le
\mathcal{M}_{\partial Q}^{m,n}\left|\frac{p}{p'}\right|^{n}|p'|^m\,,
\label{Eq:Derivatives_Qm}
\end{align}
and symmetrically for $p\leftrightarrow p'$.

\subsection{Non-local form factor}
\label{Sec:nonlocal_formfactor_bounds}
For a non-local form factor of the form
\beq
 F_{\Lambda,m}(p',p) = F_{\Lambda,m}(p')F_{\Lambda,m}(p)\,,\quad
F_{\Lambda,m}(p)=\left(\frac{\Lambda^2}{p^2+\Lambda^2}\right)^m\,,
\label{Eq:nonlocal_formfactor_m}
\eeq
the remainders $\Delta_p^{(n)}$ can be represented as
\begin{align}
&\Delta_p^{(n)} F_{\Lambda,m}(p',p)
=\left(\frac{\Lambda^2}{p'^2+\Lambda^2}
\frac{\Lambda^2}{p^2+\Lambda^2}\right)^m
\frac{p^{n+1}\sum_{\alpha+\beta=2m+n-1}\mathcal{M}_{\alpha\beta} p^\alpha \Lambda^\beta}{\Lambda^{2(n+m)}}\,.
\end{align}
Using the inequalities
\begin{align}
& |p^2+\Lambda^2|>|p|^2\,,\quad |p^2+\Lambda^2|\ge\Lambda^2\,,
\end{align}
the remainders can be bounded as 
\beqa
 \left| \Delta_p^{(n)} F_{\Lambda,m}(p',p)\right| &\le& \mathcal{M}_F^{m,n}\left|\frac{p}{\Lambda}\right|^{n+1} 
\left|F_{\Lambda,m}(p')F_{\Lambda,m}(p)\right|\nonumber\\
&\le& \mathcal{M}_F^{m,n}\left|\frac{p}{p'}\right|^{n+1} 
\left|F_{\Lambda,m-\frac{n+1}{2}}(p')\right|\,,
\quad \text{ if }|p'|>|p|\,.
\label{Eq:Delta_p_non_local_formfactor}
\eeqa

The derivatives of $F_{\Lambda,m}(p',p)$ satisfy the following inequalities:
\begin{align}
&\left|p^n \frac{\partial^n F_{\Lambda,m}(p',p)}{(\partial p)^n}\bigg|_{p=0}\right| \le \mathcal{M}_{\partial F}^{m,n}\left|\frac{p}{p'}\right|^{n} 
\left|F_{\Lambda,m-\frac{n+1}{2}}(p')\right|\,,
\label{Eq:Derivatives_non_local_formfactor}
\end{align}
including the case $n=0$.

From Eqs.~\eqref{Eq:Delta_p_non_local_formfactor},~\eqref{Eq:residue_Qm},~\eqref{Eq:Derivatives_Qm}
and the identity for the remainder of a product \eqref{Eq:subtractions_product1},
the following bound can be imposed for a function $\Phi_{k,m}(p',p)=Q_k(p',p)F_{\Lambda,m}(p',p)$,
where $Q_k(p',p)$ is an arbitrary homogeneous polynomial of degree $k$:
\begin{align}
& \left| \Delta_p^{(n)} \Phi_{k,m}(p',p)\right| \le \mathcal{M}_\Phi^{k,m,n}\left|\frac{p}{p'}\right|^{n+1} 
|p'|^k
                \left|F_{\Lambda,m-\frac{n+1}{2}}(p')\right|\,,
                \quad
\text{ if }|p'|>|p|\,,
\label{Eq:Delta_p_Phi}
\end{align}
and for the derivatives of $\Phi_{k,m}$:
\begin{align}
&\left|p^n \frac{\partial^n \Phi_{k,m}(p',p)}{(\partial p)^n}\bigg|_{p=0}\right| \le \mathcal{M}_{\partial \Phi}^{m,n}\left|\frac{p}{p'}\right|^{n} 
|p'|^k
\left|F_{\Lambda,m-\frac{n+1}{2}}(p')\right|\,,
\label{Eq:Derivatives_Phi}
\end{align}
including the case $n=0$.
All inequalities of this subsection hold also under interchange $p\leftrightarrow p'$.

\subsection{Local form factor}
\label{Sec:LocalFF_bounds}
In the case of a local regulator,
a typical function $f_\mu(p',p)$ 
\begin{align}
 f_\mu(p',p)=\frac{1}{q^2+\mu^2}=\frac{1}{p^2+p'^2-2pp'x+\mu^2}\,, \quad -1\le x\le 1\,,
  \label{Eq:f_mu}
\end{align}
appears in the one-pion-exchange and
two-pion-exchange potentials.
Its subtraction remainders are bounded as
\begin{align}
& \left| \Delta_p^{(n)} f_\mu(p',p)\right| \le
                \mathcal{M}_{f,n}\left|\frac{p}{p'}\right|^{n+1}
                \left|f_\mu(p',p)\right|\,,
                \quad
\text{ if }|p'|>|p|\,.\label{Eq:Delta_p_f}
\end{align}
Proof:
The remainder $\Delta_p^{(n)}$ can be represented as
\begin{align}
&\Delta_p^{(n)} f_\mu(p',p)
=\frac{p^{n+1} Q_{n+1}(p,p',\mu^2)}{(p^2+p'^2-2pp'x+\mu^2)(p'^2+\mu^2)^{n+1}}\,,
\end{align}
where $Q_{n+1}$ is a homogeneous polynomial
\begin{align}
&Q_{n+1}(p,p',\mu^2)=\sum_{\alpha\beta\gamma}\mathcal{M}_{\alpha\beta\gamma} p'^\alpha p^\beta\mu^\gamma\,,
\quad \alpha+\beta+\gamma=n+1\,.
\end{align}
For $|p'|>|p|$, we get:
\begin{align}
&\left|\frac{\mu}{(p'^2+\mu^2)^{1/2}}\right|<1\,,\quad
 \left|\frac{p}{(p'^2+\mu^2)^{1/2}}\right|<\left|\frac{p'}{(p'^2+\mu^2)^{1/2}}\right|<1\,, \nonumber\\
\label{Eq:inequalitity_p_ppr_M}
\end{align}
and, therefore
\beqa
\left|\frac{p'^\alpha p^\beta\mu^\gamma}{(p'^2+\mu^2)^{(n+1)}}\right|&\le&\frac{1}{|p'|^{n+1}}\,,\nonumber\\
\left| \Delta_p^{(n)} f_\mu(p',p)\right|&\le& \left|\frac{p}{p'}\right|^{n+1} \left|f_\mu(p',p)\right|
\sum_{\alpha\beta\gamma}|\mathcal{M}_{\alpha\beta\gamma}|\,,
\label{Eq:inequalitity_alpha_beta_gamma}
\eeqa
which completes the proof.
Note that coefficients $\mathcal{M}_{f,n}$ are independent of $\mu$.

For the derivatives $\frac{\partial^n f_\mu(p',p)}{(\partial p)^n}$,
we obtain
\beqa
\left|p^n\frac{\partial^n f_\mu(p',p)}{(\partial p)^n}\bigg|_{p=0}\right|
&=&\left|p^n\frac{\sum_{\alpha+\beta=n}\mathcal{M}_{n,\alpha\beta} p'^\alpha \mu^\beta}
{(p'^2+\mu^2)^{n+1}}\right|\nonumber\\
&\le&\sum_{\alpha+\beta=n}|\mathcal{M}_{n,\alpha\beta}||f_\mu(p',0)|\left|\frac{p}{p'}\right|^{n} \nn
&=&\mathcal{M}_{\partial f, n}|f_\mu(p',0)|\left|\frac{p}{p'}\right|^{n} \,.
\label{Eq:derivatives_f}
\eeqa

From the product formula~\eqref{Eq:subtractions_product2}, it follows that 
the same constraints as in Eq.~\eqref{Eq:Delta_p_f} and 
in Eq.~\eqref{Eq:derivatives_f} hold
for a product of two (or more) functions $f_{\mu_i}(p',p)$
\begin{align}
& f_{\{\mu_i\}}(p',p)=\prod_{i=1,k} f_{\mu_i}(p',p)
\label{Eq:product_f_mu_i}
\end{align}
and for any positive integer power of $f_\mu(p',p)$.

For a function 
$\psi_{k,\{\mu_i\}}(p',p)=Q_k(p',p) f_{\{\mu_i\}}(p',p)$,
which is a product of several local structures $f_{\mu_i}(p',p)$
and a homogeneous polynomial $Q_k(p',p)$,
the following bounds can be obtained:
\begin{align}
& \left| \Delta_p^{(n)} \psi_{k,\{\mu_i\}}(p',p)\right| \le \mathcal{M}_{\psi}^{k,n}\left|\frac{p}{p'}\right|^{n+1} 
\left(|\psi_{k,\{\mu_i\}}(p',p)|+\left|p'^k f_{\{\mu_i\}}(p',0)\right|\right)\,,\quad
\text{ if }|p'|>|p|\,,
\label{Eq:Delta_p_psi}
\end{align}
\begin{align}
&\left|p^n\frac{\partial^n \psi_{k,\{\mu_i\}}(p',p)}{(\partial p)^n}\bigg|_{p=0}\right|
\le\mathcal{M}_{\partial \psi}^{k,n}\left|p'^k f_{\{\mu_i\}}(p',0)\right|\left|\frac{p}{p'}\right|^{n} 
\,, \quad n\ge0\,.
\label{Eq:derivatives_psi}
\end{align}
They can be readily derived from the product formula~\eqref{Eq:subtractions_product1}
and Eqs.~\eqref{Eq:Delta_p_Phi}, \eqref{Eq:Derivatives_Phi}, \eqref{Eq:Delta_p_f}, \eqref{Eq:derivatives_f}.

Finally, from the product formula~\eqref{Eq:subtractions_product1}
and Eqs.~\eqref{Eq:residue_Qm}, \eqref{Eq:Derivatives_Qm}, \eqref{Eq:Delta_p_f}, \eqref{Eq:derivatives_f}, for a function
\begin{align}
&\Psi_{k,m,\{\mu_i\}}(p',p)=Q_k(p',p) F_{\Lambda,m}(p',p) f_{\{\mu_i\}}(p',p)\,,
\label{Eq:Psi}
\end{align}
the following bounds hold:
\begin{align}
& \left| \Delta_p^{(n)} \Psi_{k,m,\{\mu_i\}}(p',p)\right| \le \mathcal{M}_{\Psi}^{k,n}\left|\frac{p}{p'}\right|^{n+1} 
\Big(|\Psi_{k,m,\{\mu_i\}}(p',p)|
+\left|p'^k F_{\Lambda,m-\frac{n+1}{2}}(p')
                f_{\{\mu_i\}}(p',0)\right|\Big)\,,\quad \text{ if }|p'|>|p|\,,
\label{Eq:Delta_p_Psi}
\end{align}
\begin{align}
&\left|p^n\frac{\partial^n \Psi_{k,m,\{\mu_i\}}(p',p)}{(\partial p)^n}\bigg|_{p=0}\right|
\le\mathcal{M}_{\partial \Psi}^{k,n}\left|p'^k F_{\Lambda,m-\frac{n+1}{2}}(p') f_{\{\mu_i\}}(p',0)\right|\left|\frac{p}{p'}\right|^{n} \,, \quad n\ge0\,.
\label{Eq:derivatives_Psi}
\end{align}

The classes of functions $\Phi_{k,m}(p',p)$, $\psi_{k,\{\mu_i\}}(p',p)$ and $\Psi_{k,m,\{\mu_i\}}(p',p)$ cover all structures that are relevant for our study.

Notice further that all inequalities of this subsection hold also under the interchange $p\leftrightarrow p'$.

\subsection{Gaussian form factors}
The bounds for Gaussian local and non-local regulators 
and their subtraction remainders and derivatives 
can be reduced to the power-law regulators with an arbitrary power $m$.
In order to show this, we will use the properties of the exponential function
described below.

Since the function $\exp\left(-z^2\right)$ decreases at infinity for $\Re(z^2)>0$ faster than any power of $z$,
we can write:
\begin{align}
 &\left|\exp\left(-z^2\right)z^n\right|=\exp\left[-\Re(z^2)\right]|z|^n\le \mathcal{M}_{\text{exp},n}\,,
  \label{Eq:Gaussian_z2}
\end{align}
where we assume that  $\arg(z)=0$ or $\arg(z)=-\alpha_{\mathcal{C}}$.

The derivatives of the function $\exp\left(-z^2\right)$ are bounded by 
\begin{align}
 \left|\frac{\operatorname d^n\exp\left(-z^2\right)}{\operatorname dz^n}\bigg|_{z=0}\right|\le \mathcal{M}_{\text{exp},d,n}\,.
 \label{Eq:Gaussian_z2_derivatives}
\end{align}

The remainder of the function $\exp\xi$,
\begin{align}
  \Delta_\text{exp}^{(n)}(\xi)\equiv\Delta_\xi^{(n)}\exp\xi=
 \exp\xi-\sum_{i=0}^n\frac{\xi^n}{n!}\,,
\end{align}
obey the following constraints:
\begin{align}
  &\left|\Delta_\xi^{(n)}\exp\left(\xi\right)\right|\le
    \frac{|\xi|^{n+1}}{(n+1)!}\,,\quad \text{ for }\Re(\xi)\le 0\,,\nonumber\\
  &\left|\Delta_\xi^{(n)}\exp\left(\xi\right)\right|\le
    \frac{|\xi|^{n+1}}{(n+1)!}\exp\left[\Re (\xi)\right]\,,\quad \text{ for }\Re(\xi)> 0\,,
  \label{Eq:Delta_exp}
\end{align}
which can be proven by induction using the recurrence relation
\begin{align}
  \Delta_\text{exp}^{(n+1)}(\xi)=\xi\int_{0}^{1}
 \Delta_\text{exp}^{(n)}(\tau\, \xi)d\tau\,.
\end{align}

\subsection{Non-local Gaussian form factor}
The case of the non-local Gaussian form factor
\begin{align}
F_{\Lambda,\text{exp}}(p',p)
=F_{\Lambda, \text{exp}}(p')F_{\Lambda, \text{exp}}(p)=\exp\left[-\left(p^2+p'^2\right)/\Lambda^2\right]
\end{align}
can be reduced to the case of the non-local power-law form factor with an arbitrary power $m$ as will be shown below.

From Eqs.~\eqref{Eq:Gaussian_z2},~\eqref{Eq:Gaussian_z2_derivatives}, 
we can deduce the following inequalities for $F_{\Lambda, \text{exp}}(p',p)$ and its derivatives:
\begin{align}
 \left|F_{\Lambda, \text{exp}}(p',p)\right|
\le\mathcal{M}_{\text{exp},m}\left|
\frac{\Lambda^2}{p^2+\Lambda^2}\right|^m\left|\frac{\Lambda^2}{p'^2+\Lambda^2}\right|^m
=\mathcal{M}_{\text{exp},m}\left|F_{\Lambda, m}(p',p)\right|\,,
\label{Eq:bound_F_exp}
\end{align}
and
\begin{align}
&\left|p^n \frac{\partial^n F_{\Lambda,\text{exp}}(p',p)}{{\partial p}^n}\bigg|_{p=0}\right| \le
\left|\exp\left(-p'\,^2/\Lambda^2\right)\right|
\left|\frac{p'}{\Lambda}\right|^n
\mathcal{M}_{\text{exp},d,n}
\left|\frac{p}{p'}\right|^n
\le
\mathcal{M}_{\partial F,\text{exp}}^{m,n}\left|\frac{p}{p'}\right|^{n} 
\left|F_{\Lambda,m-\frac{n+1}{2}}(p')\right|\,,
\label{Eq:Derivatives_non_local_gaussian_formfactor}
\end{align}
and symmetrically for $p\leftrightarrow p'$.

The remainders $\Delta_p^{(n)} F_{\Lambda,\text{exp}}(p',p)$ 
can be bounded utilizing Eqs.~\eqref{Eq:Gaussian_z2},~\eqref{Eq:Delta_exp} as follows
\beqa
 \left| \Delta_p^{(n)} F_{\Lambda,\text{exp}}(p',p)\right| &\le& \frac{1}{[(n+1+\delta_n)/2]!}\left|\frac{p}{\Lambda}\right|^{n+1+\delta_n} 
\left|\exp\left(-p'^2/\Lambda^2\right)\right| \nn
&\le& \mathcal{M}_{F,\text{exp}}^{m,n}\left|\frac{p}{p'}\right|^{n+1+\delta_n} 
\left|F_{\Lambda,m-\frac{n+1}{2}}(p')\right|\nonumber\\
&\le& \mathcal{M}_{F,\text{exp}}^{m,n}\left|\frac{p}{p'}\right|^{n+1} 
\left|F_{\Lambda,m-\frac{n+1}{2}}(p')\right|\,, \quad \text{ if }|p'|>|p|\,,
\label{Eq:Delta_p_non_local_gauss_formfactor}
\eeqa
with $\delta_n=(n+1)\mod 2$
and, symmetrically, for $p\leftrightarrow p'$.

Equations~\eqref{Eq:bound_F_exp},~\eqref{Eq:Derivatives_non_local_gaussian_formfactor},~\eqref{Eq:Delta_p_non_local_gauss_formfactor} are identical to Eqs.~\eqref{Eq:nonlocal_formfactor_m},~\eqref{Eq:Derivatives_non_local_formfactor},~\eqref{Eq:Delta_p_non_local_formfactor} for the non-local power-law form factors from the point of view of the upper bounds.

\subsection{Local Gaussian form factor}
\label{Sec:Gaussian_Fq}
The case of the local Gaussian form factor
\begin{align}
F_{\Lambda, \text{exp}}(q)
=\exp(-q^2/\Lambda^2)=\exp\left[-\left(p^2+p'^2-2p p' x\right)/\Lambda^2\right]
\end{align}
can be reduced to the case of the local power-law form factor with an arbitrary power $m$ similarly to the non-local case.
First, let us show that 
\begin{align}
 \Re(q^2)+\Lambda^2\ge \mathcal{M}_\text{Re}\left|q^2+\Lambda^2\right|\,.
 \label{Eq:Re_q2_Lambda}
\end{align}
If $p,p'\in\mathcal{C}$, Eq.~\eqref{Eq:Re_q2_Lambda} holds with $\mathcal{M}_\text{Re}=\cos(2\alpha_{\mathcal{C}})$.
If $p'\in\mathcal{C}$ and $p=p_\text{on}$, we have (see Eq.~\eqref{Eq:estimates_re_q2}):
\beqa
 2\left|\Re(q^2+\Lambda^2)\right|-\left|\Im(q^2+\Lambda^2)\right|&=&
  2|p'|^2\cos(2\alpha_{\mathcal{C}})+
 2p_\text{on}^2-4|p'|p_\text{on}x\cos(\alpha_{\mathcal{C}})+
 2\Lambda^2\nonumber\\
 &&{}
 -s\left(|p'|^2\sin(2\alpha_{\mathcal{C}})-2|p'|p_\text{on}x\sin(\alpha_{\mathcal{C}})\right) \nn
 &\ge& -\frac{5 p_\text{on}^2 \sin^2(\alpha )}{2 \cos (2 \alpha )-s\sin (2 \alpha )}
 +2\Lambda^2\nonumber\\
 & \ge& -M_\pi^2\frac{5  \sin ^2(\alpha ) \cot ^2(2 \alpha )}{2 \cos
   (2 \alpha )-s\sin (2 \alpha )}+2\Lambda^2 \nn
 &>&0\,,
\eeqa
where $s=\operatorname{sgn}\left[|p'|^2\sin(2\alpha_{\mathcal{C}})-2|p'|p_\text{on}x\sin(\alpha_{\mathcal{C}})\right]$.
Therefore, Eq.~\eqref{Eq:Re_q2_Lambda} holds:
\begin{align}
\left|\Re(q^2+\Lambda^2)\right|\ge \frac{1}{3}
\left(\left|\Re(q^2+\Lambda^2)\right|
+\left|\Im(q^2+\Lambda^2)\right|\right)
\ge\frac{1}{3}\left|q^2+\Lambda^2\right|\,.
\end{align}

From Eq.~\eqref{Eq:Re_q2_Lambda} and Eq.~\eqref{Eq:Gaussian_z2}, we obtain the bound
\begin{align}
 \left|F_{\Lambda, \text{exp}}(q)\right|
\le\mathcal{M}_{\text{exp},q,m}\left|\frac{\Lambda^2}{q^2+\Lambda^2}\right|^m\,.
\label{Eq:bound_F_exp_q}
\end{align}
Since the derivative $\frac{\partial^n F_{\Lambda,\text{exp}}(q)}{{\partial p}^n}$ at $p=0$ is equal to 
\begin{align}
\frac{\partial^n F_{\Lambda,\text{exp}}(q)}{{\partial p}^n}
\bigg|_{p=0}=
F_{\Lambda,\text{exp}}(p')P_{\partial,n}(p'/\Lambda)/\Lambda^{n}\,,
\end{align}
where $P_{\partial,n}(z)$ is a polynomial of degree $n$, we can deduce from Eq.~\eqref{Eq:Gaussian_z2}
the following bounds:
\begin{align}
&\left|p^n \frac{\partial^n F_{\Lambda,\text{exp}}(q)}{{\partial p}^n}\bigg|_{p=0}\right| \le
\mathcal{M}_{\partial F,\text{exp},q}^{m,n}\left|\frac{p}{p'}\right|^{n} 
\left(\frac{\Lambda^2}{4|p'|^2+\Lambda^2}\right)^m
\le
\mathcal{M}_{\partial F,\text{exp},q}^{m,n}\left|\frac{p}{p'}\right|^{n} 
\left|F_{\Lambda,m}(p')\right|\,,
\label{Eq:Derivatives_local_gaussian_formfactor}
\end{align}
and
\begin{align}
&\left|p^n \frac{\partial^n F_{\Lambda,\text{exp}}(q)}{{\partial p}^n}\bigg|_{p=0}\right| \le
\mathcal{\tilde M}_{\partial F,\text{exp},q}^{(n)}\left|\frac{p}{p'}\right|^{n} 
\frac{|p'|^2}{\Lambda^2}
\,,
\label{Eq:Derivatives_local_gaussian_formfactor2}
\end{align}
including $n=0$.

Representing $F_{\Lambda,\text{exp}}(q)$ as
\begin{align}
 F_{\Lambda,\text{exp}}(q)=F_{\Lambda,\text{exp}}(p',p) F_{\Lambda,pp'}(p',p)\,,\quad
 F_{\Lambda,pp'}(p',p)=\exp\left(2pp' x/\Lambda^2\right)\,,
\end{align}
we can write down the remainders  $\Delta_p^{(n)} F_{\Lambda,\text{exp}}(p',p)$ 
using the product formula~\eqref{Eq:subtractions_product1}:
\begin{align}
 \Delta_p^{(n)} F_{\Lambda,\text{exp}}(q)= F_{\Lambda,pp'}(p',p) \Delta^{(n)}_p  F_{\Lambda,\text{exp}}(p',p) 
 +\sum_{i=0}^n\Delta^{(n-i)}_p F_{\Lambda,pp'}(p',p)\frac{p^i}{i!}
 \frac{\partial^i F_{\Lambda,\text{exp}}(p',p)}{{\partial p}^i}\bigg|_{p=0}\,.
 \label{Eq:Delta_p_F_exp_q_product}
\end{align}
The function $F_{\Lambda,pp'}(p',p)$ and the remainders 
$\Delta_p^{(n)} F_{\Lambda,pp'}(p',p)$ are bounded by (see Eq.~\eqref{Eq:Delta_exp})
\beqa
 \left| F_{\Lambda,pp'}(p',p)\right| &\le&  \exp\left[2\Re(p p')/\Lambda^2\right]\,,\nonumber\\
 \left| \Delta_p^{(n)} F_{\Lambda,pp'}(p',p)\right| &\le& 
 \exp\left[2\Re(p p')/\Lambda^2\right]\frac{|2 p p'|^{n+1}}{(n+1)!\Lambda^{2(n+1)}}\,.
 \label{Eq:Delta_F_Lambda_p_ppr}
\eeqa
From Eqs.~\eqref{Eq:Delta_p_F_exp_q_product},~\eqref{Eq:Delta_F_Lambda_p_ppr},
~\eqref{Eq:Derivatives_non_local_gaussian_formfactor},~\eqref{Eq:Delta_p_non_local_gauss_formfactor},
we obtain:
\beqa
 \left|\Delta_p^{(n)} F_{\Lambda,\text{exp}}(q)\right|&\le&
 \exp-\left[\Re\left(p'\,^2-2pp'\right)/\Lambda^2\right]
\left(\frac{1}{[(n+1+\delta_n)/2]!}\left|\frac p\Lambda\right|^{n+1+\delta_n}
+\left|\frac{p}{\Lambda}\right|^{n+1}
\sum_{i=0}^n\left|\frac{p'}{\Lambda}\right|^{n-i+1}
\frac{\mathcal{M}_{\text{exp},d,n}}{(n-i+1)!i!}
\right)\nonumber\\
&\le& \mathcal{M}_\Delta^{(m)}\exp\left[-\Re\left(p'\,^2-2pp'\right)/\Lambda^2\right]
\left|\frac p\Lambda\right|^{n+1}\left(\left|\frac p\Lambda\right|^{\delta_n}
+\left|\frac{p'}{\Lambda}\right|
P_{p',n}\left(\left|\frac{p'}{\Lambda}\right|\right)
\right)\,,
 \label{Eq:Deltap_Fq_general}
\eeqa
where $\delta_n=\left[(n+1)\mod 2\right]$ and $P_{p',n}$ is a polynomial of degree $n$ with non-negative coefficients.

Consider now the following three cases.
\begin{enumerate}
 \item $|p|\le \mathcal{M}_{pp'}|p'|$\,,\\
where $\mathcal{M}_{pp'}<1$ is some constant chosen such that 
\begin{align}
 \Re\left(p'^2-2  \mathcal{M}_{pp'}p p' |p'|/|p|\right)
\ge \mathcal{M}_{p'}|p'|^2\,.
\end{align}
For example, for $\mathcal{M}_{pp'}=1/4$, $\mathcal{M}_{p'}=\cos(2\alpha)-\cos(\alpha)/2\gtrsim1/4$
(assuming that $\alpha\le\pi/8$, which corresponds to $\left(p_\text{on}\right)_\text{max}\ge M_\pi$).

Then, Eq~\eqref{Eq:Deltap_Fq_general} becomes (after applying Eq.~\eqref{Eq:Gaussian_z2})
\beqa
 \left|\Delta_p^{(n)} F_{\Lambda,\text{exp}}(q)\right|
&\le& \mathcal{M}_\Delta^{(m)}\exp\left(-\mathcal{M}_{p'}|p'|^2/\Lambda^2\right)
\left|\frac p\Lambda\right|^{n+1}\left(\left|\frac{p'}{\Lambda}\right|^{\delta_n}
+\left|\frac{p'}{\Lambda}\right|
P_{p',n}\left(\left|\frac{p'}{\Lambda}\right|\right)
\right)\nonumber\\
&\le&
\mathcal{M}_{F,q,1}^{m,n}\left|\frac{p}{p'}\right|^{n+1}\left(\frac{\Lambda^2}{|p'|^2(1+1/\mathcal{M}_{pp'})^2+\Lambda^2}\right)^m \nn
&\le& \mathcal{M}_{F,q,1}^{m,n}\left|\frac{p}{p'}\right|^{n+1}\left|\frac{\Lambda^2}{q^2+\Lambda^2}\right|^m
\,,
 \label{Eq:Deltap_Fq_1}
\eeqa
 where we have used that 
 \begin{align}
  \left|q^2+\Lambda^2\right|\le \left|q^2\right|+\Lambda^2\le |p|^2+|p'|^2+2|p||p'|+\Lambda^2\le |p'|^2(1+1/\mathcal{M}_{pp'})^2+\Lambda^2\,.
 \end{align}

An alternative constraint has the form
 \begin{align}
 &\left|\Delta_p^{(n)} F_{\Lambda,\text{exp}}(q)\right|
\le \mathcal{\tilde M}_{F,q,1}^{(n)}\left|\frac{p}{p'}\right|^{n+1}\frac{|p'|^2}{\Lambda^2}
\,,
 \label{Eq:Deltap_Fq_1b}
\end{align}
which is correct also for $n=0$ since $\delta_0=1$.
 
 \item $\mathcal{M}_{pp'}|p'|\le|p|\le|p'|\le\Lambda$.
 
 Equation~\eqref{Eq:Deltap_Fq_general} yields
 \beqa
 \left|\Delta_p^{(n)} F_{\Lambda,\text{exp}}(q)\right|
&\le& \mathcal{M}_\Delta^{(m)}\exp(1)\left|\frac p\Lambda\right|^{n+1}\left(\left|\frac{p'}{\Lambda}\right|^{\delta_n}
+\left|\frac{p'}{\Lambda}\right|
P_{p',n}\left(\left|\frac{p'}{\Lambda}\right|\right)
\right)\nonumber\\&
\le& \frac{\mathcal{M}_{F,q,2}^{m,n}}{5^m}\left|\frac{p}{p'}\right|^{n+1}\nn
&\le& \mathcal{M}_{F,q,2}^{m,n}\left|\frac p\Lambda\right|^{n+1}\left|\frac{\Lambda^2}{q^2+\Lambda^2}\right|^m
\,,
 \label{Eq:Deltap_Fq_2}
\eeqa
where the last inequality follows from
 \begin{align}
  \left|q^2+\Lambda^2\right|\le |p|^2+|p'|^2+2|p||p'|+\Lambda^2
  \le 4|p'|^2+\Lambda^2\le 5\Lambda^2\,.
  \label{Eq:q2Lambda2_le}
 \end{align}
Alternatively,
 \begin{align}
 &\left|\Delta_p^{(n)} F_{\Lambda,\text{exp}}(q)\right|
\le \mathcal{\tilde M}_{F,q,2}^{(n)}\left|\frac p\Lambda\right|^{n+1}\frac{|p'|^2}{\Lambda^2}
\,.
 \label{Eq:Deltap_Fq_2b}
\end{align} 
 
 \item $\mathcal{M}_{pp'}|p'|\le|p|\le\Lambda\le|p'|$.
 \end{enumerate}
Using the definition of the remainders (Eq.~\eqref{Eq:remainders}) and
Eqs.~\eqref{Eq:bound_F_exp_q},~\eqref{Eq:Derivatives_local_gaussian_formfactor}
~\eqref{Eq:q2Lambda2_le}, we get:
\beqa
\left|\Delta_p^{(n)} F_{\Lambda,\text{exp}}(q)\right|
&\le& 
\left|F_{\Lambda,\text{exp}}(q)\right|+
\sum_{i=0}^{n}\left|\frac{\partial^i F_{\Lambda,\text{exp}}(q)}{{\partial p}^i}\bigg|_{p=0}\right|
\frac{|p|^i}{i!}\nonumber\\&
\le& \mathcal{M}_{\text{exp},q,m}\left|\frac{\Lambda^2}{q^2+\Lambda^2}\right|^m
+\left(\frac{\Lambda^2}{4|p'|^2+\Lambda^2}\right)^m\sum_{i=0}^{n}
\frac{\mathcal{M}_{\partial F,\text{exp},q}^{(m,i)}}{i!}\left|\frac{p}{p'}\right|^{i}
\nonumber\\
&\le& \frac{\mathcal{M}_{\text{exp},q,m}}{\left(\mathcal{M}_{pp'}\right)^{n+1}}\left|\frac{p}{p'}\right|^{n+1}
\left|\frac{\Lambda^2}{q^2+\Lambda^2}\right|^m
+\left|\frac{p}{p'}\right|^{n+1}\left|\frac{\Lambda^2}{q^2+\Lambda^2}\right|^m
\sum_{i=0}^{n}
\frac{\mathcal{M}_{\partial F,\text{exp},q}^{(m,i)}}{i!\left(\mathcal{M}_{pp'}\right)^{n-i+1}}\nonumber\\
&\le& \mathcal{M}_{F,q,3}^{m,n}\left|\frac{p}{p'}\right|^{n+1}\left|\frac{\Lambda^2}{q^2+\Lambda^2}\right|^m\,.
\label{Eq:Deltap_Fq_3}
\eeqa
For $m=0$, it gives:
 \begin{align}
 &\left|\Delta_p^{(n)} F_{\Lambda,\text{exp}}(q)\right|
\le \mathcal{M}_{F,q,3}^{(n)}\left|\frac{p}{p'}\right|^{n+1}
\le \mathcal{M}_{F,q,3}^{(n)}\left|\frac{p}{p'}\right|^{n+1}\frac{|p'|^2}{\Lambda^2}
\,.
 \label{Eq:Deltap_Fq_3b}
\end{align} 

Combining all three cases, we obtain:
\begin{align}
& \left| \Delta_p^{(n)} F_{\Lambda,\text{exp}}(q)\right| 
\le \mathcal{M}_{F,q}^{(n)}\left|\frac{p}{p'}\right|^{n+1} 
\left|\frac{\Lambda^2}{q^2+\Lambda^2}\right|\,, \quad
\text{ if }|p'|>|p|\,,
\label{Eq:Delta_p_local_gauss_formfactor}
\end{align}
or, alternatively,
\begin{align}
& \left| \Delta_p^{(n)} F_{\Lambda,\text{exp}}(q)\right| 
\le \mathcal{\tilde M}_{F,q}^{(n)}\left|\frac{p}{p'}\right|^{n+1} 
\frac{|p'|^2}{\Lambda^2}\,, \quad
\text{ if }|p'|>|p|\,,
\label{Eq:Delta_p_local_gauss_formfactor_b}
\end{align}
and, symmetrically, for $p\leftrightarrow p'$.

Equations~\eqref{Eq:bound_F_exp_q},~\eqref{Eq:Derivatives_local_gaussian_formfactor},~\eqref{Eq:Delta_p_local_gauss_formfactor} are identical to Eqs.~\eqref{Eq:f_mu},~\eqref{Eq:derivatives_f},~\eqref{Eq:Delta_p_f} for the local power-law form factors
with $\mu=\Lambda$ and their generalizations for higher powers $m>1$
from the point of view of the upper bounds.

\section{ Bounds on the plane-wave leading-order potential}
\label{Sec:AppD}
In this section, we provide bounds for leading order potential.
First, we consider the potential in the plane-wave basis.
We treat all matrix elements of the spin and isospin matrices as constants of order one,
and take care only of the dependence of the potential on the coupling constants, masses,
cutoffs, momenta, and other scales.

The locally regularized one-pion exchange potential in the spin-triplet channel
can be bounded using
Eqs.~\eqref{Eq:estimates_denominator1},\eqref{Eq:estimates_denominator2}
by the following inequality:
\beq
\left|V^{(0)}_{1\pi,\text{t},\Lambda}(\vec p\,',\vec p)\right|
\le \left|\frac{g_A^2}{4F_\pi^2}\sum_{i,j}\mathcal{M}_{\text{t},ij}\frac{q_iq_j}{q^2+M_\pi^2}\frac{\Lambda_{\text{t},1}^2-M_\pi^2}{q^2+\Lambda_{\text{t},1}^2}\right|
\le\frac{2\pi \mathcal{M}_\text{t}}{m_N
  \Lambda_V}\frac{\Lambda^2}{(|p'|-|p|)^2+\Lambda^2}
=\frac{2\pi \mathcal{M}_\text{t}}{m_N \Lambda_V}F_{\Lambda}(|p'|-|p|)
\label{Eq:bound_triplet_1pi_local}\,,
\eeq
where $\Lambda$ is the largest cutoff among all cutoffs used in the leading-order potential. 
For our estimates, it is sufficient to keep in the first inequality of Eq.~\eqref{Eq:bound_triplet_1pi_local} only one local form factor with
the cutoff $\Lambda_{\text{t},1}$, which is the smallest among $\{\Lambda_{\text{t},i}\}$.
The dimensionful factor $2\pi/m_N$ is introduced for convenience
in order to identify the typical hard scale of the leading-order potential 
$\Lambda_V$. In particular, we treat the commonly used scale $\Lambda_{NN}=16\pi F_\pi^2/ m_N /g_A^2$ as a quantity of order $\Lambda_V$.

If the triplet one-pion exchange potential is regularized by the
non-local form factor, we obtain
\begin{align}
&\left|V^{(0)}_{1\pi,\text{t},\Lambda}(\vec p\,',\vec p)\right|
\le \left|\frac{g_A^2}{4F_\pi^2}\sum_{i,j}\mathcal{M}_{\text{t},ij}\frac{q_iq_j}{q^2+M_\pi^2}\frac{\Lambda^2}{p'^2+\Lambda^2}\frac{\Lambda^2}{p^2+\Lambda^2}\right|
\le\frac{2\pi \mathcal{M}_\text{t}}{m_N \Lambda_V}\frac{\Lambda^2}{|p'|^2+\Lambda^2}=\frac{2\pi \mathcal{M}_\text{t}}{m_N \Lambda_V}F_{\Lambda}(|p'|)\,,
\label{Eq:bound_triplet_1pi_nonlocal}
\end{align}
or, symmetrically,
\begin{align}
&\left|V^{(0)}_{1\pi,\text{t},\Lambda}(\vec p\,',\vec p)\right|
\le \frac{2\pi \mathcal{M}_\text{t}}{m_N \Lambda_V}F_{\Lambda}(|p|)\,.
\end{align}

For the regularized one-pion exchange potential in the spin-singlet channel, we obtain the constraint
\begin{align}
&\left|V^{(0)}_{1\pi,\text{s},\Lambda}(\vec p\,',\vec p)\right|
\le \left|\frac{g_A^2}{4F_\pi^2}\frac{M_\pi^2}{q^2+M_\pi^2}\right|
\le\frac{2\pi \mathcal{M}_\text{s}}{m_N \Lambda_V} F_{M_\pi}(|p'|-|p|)\,
\label{Eq:bound_singlet_1pi}
\end{align}
independently of whether it is regularized by the local or the non-local form factor.

The bounds for the short-range part of the leading-order potential
are given by
\begin{align}
&\left|V^{(0)}_{\text{short},\Lambda}(\vec p\,',\vec p)\right|
\le \frac{2\pi \mathcal{M}_\text{short}}{m_N \Lambda_V}F_{\Lambda}(|p'|)
\,,\quad
\left|V^{(0)}_{\text{short},\Lambda}(\vec p\,',\vec p)\right|
\le \frac{2\pi \mathcal{M}_\text{short}}{m_N \Lambda_V}F_{\Lambda}(|p|)
\,.
\label{Eq:bound_short_range}
\end{align}
Here, we assume that the renormalized coupling constants 
$C_i$ have natural size, i.e. 
\begin{align}
&C_S=\mathcal{M}_S \frac{2\pi }{m_N \Lambda_V}\,, \quad
C_T=\mathcal{M}_T \frac{2\pi }{m_N \Lambda_V}\,
\label{Eq:natural_C_S_C_T}
\end{align}
and
\begin{align}
&C_i=\frac{\mathcal{M}_{C_i}}{\Lambda^2} \frac{2\pi }{m_N \Lambda_V}\,, \quad
i=1,..,7\,,
\label{Eq:natural_C_i}
\end{align}
assuming the cutoff regime $\Lambda \sim \Lambda_b$.
Note that the contact interactions quadratic in momenta are of order 
$\sim p^2/\Lambda^2$ and are suppressed for small momenta.
Nevertheless, for $p\sim \Lambda$ they are of order one and can be
regarded as leading order terms when iterated with the LO potential.

If the short-range part of the leading-order potential is local,
it is bounded by (except for the spin-orbit term, which will be considered below)
\begin{align}
\left|V^{(0)}_{\text{short},\Lambda}(\vec p\,',\vec p)\right|
&\le \frac{2\pi \mathcal{M}_\text{short}}{m_N \Lambda_V} F_{\Lambda}(|p'|-|p|)\,.
\label{Eq:bound_local_short_range}
\end{align}

Finally, the full leading-order potential satisfies
\begin{align}
&\left|V_0(\vec p\,',\vec p)\right|\le \frac{\mathcal{M}_{V_0}}{4\pi}
V_{0,\text{max}}(p',p)
\,,\quad
\left|V_0(\vec p\,',\vec p)\right|
\le \frac{\mathcal{M}_{V_0}}{4\pi}V_{0,\text{max}}(p,p')
\,,
\label{Eq:bound_full_LO}
\end{align}
where we have introduced
\begin{align}
&V_{0,\text{max}}(p',p)
=\frac{8\pi^2 }{m_N \Lambda_V}\Big[F_{\tilde\Lambda}(|p'|-|p|)+F_{\tilde\Lambda}(|p'|)\Big]\,.
\label{Eq:V0max}
\end{align}
 In general, $\tilde \Lambda=\Lambda$, but for the part of the potential that contributes only to the spin-singlet partial waves without short-range leading-order interactions one should choose
 $\tilde \Lambda=M_\pi$ as in Eq.~\eqref{Eq:bound_singlet_1pi}
 (note that $F_\Lambda(p',p,x)\ge F_{M_\pi}(p',p,x)$).

The remainders $\Delta_p^{(n)} V_0(\vec p\,',\vec p)$ for $|p'|>|p|$ can be estimated using 
Eqs.~\eqref{Eq:Delta_p_non_local_formfactor}, \eqref{Eq:Delta_p_Phi}, \eqref{Eq:Delta_p_psi}, \eqref{Eq:Delta_p_Psi}:
\begin{align}
&\left|\Delta_p^{(n)} V_0(\vec p\,',\vec p)\right| \le \frac{\mathcal{M}_{\Delta V_0,n}}{4\pi}
\left|\frac{p}{p'}\right|^{n+1} V_{0,\text{max}}(p',p)\,,\quad \text{ if } |p'|>|p|\,,\nonumber\\
&\left|\Delta_{p'}^{(n)} V_0(\vec p\,',\vec p)\right| \le \frac{\mathcal{M}_{\Delta V_0,n}}{4\pi}
\left|\frac{p'}{p}\right|^{n+1} V_{0,\text{max}}(p,p')\,,\quad \text{ if } |p|>|p'|\,.
\label{Eq:Delta_p_V0_constrained}
\end{align}
Using Eqs.~\eqref{Eq:Derivatives_Phi}, \eqref{Eq:derivatives_Psi} we can estimate the derivatives of the leading-order potential:
\begin{align}
&\left|p^m\frac{\partial^m V_0(\vec p\,',\vec p)}{(\partial p)^m}\bigg|_{p=0}\right|
\le \frac{2\pi \mathcal{M}_{\partial V_0,m}}{m_N \Lambda_V}F_{\tilde\Lambda}(|p'|)\left|\frac{p}{p'}\right|^m
\le \frac{\mathcal{M}_{\partial V_0,m}}{4\pi}\left|\frac{p}{p'}\right|^mV_{0,\text{max}}(p',p)
\,,\label{Eq:V0_derivatives_a}\\
&\left|p'^m\frac{\partial^m V_0(\vec p\,',\vec p)}{(\partial p')^m}\bigg|_{p'=0}\right|
\le \frac{2\pi \mathcal{M}_{\partial V_0,m}}{m_N \Lambda_V}F_{\tilde\Lambda}(|p|)\left|\frac{p'}{p}\right|^m
\le \frac{\mathcal{M}_{\partial V_0,m}}{4\pi}\left|\frac{p'}{p}\right|^m V_{0,\text{max}}(p,p')\,,
\label{Eq:V0_derivatives_b}
\end{align}
including the case $m=0$.
Applying Eq.~\eqref{Eq:V0_derivatives_a} (Eq.~\eqref{Eq:V0_derivatives_b})
to the definition of $\Delta_p^{(n)} V_0(\vec p\,',\vec p)$
($\Delta_{p'}^{(n)} V_0(\vec p\,',\vec p)$) in Eq.~\eqref{Eq:remainders}
for $|p|>|p'|$ ($|p'|>|p|$), and combining it with Eq.~\eqref{Eq:Delta_p_V0_constrained}, we obtain the following bounds for the remainders:
\begin{align}
&\left|\Delta_p^{(n)} V_0(\vec p\,',\vec p)\right| \le \frac{\mathcal{M}_{\Delta V_0,n}}{4\pi}
\left|\frac{p}{p'}\right|^{n+1} V_{0,\text{max}}(p',p)\,,\nonumber\\
&\left|\Delta_{p'}^{(n)} V_0(\vec p\,',\vec p)\right| \le \frac{\mathcal{M}_{\Delta V_0,n}}{4\pi}
\left|\frac{p'}{p}\right|^{n+1} V_{0,\text{max}}(p,p')\,,
\label{Eq:Delta_p_V0}
\end{align}
which are valid for all considered $p$ and $p'$.

To make the general formulae simpler, we did not include in our
estimates so far the case of the locally regulated
spin-orbit term
\begin{align}
V^{(0)}_{C_5}(\vec p\,',\vec p)=C_5 \frac{i}{2}( \vec{\sigma}_1 + \vec{\sigma}_2)
\cdot ( \vec{k} \times \vec{q})\left(\frac{\Lambda_5^2}{q^2+\Lambda_5^2}\right)^{n_5}\,, 
\end{align}
with $n_5>1$ (or with the Gaussian form factor).
This potential has a more singular behavior at large momenta, because
$\left|(\vec{k} \times \vec{q})_i/(q^2+\Lambda_5^2)\right|$
is bounded by a factor $\sim 1/\sin(\theta)$ as follows from Eq.~\eqref{Eq:bounds_qi_kj_over_q2}.
This factor is, however, canceled when performing the partial-wave projection of the potential, see Sec.~\ref{Sec:LO_bounds_Swaves}.
We rewrite $V^{(0)}_{C_5}$ as follows:
\beqa
V^{(0)}_{C_5}(\vec p\,',\vec p)&=&\tilde V^{(0)}_{C_5}(\vec p\,',\vec p)
\frac{i}{2}( \vec{\sigma}_1 + \vec{\sigma}_2)\cdot \vec n_\phi/\sin\theta\,,\nonumber\\
\tilde V^{(0)}_{C_5}(\vec p\,',\vec p)&=&C_5\Lambda_5^2
\frac{pp'\sin^2\theta}{q^2+\Lambda_5^2}
\left(\frac{\Lambda_5^2}{q^2+\Lambda_5^2}\right)^{n_5-1}\,,
\label{Eq:V_C5_tilde}
\eeqa
where $\vec n_\phi=(-\sin\phi,\cos\phi,0)$.
It is straightforward to derive, taking into account Eq.~\eqref{Eq:bounds_qi_kj_over_q2},
that $\tilde V^{(0)}_{C_5}(\vec p\,',\vec p)$ obeys the same bounds as other local short-range terms and
\beqa
\left|\tilde V^{(0)}_{C_5}(\vec p\,',\vec p)\right|&\le& \frac{\mathcal{M}_{C_5}}{4\pi}
V_{0,\text{max}}(p',p)
\,,
\left|\tilde V^{(0)}_{C_5}(\vec p\,',\vec p)\right|
\le \frac{\mathcal{M}_{C_5}}{4\pi}V_{0,\text{max}}(p,p')\,,\nonumber\\
 \left|\Delta_p^{(n)} \tilde V^{(0)}_{C_5}(\vec p\,',\vec p)\right| &\le& \frac{\mathcal{M}_{{C_5},n}}{4\pi}
\left|\frac{p}{p'}\right|^{n+1} V_{0,\text{max}}(p',p)\,,
\label{Eq:bound_V_LS_pm}
\eeqa
and symmetrically for $p\leftrightarrow p'$.

\section{Bounds on the plane-wave next-to-leading-order potential}
\subsection{Bounds for the loop function}
The loop functions $L(q)$, $\tilde L(q)$ are defined as 
\begin{align}
&L(q)=\frac{1}{q} \sqrt{4 \mathcal{M}_{\pi}^{2}+q^{2}} \log \frac{\sqrt{4 \mathcal{M}_{\pi}^{2}+q^{2}}+q}{2 \mathcal{M}_{\pi}}\,,\
L(0)=1\,, \nonumber\\
&\tilde L(q)=L(q)-L(0)=
q^2\int_{2M_\pi}^\infty  \frac{\rho_L(\mu)d\mu}{q^2+\mu^2}\,, \quad 
\rho_L(\mu)=\frac{\sqrt{\mu^2-4M_\pi^2}}{\mu^2}\,.
\end{align}

The function $\tilde L(q)$ is bounded by (see Eq.~\eqref{Eq:estimates_denominator1})
\beqa
 |\tilde L(q)|&\le& \mathcal{M}_{\text{den},1}^{-1}\tilde L(|q|) 
=\mathcal{M}_{\text{den},1}^{-1}|q^2|\int_{2M_\pi}^\infty  \frac{\rho_L(\mu)d\mu}{|q^2|+\mu^2}\nonumber\\
&\le& \mathcal{M}_{\text{den},1}^{-1}|q^2|\int_{2M_\pi}^\infty \frac{d\mu}{\mu} \frac{1}{|q^2|+\mu^2}
=\frac{1}{2}\mathcal{M}_{\text{den},1}^{-1}\log\left(1+\frac{|q^2|}{4M_\pi^2}\right)\nonumber\\
&\le& \frac{1}{2}\mathcal{M}_{\text{den},1}^{-1}\log\left(1+\frac{|p|^2+|p'|^2}{4M_\pi^2}\right)
\le \frac{1}{2}\mathcal{M}_{\text{den},1}^{-1}\log\left[\left(1+\frac{|p|^2}{4M_\pi^2}\right)\left(1+\frac{|p'|^2}{4M_\pi^2}\right)\right]\nonumber\\
&=&\frac{1}{2}\mathcal{M}_{\text{den},1}^{-1}\log\left(1+\frac{|p|^2}{4M_\pi^2}\right)
+\frac{1}{2}\mathcal{M}_{\text{den},1}^{-1}\log\left(1+\frac{|p'|^2}{4M_\pi^2}\right)
\le \mathcal{M}_{L,1}f_\text{log}(p',p)\,,
\label{Eq:bound_Lq1}
\eeqa
with
\begin{align}
 f_\text{log}(p',p)=\theta(|p|-M_\pi)\log\frac{|p|}{M_\pi}
+\theta(|p'|-M_\pi)\log\frac{|p'|}{M_\pi}+\log\frac{\tilde\Lambda}{M_\pi}+1\,,
\label{Eq:f_log}
\end{align}
where we have used
\beq
\log\left(1+\frac{|p|^2}{4M_\pi^2}\right)\le \theta(M_\pi-|p|)\log\frac{5}{4}
+\theta(|p|-M_\pi)\log\frac{5|p|^2}{4M_\pi^2}
=\log\frac{5}{4}+2\theta(|p|-M_\pi)\log\frac{|p|}{M_\pi}\,.
\eeq
The bound~\eqref{Eq:bound_Lq1} obviously also remains true if the spectral integral is
regularized by some cutoff $\Lambda_\text{disp}$.
Another bound that follows from Eq.~\eqref{Eq:bound_Lq1} is
\begin{align}
&|\tilde L(q)|\le \mathcal{M}_{\text{den},1}^{-1}\tilde L(|q|)\le 
\mathcal{M}_{L,2}\frac{|p|^2+|p'|^2}{M_\pi^2}\le \mathcal{M}_{L,2}\frac{|p|^2+|p'|^2}{M_\pi^2}f_\text{log}(p',p)\,,
\end{align}
which is a consequence of the inequality $\log(1+|\xi|)\le |\xi|$.

Note that $\tilde L(q)$ does not depend on $\Lambda$, but we 
introduce the $\log\Lambda$ term into function $f_\text{log}(p',p)$ 
for convenience.  Such terms are generated in loop integrals at higher orders in leading-order-potential $V_0$ anyway.

The following remainders for $|p'|>|p|$ can be estimated using 
Eq.~\eqref{Eq:Delta_p_psi},
and the dispersive representation for $\tilde L(q)$:
\beqa
\left|\Delta_p^{(n)} q^2 \tilde L(q)\right| &\le&  \mathcal{M}_{\Delta L}^{2,n}\left|\frac{p}{p'}\right|^{n+1} 
\left(|q^2\tilde L(q)|+|p'^2\tilde L(|p'|, 0)|\right)\nonumber\\
&\le& \frac{\mathcal{M}_{\Delta L,1}}{2}\left|\frac{p}{p'}\right|^{n+1} (|p|^2+|p'|^2)f_\text{log}(p',p)\nn
&\le& \mathcal{M}_{\Delta L,1}\left|\frac{p}{p'}\right|^{n+1} |p'|^2f_\text{log}(p',p)\,,
\label{Eq:Delta_L1}
\left|\Delta_p^{(n)} q_i q_j  \tilde L(q)\right| \nn
&\le& \mathcal{M}_{\Delta L,2}\left|\frac{p}{p'}\right|^{n+1} |p'|^2f_\text{log}(p',p)\,,
\label{Eq:Delta_L2}
\\
\left|\Delta_p^{(n)} M_\pi^2 \tilde L(q)\right|
&\le& \mathcal{M}_{\Delta L,3}\left|\frac{p}{p'}\right|^{n+1} |p'|^2f_\text{log}(p',p)\,,
\label{Eq:Delta_L3}
\\
\left|\Delta_p^{(n)} \frac{M_\pi^4}{M_\pi^2+q^2} \tilde L(q)\right|
&\le& \mathcal{M}_{\Delta L,4}\left|\frac{p}{p'}\right|^{n+1} |p'|^2f_\text{log}(p',p)\,.
\label{Eq:Delta_L4}
\eeqa

\subsection{Bounds on the cutoff corrections}
\label{Sec:Bounds_deltaLambdaV}
In this subsection, we provide  bounds for  $\delta_\Lambda V_0=V_{\Lambda=\infty}^{(0)}-V_{\Lambda}^{(0)}$ and its subtraction remainders. 
Note that the part of $\delta_\Lambda V_0$ corresponding to 
the short range interactions quadratic in momenta (see Eq.~\eqref{Eq:short_range_nonlocal})
\begin{align}
 C_i V_{C_i}\delta_{\Lambda} \left[F_{\Lambda_i, n_i}( p\,', p)\right]\,,\quad i=1,..,7\,,
\end{align}
is of order $\mathcal{O}(Q^4)$
(as follows from Eq.~\eqref{Eq:natural_C_i} and the fact that 
$\delta_{\Lambda} \left[F_{\Lambda_i, n_i}( p\,', p)\right]\sim \frac{q^2}{\Lambda^2}\sim\frac{q^2}{\Lambda_b^2}$,
which will be shown below), and should not
be included at next-to-leading order.
The same is true for the local regulator.
Therefore, we have to consider in $\delta_\Lambda V_0$ 
only the short-range potential with non-derivative contact interactions and the one-pion-exchange potential.
Below, we will show that the cutoff corrections to the leading-order potential and their subtraction remainders
satisfy the following inequalities:
\beqa
 \left|\delta_\Lambda \tilde V^{(0)}_{i,\Lambda}(\vec p\,',\vec p)\right|
 &=&\left|\delta_\Lambda V^{(0)}_{i,\Lambda}(\vec p\,',\vec p)
 -\delta_\Lambda V^{(0)}_{i,\Lambda}(0,0)\right|
 \le \frac{2\pi \mathcal{M}_{\delta\Lambda,i}}{m_N \Lambda_V} \frac{|p|^2+|p'|^2}{\Lambda_{b}^2}\,,\nonumber\\
 \left|\Delta_p^{(n)}\left[\delta_\Lambda V^{(0)}_{i,\Lambda}(\vec p\,',\vec p)\right]\right|
 &\le& \frac{2\pi\mathcal{M}_{\delta\Lambda,i}}{m_N \Lambda_V}
 \left|\frac{p}{p'}\right|^{n+1}\frac{|p'|^2}{\Lambda_{b}^2}\,,\quad \text{ if }|p'|>|p|\,,\nonumber\\
 \left|\Delta_{p'}^{(n)}\left[\delta_\Lambda V^{(0)}_{i,\Lambda}(\vec p\,',\vec p)\right]\right|
 &\le& \frac{2\pi\mathcal{M}_{\delta\Lambda,i}}{m_N \Lambda_V}
 \left|\frac{p'}{p}\right|^{n+1}\frac{|p|^2}{\Lambda_{b}^2}\,,\quad \text{ if }|p|>|p'|\,,
 \label{Eq:Bounds_deltaLambdaVi}
\eeqa
where $V^{(0)}_{i,\Lambda}$ is $V^{(0)}_{\text{short},\Lambda}$ ($C_S$ and $C_T$ terms), 
$V^{(0)}_{1\pi,\text{t},\Lambda}$, or $V^{(0)}_{1\pi,\text{s},\Lambda}$.
The role of $\Lambda_b$ in Eq.~\eqref{Eq:Bounds_deltaLambdaVi} is
played by $\Lambda$ --
the smallest cutoff mass among the ones contained in the LO potential.

From the explicit expression for $\delta_\Lambda F_\Lambda(p)$,
\begin{align}
 \delta_\Lambda F_\Lambda(p)=1-F_\Lambda(p)=\frac{p^2}{p^2+\Lambda^2}\,,
\end{align}
and the bounds~\eqref{Eq:bounds_qi},~\eqref{Eq:estimates_denominator1},
it follows:
\begin{align}
  \left|\delta_\Lambda F_\Lambda(p)\right|\le
\mathcal{M}_{\delta F}\frac{|p^2|+|p'^2|}{\Lambda^2}\,,\quad
  \left|\delta_\Lambda F_\Lambda(q)\right|\le
  \mathcal{M}_{\delta F}\frac{|p^2|+|p'^2|}{\Lambda^2}\,.
\end{align}
Utilizing the product formula~\eqref{Eq:subtractions_product1} (with the role of $p$ played by $1/\Lambda$),
we conclude that
\begin{align}
 \left|\delta_\Lambda F_{\Lambda,n}(p',p)\right|\le
\mathcal{\tilde M}_{\delta F}\frac{|p^2|+|p'^2|}{\Lambda^2}\,,\quad
  \left|\delta_\Lambda F_{q,\Lambda,n}(\vec p\,',\vec p)\right|\le
  \mathcal{\tilde M}_{\delta F}\frac{|p^2|+|p'^2|}{\Lambda^2}\,.
\end{align}
The same bounds for the Gaussian form factors follow from Eq.~\eqref{Eq:Delta_exp}:
\begin{align}
 \left|\delta_\Lambda F_{\Lambda,\text{exp}}(p',p)\right|\le
\mathcal{\tilde M}_{\delta F}\frac{|p^2|+|p'^2|}{\Lambda^2}\,,\quad
  \left|\delta_\Lambda F_{\Lambda,\text{exp}}(q)\right|\le
  \mathcal{\tilde M}_{\delta F}\frac{|p^2|+|p'^2|}{\Lambda^2}\,,
\end{align}
where for the latter inequality, we have used that $\Re(q^2)/\Lambda^2\ge-\mathcal{M}_{\alpha_{\mathcal{C}},1} M_\pi^2/\Lambda^2$ 
(see Eq.~\eqref{Eq:estimates_re_q2}).

From Eqs.~\eqref{Eq:Delta_p_non_local_formfactor},~\eqref{Eq:Derivatives_non_local_formfactor},
we obtain the bounds for the remainders
\begin{align}
 \left| \Delta_p^{(n)}\left[\delta_\Lambda F_{\Lambda,m}(p',p)\right]\right|
= \left| \Delta_p^{(n)} F_{\Lambda,m}(p',p)\right|\le
\mathcal{M}_{\delta F}^{m,n}\left|\frac{p}{p'}\right|^{n+1} \frac{|p'|^2}{\Lambda^2}
\,,\quad \text{ if }|p'|>|p|\,,
\label{Eq:Delta_p_delta_Lambda_F}
\end{align}
and the derivatives
\begin{align}
&\left|p^n \frac{\partial^n \delta_\Lambda F_{\Lambda,m}(p',p)}{{\partial p}^n}\bigg|_{p=0}\right|=
\left|p^n \frac{\partial^n F_{\Lambda,m}(p',p)}{{\partial p}^n}\bigg|_{p=0}\right| \le \mathcal{M}_{\partial \delta F}^{m,n}\left|\frac{p}{p'}\right|^{n} 
 \frac{|p'|^2}{\Lambda^2}\,,
\label{Eq:Derivatives_delta_Lambda_F}
\end{align}
and, symmetrically, for $p\leftrightarrow p'$.
Equation~\eqref{Eq:Delta_p_delta_Lambda_F} is also true  for $n=0$,
and Eq.~\eqref{Eq:Derivatives_delta_Lambda_F} is also true for $n=1$ since 
$\frac{\partial{F_{\Lambda,m}(p',p)}}{\partial{p}}\big|_{p=0}=0$ and
$ \Delta_p^{(0)} F_{\Lambda,m}(p',p)=\Delta_p^{(1)} F_{\Lambda,m}(p',p)$.
For the Gaussian non-local form factor, the same relations for the
remainders and the derivatives hold true due to 
Eqs.~\eqref{Eq:Delta_p_non_local_gauss_formfactor},~\eqref{Eq:Derivatives_non_local_gaussian_formfactor}.

The analogous bounds for the remainders of the local form factor,
\begin{align}
 \left| \Delta_p^{(n)}\left[\delta_\Lambda F_{q,\Lambda,1}(\vec p\,',\vec p)\right]\right|
= \left| \Delta_p^{(n)} F_{q,\Lambda,1}(\vec p\,',\vec p)\right|\le
\mathcal{M}_{\delta F,q}^{1,n}\left|\frac{p}{p'}\right|^{n+1} \frac{|p'|^2}{\Lambda^2}
\,,\quad \text{ if }|p'|>|p|\,,
\label{Eq:Delta_p_delta_Lambda_Fq}
\end{align}
and its derivatives,
\begin{align}
 &\left|p^n \frac{\partial^n \delta_\Lambda F_{q,\Lambda,1}(\vec p\,',\vec p)}{{\partial p}^n}\bigg|_{p=0}\right|
 \le \mathcal{M}_{\partial \delta F,q}^{1,n}\left|\frac{p}{p'}\right|^{n} 
 \frac{|p'|^2}{\Lambda^2}\,,
\label{Eq:Derivatives_delta_Lambda_Fq}
\end{align}
and, symmetrically, for $p\leftrightarrow p'$,
can be obtained by a slight modification of 
the proof of Eqs.~\eqref{Eq:Delta_p_f},~\eqref{Eq:derivatives_f} in Sec.~\ref{Sec:LocalFF_bounds}
(replacing one factor of $\left|p'^2/(p'^2+\Lambda^2)\right|$ not by $1$ but rather by
$|p'|^2/\Lambda^2$).
The correctness of Eq.~\eqref{Eq:Delta_p_delta_Lambda_Fq} for $n=0$
and Eq.~\eqref{Eq:Derivatives_delta_Lambda_Fq} for $n=1$
follows from the inequalities
\beqa
\left| \Delta_p^{(0)} F_{q,\Lambda,1}(\vec p\,',\vec p)\right|
&=&\left|
\frac{\Lambda ^2 p (2 p' x-p)}{\left(p'^2+\Lambda^2\right) \left(q^2+\Lambda^2\right)}\right|
\le 3\mathcal{M}_{\text{den},1}^{-1}\frac{|p||p'|}{\Lambda^2}
\,,\quad \text{ if }|p'|>|p|\,,\nonumber\\
 \left|p\frac{\partial\left[\delta_\Lambda F_{q,\Lambda,1}(\vec p\,',\vec p)\right]}{\partial p}\bigg|_{p=0}\right|
&=&\left|\frac{2pp'x\Lambda^2}{\left(p'\,^2+\Lambda^2\right)^2}\right|\le \frac{2|p||p'|}{\Lambda^2}\,.
\eeqa

The corresponding bounds for $F_{q,\Lambda,m}(\vec p\,',\vec p)$ 
for $m>1$ (they have the same form as in Eqs~\eqref{Eq:Delta_p_delta_Lambda_Fq},~\eqref{Eq:Derivatives_delta_Lambda_Fq})
can be easily derived using the product formula~\eqref{Eq:subtractions_product1}.

The same inequalities for the remainders and the derivatives of the Gaussian local form factor
were derived in Sec.~\ref{Sec:Gaussian_Fq}, see Eqs.~\eqref{Eq:Delta_p_local_gauss_formfactor_b},~\eqref{Eq:Derivatives_local_gaussian_formfactor2}.

Thus, we conclude that the cutoff corrections to the short-range leading order potential
satisfy the inequalities~\eqref{Eq:Bounds_deltaLambdaVi} (see Eq.~\eqref{Eq:natural_C_S_C_T})
for all types of the regulator.

Now, consider the cutoff corrections to the leading-order one-pion-exchange potential.
The bounds on the unregulated singlet and triplet one-pion-exchange potential, its subtraction remainders
and its derivatives can be obtained from the general bounds
 on the leading-order potential 
 (Eqs.~\eqref{Eq:bound_full_LO},~\eqref{Eq:Delta_p_V0_constrained},~\eqref{Eq:V0_derivatives_b})
 replacing $F_{\tilde\Lambda}\to 1$:
\beqa
\left|V_{1\pi,i}(\vec p\,',\vec p)\right|
&\le& \frac{2\pi \mathcal{M}_i}{m_N \Lambda_V}\,,\nonumber\\
\left|\Delta_p^{(n)} V_{1\pi,i}(\vec p\,',\vec p)\right| &\le& \frac{2\pi \mathcal{M}_{\Delta V_i,n}}{m_N \Lambda_V}
\left|\frac{p}{p'}\right|^{n+1}\,, \quad \text{ if } |p'|>|p|\,,\nonumber\\
\left|p^n\frac{\partial^n V_0(\vec p\,',\vec p)}{\partial p^n}\bigg|_{p=0}\right|
&\le& \frac{2\pi \mathcal{M}_{\partial V_i,n}}{m_N \Lambda_V}
\mathcal{M}_{\partial V_0,n}\left|\frac{p}{p'}\right|^n\,,\quad
i=\text{s}\,,\text{t}\,,
\eeqa
and, symmetrically, for $p\leftrightarrow p'$.

Combining these bounds with Eqs.~\eqref{Eq:Delta_p_delta_Lambda_F},~\eqref{Eq:Derivatives_delta_Lambda_F}
and applying the product formula~\eqref{Eq:subtractions_product1}, we derive the inequalities~\eqref{Eq:Bounds_deltaLambdaVi}
for the case of the non-local regulator of the one-pion-exchange potential including the Gaussian regulator.

For the local regulator, a minor modification is needed due to the fact that 
$F_{q,1\pi,\{\Lambda_k\}}(0,0)\ne 1$ as follows from Eq.~\eqref {Eq:local_formfactor_1pi}.
In this case, we split $\delta_\Lambda F_{q,1\pi,\{\Lambda_k\}}(\vec p\,',\vec p)$ into
\beqa
\delta_\Lambda F_{q,1\pi,\{\Lambda_k\}}(\vec p\,',\vec p)&=&
 \delta_\Lambda \tilde F_{q,1\pi,\{\Lambda_k\}}(\vec p\,',\vec p)
 +\delta_\Lambda F_{q,1\pi,\{\Lambda_k\}}(0,0)\,,\nonumber\\
\delta_\Lambda \tilde F_{q,1\pi,\{\Lambda_k\}}(\vec p\,',\vec p)
&=&-\tilde F_{q,1\pi,\{\Lambda_k\}}(\vec p\,',\vec p)
=F_{q,1\pi,\{\Lambda_k\}}(0,0)-F_{q,1\pi,\{\Lambda_k\}}(\vec p\,',\vec p)\,,\nonumber\\
\delta_\Lambda F_{q,1\pi,\{\Lambda_k\}}(0,0)&=&1-F_{q,1\pi,\{\Lambda_k\}}(0,0)\,.
\eeqa
Thus,
\begin{align}
 \delta_\Lambda \tilde V_{1\pi,i,\Lambda}(\vec p\,',\vec p)=V_{1\pi,i}(\vec p\,',\vec p)
 \delta_\Lambda \tilde F_{q,1\pi,\{\Lambda_k\}}(\vec p\,',\vec p)+
 \tilde V_{1\pi,i}(\vec p\,',\vec p)\delta_\Lambda F_{q,1\pi,\{\Lambda_k\}}(0,0)\,,\quad i=\text{s}\,,\text{t}\,,
\end{align}
with $\tilde V_{1\pi,i}(\vec p\,',\vec p)=V_{1\pi,i}(\vec p\,',\vec p)-\tilde V_{1\pi,i}(0,0)$.

Since $\delta_\Lambda \tilde F_{q,1\pi,\{\Lambda_k\}}(\vec p\,',\vec p)$
obeys the same constraints as $\delta_\Lambda F_{q,\Lambda,n}(\vec p\,',\vec p)$,
applying the product formula~\eqref{Eq:subtractions_product1} leads to
Eq.~\eqref{Eq:Bounds_deltaLambdaVi} for 
the part $V_{1\pi,i}(\vec p\,',\vec p)\delta_\Lambda \tilde F_{q,1\pi,\{\Lambda_k\}}(\vec p\,',\vec p)$
of $\delta_\Lambda \tilde V_{1\pi,i,\Lambda}(\vec p\,',\vec p)$.
The generalization to the local Gaussian regulator is obvious because
\begin{align}
 F_{q,1\pi,\text{exp},\Lambda}(\vec p\,',\vec p)=\exp(-M_\pi^2/\Lambda^2)F_{\Lambda,\text{exp}}(q)\,, \quad
\exp(-M_\pi^2/\Lambda^2)<1\,,
\end{align}
and the case of $F_{\Lambda,\text{exp}}(q)$ was considered above.

In order to estimate the remaining part of $\delta_\Lambda \tilde V_{1\pi,i,\Lambda}(\vec p\,',\vec p)$,
namely, $\tilde V_{1\pi,i}(\vec p\,',\vec p)\delta_\Lambda F_{q,1\pi,\{\Lambda_k\}}(0,0)$,
we first note that $\delta_\Lambda F_{q,1\pi,\{\Lambda_k\}}(0,0)$
is a constant bounded by
\begin{align}
 \left| \delta_\Lambda F_{q,1\pi,\{\Lambda_k\}}(0,0) \right|
 \le\mathcal{M}_{\delta\Lambda,0,0}\frac{M_\pi^2}{\Lambda^2}\,,
\end{align}
which follows from the explicit form of the remainder
\begin{align}
  \delta_\Lambda F_{q,1\pi,\Lambda}(0,0)=\frac{M_\pi^2}{\Lambda^2}\,,
\end{align}
and the product formula~\eqref{Eq:subtractions_product1} (with the role of $p$ played by $M_\pi^2$).
For the Gaussian regulator, the analogous bound 
\begin{align}
 \left| \delta_\Lambda F_{\Lambda,\text{exp}}(q=0) \right|=
 \left|1- \exp(-M_\pi^2/\Lambda^2)\right|
 \le\mathcal{M}_{\delta\Lambda,0,0}\frac{M_\pi^2}{\Lambda^2}\,,
\end{align}
follows from Eq.~\eqref{Eq:Delta_exp}.

Therefore, we get (see Eqs.~\eqref{Eq:V_1pi_0},~\eqref{Eq:V_1pi},~\eqref{Eq:bound_triplet_1pi_local},~\eqref{Eq:bound_singlet_1pi}):
\begin{align}
 \tilde V_{1\pi,i}(\vec p\,',\vec p)\delta_\Lambda F_{q,1\pi,\{\Lambda_k\}}(0,0)=
 \sum_{i,j}\frac{2\pi\mathcal{M}_{i,jk}}{m_N\Lambda_V}
 \frac{q_jq_k}{\Lambda^2}\frac{M_\pi^2}{q^2+M_\pi^2}\,.
\end{align}
Equation~\eqref{Eq:Bounds_deltaLambdaVi} for 
$\tilde V_{1\pi,i}(\vec p\,',\vec p)\delta_\Lambda F_{q,1\pi,\{\Lambda_k\}}(0,0)$ follows immediately after implementing Eqs.\eqref{Eq:Delta_p_Psi},~\eqref{Eq:derivatives_Psi},~\eqref{Eq:estimates_denominator1}.

\subsection{Bounds on the full next-to-leading-order potential}
\label{Sec:Bounds_NLO}
We split the NLO potential into two parts:
\begin{align}
 V_2(\vec p\,',\vec p)=\hat V_2(\vec p\,',\vec p)+\tilde V_2(\vec p\,',\vec p)\,,
 \label{Eq:V2_tilde}
\end{align}
with 
\begin{align}
\hat V_2(\vec p\,',\vec p)=V_2(0,0)\,,\quad \tilde V_2(\vec p\,',\vec p)=V_2(\vec p\,',\vec p)-V_2(0,0)\,.
\end{align}

The constant term $\hat V_2(\vec p\,',\vec p)$ is determined by the structures 
$C_{2,S} V_{C_S}+C_{2,T} V_{C_T}$ (see Eq.~\eqref{Eq:V2_short_range}) and is bounded by
\begin{align}
 \left|\hat V_2(\vec p\,',\vec p)\right|\le \mathcal{\hat M}_{V_2} \frac{2\pi}{m_N \Lambda_V}\frac{M_\pi^2}{\Lambda_{b}^2}\,,
 \label{Eq:bound_V_2_0_hat_plane_wave}
\end{align}
under the naturalness assumption for the renormalized LECs:
\begin{align}
 C_{2,S}\sim C_{2,T}\sim  \frac{2\pi}{m_N \Lambda_V}\frac{m_\pi^2}{\Lambda_{b}^2}\,.
\end{align}

The full subtracted next-to-leading-order potential $\tilde V_2(\vec p\,',\vec p)$
consists of the terms proportional to the following structures:
\begin{align}
&Q_2(p',p)\,, \quad M_\pi^2\tilde L(q)\,, \quad q^2\tilde L(q)\,, \quad q_iq_j\tilde L(q)\,,
\quad \frac{M_\pi^4\tilde L(q)}{q^2+M_\pi^2}\,,\quad \frac{M_\pi^2 q^2}{q^2+M_\pi^2}\,,\quad V_0-V_0^{\Lambda}\,,
\end{align}
where $Q_2(p',p)$ is a homogeneous polynomial of order $2$.

Thus, we obtain the following bounds using the results of the previous subsection:
\begin{align}
& \left|\tilde V_2(\vec p\,',\vec p) \right|
\le  \frac{2\pi \mathcal{M}_{V_2}}{m_N \Lambda_V} \frac{|p|^2+|p'|^2}{\Lambda_{b}^2}f_\text{log}(p',p)
=\frac{\mathcal{M}_{V_2}}{4\pi}\left(|p|^2+|p'|^2\right)\tilde f_\text{log}(p',p)\,,
 \label{Eq:bounds_V2_full}
\end{align}
with
\begin{align}
\tilde f_\text{log}(p',p)=\frac{8\pi^2}{m_N \Lambda_V \Lambda_{b}^2} f_\text{log}(p',p)\,,
\label{Eq:f_log_tilde}
\end{align}
where $\Lambda_{b}$ is the chiral symmetry breaking scale.
In particular, we treat
\begin{align}
&4\pi F_\pi \sim \Lambda_{b}\,,\quad C_{2,i}\sim  \frac{2\pi}{m_N \Lambda_V}\frac{1}{\Lambda_{b}^2}\,, \quad i=1,..,7\,.
\label{Eq:Lambda_b}
\end{align}

Analogously, for the subtraction remainders, we get
\begin{align}
 &\left|\Delta_p^{(n)}\tilde V_2(\vec p\,',\vec p)\right|\le \frac{\mathcal{M}_{V_2,n}}{4\pi}\left|\frac{p}{p'}\right|^{n+1}|p'|^2 \tilde f_\text{log}(p',p)
 \,,\quad\text{ if } |p'|>|p|\,,\nonumber\\
  &\left|\Delta_{p'}^{(n)}\tilde V_2(\vec p\,',\vec p) \right|\le \frac{\mathcal{M}_{V_2,n}}{4\pi}
\left|\frac{p'}{p}\right|^{n+1}|p|^2 \tilde f_\text{log}(p',p)
 \,,\quad \text{ if } |p|>|p'|\,.
\label{Eq:bounds_Delta_V2}
\end{align}

Multiplying the NLO potential by a local or a non-local form factor
does not modify the obtained bounds as follows from the results of Secs.~\ref{Sec:nonlocal_formfactor_bounds},~\ref{Sec:LocalFF_bounds}.

\section{Bounds on the partial-wave potentials}
\label{Sec:PW_bounds}
The partial-wave potential is obtained from the plain-wave potential via~\cite{Erkelenz:1971caz}
\begin{align}
 &V_{l', l}^{s j}(p', p) = \sum_{ \lambda_{1}, \lambda_{2},\lambda_{1}^{\prime}, \lambda_{2}^{\prime}}
\int d\Omega\
\langle jl's|\lambda_{1}^{\prime} \lambda_{2}^{\prime}\rangle
\langle \lambda_{1}^{\prime} \lambda_{2}^{\prime}|V(\vec p\,',\vec p)| \lambda_{1} \lambda_{2}\rangle
\langle \lambda_{1} \lambda_{2}|jls \rangle\,
d_{\lambda_{1}-\lambda_{2}, \lambda_{1}^{\prime}- \lambda_{2}^{\prime}}^{j}(\theta)\,,\nonumber\\
&\langle \lambda_{1} \lambda_{2}|jls\rangle
=\left(\frac{2 l+1}{2 j+1}\right)^{\frac{1}{2}} C(l\,, s\,, j ; 0\,, \lambda_{1}-\lambda_{2}) 
C\left(1/2\,, 1/2\,, s ; \lambda_{1},-\lambda_{2}\right)\,.
\end{align}
Due to unitarity of the transformation, the following 
constraints hold:
\begin{align}
& |\langle \lambda_{1} \lambda_{2}|jls\rangle|\le 1\,,\quad
|\langle 1/2\,,s_z|\lambda\rangle|\le 1\,,\quad
|d_{\lambda, \lambda^{\prime}}^{j}(\theta)|\le 1\,.
 \nonumber\\
\end{align}
Therefore, if the plain-wave potential (each matrix element in the spin-isospin space)
is bounded by some function $f(p',p)$,
\begin{align}
&|V(\vec p\,',\vec p)|\le \mathcal{M}_V f(p',p)\,,
\end{align}
where $\mathcal{M}_V$ is some constant of order one,
we can claim that the partial wave potential is bounded by some
other constant $\mathcal{\tilde M}_V$ of order one:
\begin{align}
&|V_{l', l}^{s j}(p', p)|\le 4\pi \mathcal{\tilde M}_V f(p',p)\,,
\label{Eq:bound_PW}
\end{align}
where we have introduced, for convenience,
a factor of $4\pi$ coming from the angular integration.

For the special case of the locally regulated spin-orbit contact interaction (see Eq.~\eqref{Eq:V_C5_tilde}),
the explicit partial-wave-projection angular integral can be written as
\begin{align}
 \left(V^{(0)}_{C_5}\right)_{l, l}^{1 j}(p', p) =\, 2 \pi \int_{-1}^{1} d x   \tilde V^{(0)}_{C_5}(\vec p\,',\vec p)
 a_{j} P^1_l(x)\,,
 \end{align}
with
\begin{align}
 a_j=\left\{\begin{array}{ll}\frac{1}{j+1}\,,&j=l-1\\[6pt]
 \frac{1}{j(j+1)}\,,&j=l\\[6pt]
 -\frac{1}{j}\,,&j=l+1\end{array}\right.
 \label{Eq:a_j}
 \end{align}
 and the associated Legendre polynomials $ P^1_l(x)$, which are bounded by \cite{LOHOFER1998178}
 \begin{align}
  \left|P^1_l(x)\right|\le\sqrt{\frac{l(l+1)}{2}}\,, \quad\text{ for }-1\le x\le 1\,.
  \label{Eq:P_1_l}
 \end{align}
The factor $1/\sin\theta$ in Eq.~\eqref{Eq:V_C5_tilde} disappears after the
partial-wave projection. 
Moreover, as follows from Eqs.~\eqref{Eq:a_j},~\eqref{Eq:P_1_l}, the product $a_j P^1_l(x)$
is bounded by a constant of order one:
\begin{align}
 \left|a_j P^1_l(x)\right|\le\mathcal{M}_{l,j}<1\,, \quad\text{ for }-1\le x\le 1\,.
\end{align}
Therefore, the bounds given in Eq.~\eqref{Eq:bound_V_LS_pm}
allow one to treat this type of a contact interaction on the same footing as all other short-range terms.

\subsection{Bounds on the partial-wave leading-order potential}
\label{Sec:LO_bounds}
\subsubsection{$S$-wave}                                                                                                            
\label{Sec:LO_bounds_Swaves}
Using the bounds for the plane-wave leading-order potential
in Eq.~\eqref{Eq:bound_full_LO}
and performing the partial-wave projection according to Eq.~\eqref{Eq:bound_PW}, we obtain
for $l=0$ (for the coupled partial waves, we mean by $l$ the lowest orbital angular momentum):                                                                                                                                                      
\begin{align}
&\left|V_0(p',p)\right|< \mathcal{M}_{V_0,0}
V_{0,\text{max}}(p',p)\,,\quad
\left|V_0(p',p)\right|< \mathcal{M}_{V_0,0}
V_{0,\text{max}}(p,p')\,.
\label{Eq:bounds_V0_l_0}
\end{align}
The above inequalities are meant to hold for all matrix elements
of  $V_0(p',p)$ in $l\,,l'$ space.

From parity conservation, it follows that the first derivative 
in $p$ ($p'$) of $V_0(p',p)$ at $p=0$ ($p'=0$) vanishes, and
\begin{align}
\Delta_p V_0(p',p)\equiv \Delta_p^{(0)} V_0(p',p)
=\Delta_p^{(1)} V_0(p',p)\,,
\end{align}
and, analogously, for $\Delta_{p'} V_0(p',p)$.

Therefore, for the subtraction remainders, we obtain from 
Eq.~\eqref{Eq:Delta_p_V0}
\beqa
\left|\Delta_p V_0(p',p)\right|&\le&
\mathcal{M}_{V_0,0}\left|\frac{p}{p'}\right|^2V_{0,\text{max}}(p',p)\,
\,,\label{Eq:bounds_Delta_V0_l_0a}\\
\left|\Delta_{p'} V_0(p',p)\right|&\le&
\mathcal{M}_{V_0,0}\left|\frac{p'}{p}\right|^2V_{0,\text{max}}(p,p')\,
\,.
\label{Eq:bounds_Delta_V0_l_0b}
\eeqa
On the other hand, from Eqs.~\eqref{Eq:bounds_Delta_V0_l_0a},~\eqref{Eq:bounds_Delta_V0_l_0b}, it  also follows
\begin{align}
&\left|\Delta_p V_0(p',p)\right|\le
\mathcal{M}_{V_0,0}V_{0,\text{max}}(p',p)\,
\,,\quad
\left|\Delta_{p'} V_0(p',p)\right|\le
\mathcal{M}_{V_0,0}V_{0,\text{max}}(p,p')\,
\,.
\label{Eq:bounds_Delta_V0_l_0_2}
\end{align}
For simplicity, we use the same constant $\mathcal{M}_{V_0,0}$ in Eqs.~\eqref{Eq:bounds_V0_l_0},  \eqref{Eq:bounds_Delta_V0_l_0a}, \eqref{Eq:bounds_Delta_V0_l_0b}, which can be chosen to be the largest of the three.

\subsubsection{Higher partial waves}
\label{Sec:bounds_LO_higher_PW}
For $l>0$,
we can use the fact that for $m<l$,
\begin{align}
&\frac{\partial^m V_0(p\,',p)}{(\partial p)^m}\bigg|_{p=0}=
\frac{\partial^m V_0(p\,',p)}{(\partial p')^m}\bigg|_{p'=0}=0\,,
\end{align}
and derive
after performing the partial-wave projection according to Eq.~\eqref{Eq:bound_PW}
\begin{align}
&\left| V_0(p',p)\right|\le \mathcal{M}_{V_0,\tilde l}
\left|\frac{p}{p'}\right|^{\tilde l}V_{0,\text{max}}(p',p) \,
\,,
\label{Eq:bounds_V0_la}\\
&\left| V_0(p',p)\right|\le \mathcal{M}_{V_0,\tilde l}
\left|\frac{p'}{p}\right|^{\tilde l}V_{0,\text{max}}(p,p') \,
\,,
\label{Eq:bounds_V0_lb}
\end{align}
with $0\le \tilde l \le l$.

\subsection{Bounds on the partial-wave next-to-leading-order potential}
\label{Sec:NLO_bounds}
\subsubsection{$S$-wave}
Using the bounds for the plane-wave leading-order potential
in Eqs.~\eqref{Eq:bound_V_2_0_hat_plane_wave},~\eqref{Eq:bounds_V2_full}
and performing the partial-wave projection according to Eq.~\eqref{Eq:bound_PW}, we obtain
for $l=0$: 
\begin{align}
 \left|\hat V_2(p', p)\right|\le \mathcal{\hat M}_{V_2,0} \frac{8\pi^2 }{m_N \Lambda_V}\frac{M_\pi^2}{\Lambda_{b}^2}\,,
 \label{Eq:bound_V_2_0_hat}
\end{align}
and
\begin{align}
&\left|\tilde V_2(p',p)\right|\le \mathcal{M}_{V_2,0}
\left(|p|^2+|p'|^2\right)\tilde f_\text{log}(p',p) \,.
\label{Eq:bounds_V2_l_0}
\end{align}

As in the case of the leading-order potential,
it follows from parity conservation that 
\begin{align}
\Delta_p \tilde V_2(p',p)\equiv \Delta_p^{(0)} \tilde V_2(p',p)
=\Delta_p^{(1)} \tilde V_2(p',p)\,,
\end{align}
and, analogously, for $\Delta_{p'} \tilde V_2(p',p)$.
Therefore, for the subtraction remainders, we obtain
\begin{align}
&\left|\Delta_p\tilde V_2(p',p)\right|\le \mathcal{M}_{V_2,0}
|p|^2\tilde f_\text{log}(p',p) \,,
\label{Eq:bounds_Delta_V2_l_0a}\\
&\left|\Delta_{p'}\tilde V_2(p',p)\right|\le \mathcal{M}_{V_2,0}
|p'|^2\tilde f_\text{log}(p',p) \,.
\label{Eq:bounds_Delta_V2_l_0b}
\end{align}
Equation~\eqref{Eq:bounds_Delta_V2_l_0a} (Equation~\eqref{Eq:bounds_Delta_V2_l_0b}) follows from 
Eq.~\eqref{Eq:bounds_Delta_V2} for $|p'|>|p|$ ($|p|>|p'|$) and from
Eq.~\eqref{Eq:bounds_V2_l_0} for $|p|>|p'|$ ($|p'|>|p|$).

\subsubsection{Higher partial waves}
For $l>0$, we get an analog of Eq.~\eqref{Eq:bounds_V2_l_0}:
\begin{align}
&\left|\tilde V_2(p',p)\right|\le \mathcal{M}_{V_2,l}
\left(|p|^2+|p'|^2\right)\tilde f_\text{log}(p',p) \,.
\label{Eq:bounds_V2_l}
\end{align}
Moreover, as in the case of the leading-order potential (see Sec.\ref{Sec:bounds_LO_higher_PW}),
we can use vanishing of the derivatives at $p=0$ and $p'=0$
to derive
\begin{align}
&\left|\tilde V_2(p',p)\right|\le \mathcal{M}_{V_2,\tilde l}
\left|\frac{p}{p'}\right|^{\tilde l}|p'|^2\tilde f_\text{log}(p',p) \,,
\label{Eq:bounds_V2_l_1a}\\
&\left|\tilde V_2(p',p)\right|\le \mathcal{M}_{V_2,\tilde l}
\left|\frac{p'}{p}\right|^{\tilde l}|p|^2\tilde f_\text{log}(p',p) \,,
\label{Eq:bounds_V2_l_1b}
\end{align}
where $0\le\tilde l \le l$.

For $\tilde l=1$, both above equations are the same and give
\begin{align}
&\left|\tilde V_2(p',p)\right|\le \mathcal{M}_{V_2,1}
|p'||p|\tilde f_\text{log}(p',p) \,,
\label{Eq:bounds_V2_l_2}
\end{align}
and follow from Eq.~\eqref{Eq:bounds_Delta_V2} for both cases $|p|>|p'|$ and $|p'|>|p|$.
For $\tilde l\ge 2$, Eq.~\eqref{Eq:bounds_V2_l_1a} (Eq.~\eqref{Eq:bounds_V2_l_1b}) 
follows from Eq.~\eqref{Eq:bounds_Delta_V2} for $|p'|>|p|$ ($|p|>|p'|$) and from
Eq.~\eqref{Eq:bounds_V2_l} for $|p|>|p'|$ ($|p'|>|p|$).

\section{Bounds on typical integrals}
\label{Sec:integrals}
The typical integral $I_{f_\text{log}}$ that appears in our calculations
\begin{align}
 &I_{f_\text{log}}(p',p{;p_\text{on}})=\int \frac{|p''|^2 d|p''|}{(2\pi)^3} \tilde f_\text{log}(p',p'') 
\left|G(p''{;p_\text{on}}) \right|
 V_{0,\text{max}}(p'',p)\,,
\end{align}
is bounded by
\begin{align}
 I_{f_\text{log}}(p',p{;p_\text{on}})\le \mathcal{M}_{f_\text{log}}\frac{{\tilde \Lambda}}{\Lambda_V}
 \tilde f_\text{log}(p',p) \,,
 \label{Eq:bound_I_f_log}
\end{align}
which we will prove below.

First, we write using the definition of $V_{0,\text{max}}$ (Eq.~\eqref{Eq:V0max}) and the  bound on the two-nucleon propagator (Eq.~\eqref{Eq:bound_on_G}):
\begin{align}
 I_{f_\text{log}}(p',p{;p_\text{on}})&= 
 \tilde I_{f_\text{log}}(p',p{;p_\text{on}})
 + \tilde I_{f_\text{log}}(p',0{;p_\text{on}})\,,\nonumber\\
 \tilde I_{f_\text{log}}(p',p{;p_\text{on}})&=
 \frac{8\pi^2 }{m_N \Lambda_V}\int \frac{|p''|^2 d|p''|}{(2\pi)^3} \tilde f_\text{log}(p',p'') \left|G(p''{;p_\text{on}}) \right|
 F_{\tilde\Lambda}(|p''|-|p|)\nonumber\\
 &\le \frac{\mathcal{M}_G}{ \Lambda_V}\int \frac{d|p''|}{\pi} \tilde f_\text{log}(p',p'') 
 \frac{\tilde\Lambda^2}{(|p|-|p''|)^2+\tilde\Lambda^2}\,.
 \end{align}
The last integral is a linear combination of the integrals
(see the definition of $\tilde f_\text{log}$ in 
Eqs.~\eqref{Eq:f_log},~\eqref{Eq:f_log_tilde})
\begin{align}
&I_0(p,\tilde\Lambda)=\int \frac{d|p''|}{\pi} \frac{\tilde\Lambda^2}{(|p|-|p''|)^2+\tilde\Lambda^2}\,,\nonumber\\
&I_\text{log}(p,\tilde\Lambda)=\int \frac{d|p''|}{\pi}\log\frac{|p''|}{M_\pi}\theta(|p''|-M_\pi)
 \frac{\tilde\Lambda^2}{(p-p'')^2+\tilde\Lambda^2}\,,
\label{Eq:integrals_0_log}
\end{align}
for which the following bounds hold:
\begin{align}
&I_0(p,\tilde\Lambda)
\le \tilde\Lambda\,,
\label{Eq:bound_I0}\\
&I_\text{log}(p,\tilde\Lambda)\le \mathcal{M}_{\text{log}}\tilde\Lambda\left(\log\frac{|p|}{M_\pi}\theta(|p|-M_\pi)
+\log\frac{\tilde\Lambda}{M_\pi}+1\right)\,.
\label{Eq:bound_Ilog}
\end{align}
The bound for $I_0$ is trivial. To derive the bound for $I_\text{log}$,
we split the integration region in Eq.~\eqref{Eq:integrals_0_log} into two parts
$|p''|<a=2\max(p,\tilde\Lambda)$ and $|p''|>a$:
\beqa
I_\text{log}(p,\tilde\Lambda)
&\le& \log\frac{a}{M_\pi}\int_0^a d|p''|\frac{\tilde\Lambda^{2}}{(|p''|-|p|)^2+\tilde\Lambda^2}
+\int_a^\infty d|p''|\frac{4\tilde\Lambda^2}{|p''|^{2}}\log\frac{|p''|}{M_\pi}\nonumber\\
&\le& \pi\tilde\Lambda \log\frac{a}{M_\pi}
+4\tilde\Lambda\left(\log\frac{a}{M_\pi}+1\right)\nn
&\le&\mathcal{\tilde M}_{\text{log}}\tilde\Lambda\left(\log\frac{|p|}{M_\pi}\theta(|p|-M_\pi)
+\log\frac{\tilde\Lambda}{M_\pi}+1\right)\,,
\label{Eq:I_log_p}
\eeqa
where we have used 
\begin{align}
 &\log\frac{\max(|p|,\tilde\Lambda)}{M_\pi}=\log\frac{\tilde\Lambda}{M_\pi}\theta(\tilde\Lambda-|p|)
+\log\frac{|p|}{M_\pi}\theta(|p|-\tilde\Lambda)
\le\log\frac{\tilde\Lambda}{M_\pi}
+\log\frac{|p|}{M_\pi}\theta(|p|-M_\pi)\,,
\end{align}
and
\begin{align}
 1/\max(p,\tilde\Lambda)\le 1/\tilde\Lambda\,.
\end{align}
Eq.~\eqref{Eq:bound_I_f_log} follows immediately from Eqs.\eqref{Eq:bound_I0},~\eqref{Eq:bound_Ilog}.

In the same manner, one can estimate integral ${I_{V_0}}$ that enters the evaluation of the leading order amplitude,
\begin{align}
&I_{V_0}(p',p;p_\text{on})=
\int \frac{|p''|^2 d|p''|}{(2\pi)^3}
\left|G(p'';p_\text{on})\right|V_{0,\text{max}}(p'',p)
\le \mathcal{M}_{I_{V_0}}
\frac{\tilde \Lambda}{\Lambda_V}\,.
\end{align}

\bibliography{4.3}
\bibliographystyle{apsrev}

\end{document}